\documentclass[a4paper,aps,prb,superscriptaddress, twocolumn,preprintnumbers,amsmath,amssymb,nobalancelastpage,10pt,longbibliography]{revtex4-2}

\usepackage{amsmath}
\usepackage{bm}
\usepackage{booktabs}
\usepackage{braket}
\usepackage{capt-of}
\usepackage{color}
\usepackage{dcolumn}
\usepackage{dsfont}
\usepackage{float}
\usepackage{graphicx}
\usepackage[colorlinks=true,citecolor=blue,linkcolor=blue,urlcolor=blue]{hyperref}
\usepackage[utf8]{inputenc}
\usepackage{mhchem}
\usepackage{soul}
\usepackage{subfigure}
\usepackage{varwidth}
\usepackage{verbatim}
\usepackage[dvipsnames]{xcolor}
\usepackage{yfonts}
\usepackage{physics}
\usepackage{color}

\newcommand{\cop}{\hat{c}}
\newcommand{\cdop}{\hat{c}^\dag}
\newcommand{\ci}{\cop_{i}}
\newcommand{\cdi}{\cdop_{i}}
\newcommand{\cj}{\cop_{j}}
\newcommand{\cdj}{\cdop_{j}}
\newcommand{\cda}{\cdop_{\alpha}}
\newcommand{\cb}{\cop_{\beta}}
\newcommand{\alatt}{a_{\mathrm{latt}}}

\newcommand{\epl}{E_p}
\newcommand{\geff}{\Gamma_{\textrm{eff}}}
\newcommand{\hP}{\hat{P}}
\newcommand{\hQ}{\hat{Q}}
\newcommand{\hU}{\hat{U}}
\newcommand{\hG}{\hat{G}}
\newcommand{\hR}{\hat{R}}
\newcommand{\hc}{\mathrm{H.c.}}
\newcommand{\Hop}{\hat{H}}
\newcommand{\Hzero}{\Hop_0}
\newcommand{\HHF}{\Hop_{\textrm{HF}}}
\newcommand{\im}{\mathrm{i}}
\newcommand{\kv}{\mathbf{k}}
\newcommand{\nop}{\hat{n}}
\newcommand{\nopi}{\nop_{i}}
\newcommand{\nopj}{\nop_{j}}

\newcommand{\VRyd}{V}
\newcommand{\dVRyd}{V(r)}
\newcommand{\Vgen}{\tilde{V}}

\newcommand{\Hvv}{\Hop_{\mathrm{VV}}}
\newcommand{\uvw}{U_{\mathrm{vdW}}}
\newcommand{\uvwr}{\uvw(r)}
\newcommand{\Gz}{\hG^{(0)}}
\newcommand{\Gu}{\hG^{(1 )}}
\newcommand{\Gd}{\hG^{(2 )}}
\newcommand{\Gt}{\hG^{(3 )}}
\newcommand{\Gq}{\hG^{(4 )}}
\newcommand{\dis}[1]{d_{#1}}

\begin{document} 

\title{Accessing the topological Mott insulator in cold atom quantum simulators with realistic Rydberg dressing}
\author{Lorenzo Cardarelli}
\affiliation{Peter Gr\"unberg Institute, Theoretical Nanoelectronics, Forschungszentrum J\"ulich, D-52428 J\"ulich, Germany}
\affiliation{Institute for Quantum Information, RWTH Aachen University, D-52056 Aachen, Germany}
\author{Sergi Juli\`a-Farr\'e}
\affiliation{ICFO - Institut de Ciencies Fotoniques, The Barcelona Institute of Science and Technology, Av. Carl Friedrich Gauss 3, 08860 Castelldefels (Barcelona), Spain}
\author{Maciej Lewenstein}
\affiliation{ICFO - Institut de Ciencies Fotoniques, The Barcelona Institute of Science and Technology, Av. Carl Friedrich Gauss 3, 08860 Castelldefels (Barcelona), Spain}
\affiliation{ICREA, Pg. Lluis Companys 23, 08010 Barcelona, Spain}
\author{Alexandre Dauphin}
\affiliation{ICFO - Institut de Ciencies Fotoniques, The Barcelona Institute of Science and Technology, Av. Carl Friedrich Gauss 3, 08860 Castelldefels (Barcelona), Spain}
\author{Markus M\"uller}
\affiliation{Peter Gr\"unberg Institute, Theoretical Nanoelectronics, Forschungszentrum J\"ulich, D-52428 J\"ulich, Germany}
\affiliation{Institute for Quantum Information, RWTH Aachen University, D-52056 Aachen, Germany}

\begin{abstract}
The interplay between many-body interactions and the kinetic energy gives rise to rich phase diagrams hosting, among others, interaction-induced topological phases. These phases are characterized by both a local order parameter and a global topological invariant, and can exhibit exotic ground states such as self-trapped polarons and interaction-induced edge states.
In this work, we investigate a realistic scenario for the quantum simulation of such systems using cold Rydberg-dressed atoms in optical lattices.
We consider spinless fermions on a checkerboard lattice, interacting via the tunable-range effective potential induced by the Rydberg dressing.
We perform a detailed analysis of the phase diagram at half- and incommensurate fillings, in the mean-field approximation.
We furthermore study the stability of the phases with respect to temperature within the mean-field approximation and with respect to quantum fluctuations using the density matrix renormalization group method.
Finally, we propose an implementation protocol, and in particular identify attainable regimes of experimental parameters in which the topological properties of the model become accessible. Our work thereby opens a realistic pathway to the outstanding experimental observation of this predicted phase in state-of-the-art cold atom quantum simulators. 
\end{abstract}

\maketitle


\section{Introduction}

Quantum simulators offer a powerful avenue for the study of many-body physics. These quantum systems mimic the dynamics of complex quantum matter in a highly controllable environment. They are in fact ideal candidates to solve many-body problems whose computational cost on classical computers scales exponentially with the system size. Theoretically proposed in the 80s~\cite{Feynman_1982}, they are nowadays a reality and can be realized in various physical systems such as photonics, superconducting qubits, and cold ions or neutral atoms~\cite{Trabesinger2012,lewenstein2017,Brierley2021}. 
Here, we focus on cold atomic simulators based on atoms excited to Rydberg states~\cite{Santos2000,*PhysRevLett.88.139904,saffman-rmp-82-2313}, which offer rich opportunities for quantum information processing, owing to their long-lived nature and strong long-range interactions leading to the paradigmatic Rydberg blockade effect~\cite{heidemann_evidence_2007}. Furthermore, individually controlled Rydberg atoms in optical tweezers~\cite{browaeys_many-body_2020} have emerged as a powerful platform for quantum computation~\cite{isenhower_demonstration_2010,wilk_entanglement_2010,levine_high-fidelity_2018, omran_generation_2019, henriet_quantum_2020, cong_hardware-efficient_2021}, and for the simulation of quantum spin models~\cite{endres_atom-by-atom_2016,de_leseleuc_observation_2019,bernien_probing_2017, keesling_quantum_2019,labuhn_tunable_2016, scholl_quantum_2021}, as highlighted by recent observations of 2D spin liquid phases~\cite{ebadi_quantum_2021,semeghini_probing_2021}.

While the strong interactions in Rydberg arrays are typically well captured by spin Hamiltonians, in which kinetic terms accounting for the itinerant nature of the particles can effectively be neglected, one of the challenges in the field is to achieve comparable kinetic and interaction energy scales in order to observe the interplay of interaction and motional effects. This can be achieved in a variety of platforms that feature long-range interactions, such as dipolar quantum gases~\cite{Chomaz2016,Schmitt2016,Boettcher2019,Tanzi2019a,Chomaz2019,Norcia2021}, or polar molecules~\cite{deMarco2019,Bohn2017}, and in the presence of an optical lattice this allows one to simulate extended Hubbard Hamiltonians with both non-local interactions and tunneling terms~\cite{Trefzger2011,dePaz2013,Dutta2015,Baier2016,Lepoutre2019,Patscheider2020}. Rydberg dressing~\cite{Pupillo-2010,Nath-2010,Johnson2010} has emerged as a powerful alternative in this context. In this approach, instead of exciting the atoms resonantly to a highly excited Rydberg state, in which the energy scale of the strong dipole-dipole interactions dominates over the itinerant dynamics, the atomic gas in the ground state is coupled off-resonantly to the Rydberg state, thereby admixing a reduced amount of Rydberg character to the electronic ground state. Compared to other techniques, Rydberg dressing offers the possibility to tune the strength and shape of interactions, which can be highly adjusted by a proper choice of the atomic and laser parameters of the underlying dressing protocol. Such degree of control has allowed to generate Bell pairs in optical tweezers~\cite{Jau2016}, to engineer long-range~\cite{Zeiher2016,Zeiher2017,Borish2020} or even distance-selective~\cite{Hollerith2021} interactions in Ising Hamiltonians, and to realize extended Fermi-Hubbard Hamiltonians~\cite{guardado-sanchez_quench_2021} with interaction strengths and kinetic
terms of the same order of magnitude. The latter has led to the observation of quench dynamics of a Fermi gas with long-range interactions~\cite{guardado-sanchez_quench_2021}, paving the way for the simulation of other novel phases of quantum matter resulting from the interplay between non-local interactions and the kinetic energy.
In particular, this Rydberg dressing toolbox is perfectly suited for the simulation of interaction-induced topological insulators~\cite{Rachel_2018}, which requires a high control over the ratio of interactions in the presence of a finite tunneling term. 

Topological insulators constitute a new paradigm of quantum matter~\cite{RevModPhys.82.3045,Qi2011}: characterized by a global topological invariant, they escape the standard classification of phases of matter and are very robust against local perturbations such as disorder or interactions. While these phases have been realized in quantum simulators~\cite{aidelsburger_realization_2013,Aidelsburger_2014,Jotzu_2014,Mancini2015,Asteria_2019,mancini_2015,Stuhl_2015}, they generally require the engineering of an external gauge field~\cite{Goldman_2014,Cooper_2019}. Alternatively, topological insulators can also arise solely from interactions through a symmetry breaking mechanism. In a seminal work~\cite{raghu_topological_2008}, it was shown that such an interaction-induced topological insulator, also called topological Mott insulator, can arise for fermions on a hexagonal lattice, with sufficiently strong inter-site interactions. In particular, next-nearest neighbor interactions can give rise to a ground state which breaks the time-reversal symmetry and is characterized by a non-zero topological invariant, the Chern number. Subsequent studies also found topological Mott insulators in other lattice geometries~\cite{Sun2009,zhu_interaction-driven_2016,Sun2009,dauphin_rydberg-atom_2012,zeng_tuning_2018,Sur2018,PhysRevLett.117.066403}. Interaction-induced topological phases are quite different from externally induced topological phases~\cite{Rachel_2018}. One of the most striking differences is the ground-state degeneracy. In the case of externally induced topological phases, the ground state is non-degenerate, whereas the ground state of the topological Mott insulator is two-fold degenerate, with each of its two sectors being characterized by opposite-valued Chern numbers. These two degenerate ground states with opposite Chern numbers can give rise to interesting effects around half filling such as the appearance of self-trapped polarons or interaction-induced topologically protected edge states, discussed in a previous work by some of us~\cite{julia-farre_self-trapped_2020}.

In this work, we address the timely question of whether the TMI phase can be accessed in quantum simulators based on dressed Rydberg atoms in an optical lattice, under realistic experimental conditions. To this end, we go beyond previous models~\cite{zeng_tuning_2018,PhysRevLett.117.066403,Sur2018,julia-farre_self-trapped_2020} relying on the simplified assumption of only nearest and next-nearest neighbors interactions, and for the first time properly account for the long-range nature of the Rydberg potential up to fourth order neighbors. Furthermore, we examine the sensitivity of the TMI phase with regard to finite temperature. Our extensive numerical analysis combines mean-field and density-matrix-renormalization group techniques, and is complemented by a thorough discussion of an experimental implementation proposal. Thereby, our study clearly establishes this phase in a robust parameter window, and furthermore provides a clear and experimentally feasible route towards the quantum simulation of the considered topological Mott insulator phase. 

The article is organized as follows.
In Section~\ref{sec:tmiqbt}, we review the phase diagram of the model featuring interactions up to next-nearest neighbors, and we introduce the different order parameters characterizing the charge orders and the quantum anomalous Hall (QAH) phase.
In Section~\ref{sec:qs}, we present a scheme based on dressed Rydberg atoms for the quantum simulation of the model.
We review the ingredients required, crucially observing that all such elements have been demonstrated in state-of-the-art setups.
We then perform an in-depth study of the phase diagram in Section~\ref{sec:phase_diag}. We discuss the impact of longer-range interactions, present in the Rydberg dressing scheme, on the interaction-induced QAH phase. In particular, we show that these can stabilize the QAH phase. We then study how this interaction profile affects the phases at incommensurate fillings around half filling. We additionally probe the stability of the phases at finite temperature. Furthermore, we confirm the stability of the phases with respect to quantum fluctuations with the help of a density matrix renormalization group analysis. Finally, in Section~\ref{sec:exp_param}, we discuss possible parameter regimes, accessible in state-of-the-art experiments, where the QAH phase can realistically be observed.


\section{Topological Mott insulator in quadratic band touching systems}
\label{sec:tmiqbt}

\subsection{Model}
\label{sec:model}

The emergence of a TMI phase has been extensively studied~\cite{raghu_topological_2008,Sun2009,zhu_interaction-driven_2016,Sun2009,dauphin_rydberg-atom_2012,PhysRevA.93.043611,zeng_tuning_2018,Sur2018,PhysRevLett.117.066403,julia-farre_self-trapped_2020} in lattice systems of spinless fermions described by the extended Fermi-Hubbard Hamiltonian,
\begin{equation}
\Hop_{\textrm{EFH}}  = \sum_{<ij>}[(t_{ij}\cdop_i\cop_j + \hc) + \Vgen_{ij} \nop_i\nop_j].
 \end{equation}
The first term of the equation describes spinless fermions hopping on a two-dimensional lattice, with  $\cdop_i$ ($\cop_i$) being the fermionic creation (annihilation) operator at lattice site $i$. The second term represents repulsive interactions, $\Vgen_{ij}>0$, between fermions on different lattice sites, with local particle number operators $\hat{n}_i \equiv \cdop_i\cop_i$.

In the original proposal, Raghu et al.~\cite{raghu_topological_2008} considered the honeycomb lattice at half filling, for which the non-interacting band structure obtained from the hopping matrix $t_{ij}$ is topologically trivial and exhibits a linear band touching, i.e., Dirac cones.
The authors showed that, in the mean-field approximation, the repulsive interactions open a topological gap, leading therefore to an interaction-induced topological phase that they termed Topological Mott Insulator.
Subsequent exact diagonalization and DMRG studies of Dirac semimetals, including the semimetallic model of the initial proposal, showed that, beyond the mean-field approximation, interactions favor trivial charge orders with lower energy than the TMI phase~\cite{Garcia-Martinez2013,Jia2013,Daghofer2014,Guo_2014,Motruk2015,capponi_phase_2015,Scherer2015}.

In parallel, perturbative analyses in several models, for which the non-interacting fermionic band of $\Hop_{\textrm{EFH}}$ exhibits instead a quadratic band touching (QBT) also suggested the appearance of a TMI phase~\cite{Sun2008TRS,Sun2009,Vafek2010,Dora2014}. In the perturbative limit, the TMI phase of such QBT systems was shown to be more stable that in Dirac semimetals, which are more robust with respect to instabilities driven by small symmetry-preserving interactions~\cite{Sun2009}. More recently, researchers have confirmed, using non-perturbative numerical methods such as DMRG or exact diagonalization, the existence of the TMI phase in many of these QBT systems both for weak and intermediate values of the interactions. This is for example the case for the kagome lattice at $1/3$ filling~\cite{PhysRevLett.117.066403,zhu_interaction-driven_2016}, or the checkerboard lattice at $1/2$ filling~\cite{zeng_tuning_2018,Sur2018,PhysRevLett.117.066403}. 

In this work, we focus on this latter case, that is, we consider a checkerboard lattice with a Hamiltonian
\begin{equation}
    \Hop=\Hop_0+\Hop_{\textrm{int}},
    \label{eq:full_ham}
\end{equation}
\begin{figure}[t]
\includegraphics[width=\columnwidth]{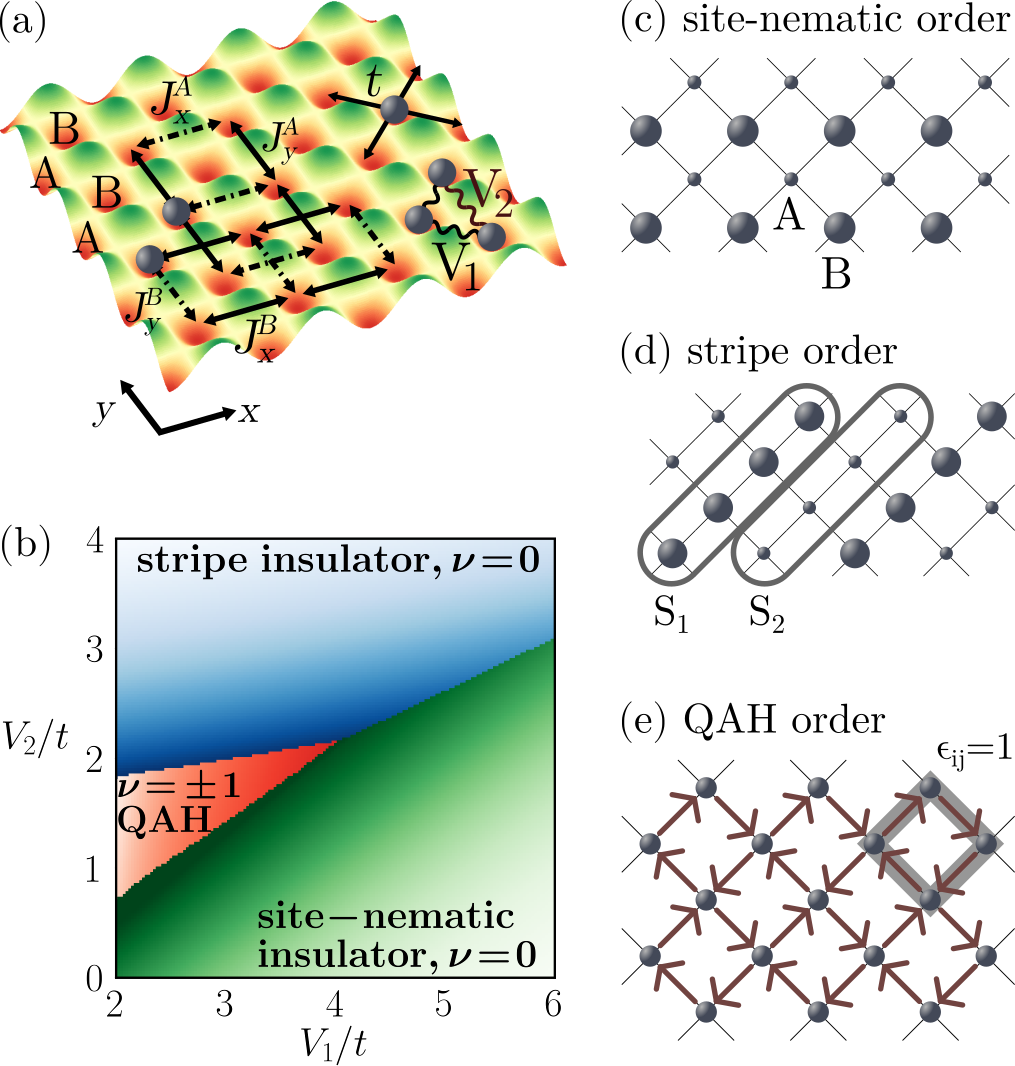}
\caption{
Topological Mott insulator in an extended Fermi-Hubbard model.
(a) Checkerboard optical lattice.
The Hamiltonian comprises a NN interaction $\Vgen_1$ and hopping amplitude $t$, and a NNN interaction $\Vgen_2$ and hopping amplitudes $J_x^A/t=J_y^B/t$ (dashed arrows) and $J_y^A/t=J_x^B/t=-J_x^A/t$ (straight arrows).
(b) Mean-field phase diagram of model~\eqref{eq:hamiltonianmf} with $J_x^A/t=0.5$ and $M=2$, at half-filling and at zero temperature. The Chern number $\nu$ is $0$ for trivial Mott insulating phases and $\nu=\pm 1$ in the TMI phase, indicating a ground-state double degeneracy; brighter blue/green corresponds to larger stripe/site-nematic charge order, darker red to larger current loop order.
(c-e) Instances of ordered states with spontaneously broken symmetry:
(c) site-nematic order, imbalanced density between $A$ and $B$ (two configurations);
(d) stripe order, alignment of the particles along $S_1$ or $S_2$, or along their orthogonal axes (four configurations);
(e) QAH order: average half-filling and current loops with chirality $\epsilon_{ij}=+1$ on square plaquettes of nearest-neighbors, see Eq.~\eqref{eq:loop}; for $\epsilon_{ij}=-1$, the loops has opposite chirality.}
\label{fig:01_intro}
\end{figure}
where $\Hop_{0}$ is the non-interacting Hamiltonian~\cite{zeng_tuning_2018,PhysRevLett.117.066403,Sur2018,julia-farre_self-trapped_2020} and reads~\footnote{we hereafter set $\hbar=1$}:
\begin{equation} \label{eq:hamiltonianfree}
\begin{split}
\Hop_{0} = -t&\sum_{\langle ij\rangle}(\cdop_{i,A}\cop_{j,B}+\rm{H.c.}) \\
+&\sum_{i}\sum_{\substack{\alpha=A,B \\ \eta=x,y}}(J_{\eta}^{\alpha}\cdop_{i,\alpha}\cop_{i+2\eta,\alpha}+\mathrm{H.c.})-\mu \sum_{i} \hat{n}_i.
\end{split}
\end{equation}
Here, $\mu$ is the chemical potential and fixes the particle number in the grand-canonical ensemble at temperature $T$, $t$ is the nearest-neighbors (NN) hopping amplitude, and $J_{\eta}^{\alpha}$ is the next-nearest-neighbors (NNN) hopping amplitude, which depends on the sub-lattice $\alpha\in(A,B)$ and hopping direction $\eta \in (x,y)$ [see Fig.~\ref{fig:01_intro}(a)].
The non-interacting band structure exhibits a quadratic band touching for the choice of NNN hopping $J_x^A=J_y^B=0.5t$ and $J_y^A=J_x^B=-0.5t$, corresponding to a $\pi$-flux through the unit cell of both sub-lattices. For more general designs, especially in the case of homogeneous and isotropic NNN hopping, the dispersion is linear, as shown in Appendix~\ref{sec:app01_qbt}.
For the interaction, we consider a general Hamiltonian with repulsive interactions of the density-density type, which reads
\begin{equation} \label{eq:hamiltonianint}
\Hop_{\text{int}} = \sum_{m\leq M}\sum_{\langle{ij}\rangle_m} \Vgen_m\hat{n}_i\hat{n}_j.
\end{equation}
Here the second sum is performed over the $m$-th order neighbors $\langle{ij}\rangle_m$ of the checkerboard lattice, e.g., $\langle ij \rangle_1$ corresponds to NN terms.
The isotropic repulsive interaction between $m$-th neighbors is then parametrized by the potential $\Vgen_m>0$.
As will be discussed in Sec.~\ref{sec:qs}, in this work we consider the Hamiltonian in Eq.~\eqref{eq:hamiltonianint} with interactions up to $M=4$, which faithfully describes the repulsive interactions experienced by dressed Rydberg atoms in an optical checkerboard lattice.
To provide background, we begin our analysis by first reviewing some known results~\cite{zeng_tuning_2018,PhysRevLett.117.066403,Sur2018,julia-farre_self-trapped_2020} for $M=2$.

\subsection{Half-filling interacting phases}
\label{sec:M2}

The TMI phase is captured already at the mean-field level.
By means of a standard Hartree-Fock decoupling, the repulsive density-density interactions of amplitude $\Vgen_1$ and $\Vgen_2$ are approximated as
\begin{equation} 
\begin{split}
\hat{n}_i\hat{n}_j \simeq &-\xi_{ij}\cdop_j\cop_i-\xi_{ij}^*\cdop_i\cop_j+\abs{\xi_{ij}}^2\\&+\bar{n}_i\nop_j+\bar{n}_j\nop_i-\bar{n}_i\bar{n}_j,
\end{split}
\label{eq:wicks}
\end{equation}
with $\xi_{ij}\equiv\langle\cdop_i\cop_j\rangle$ and $\bar{n}_i\equiv\langle\nop_i\rangle$, leading to the Hartree-Fock Hamiltonian 
\begin{equation}
\begin{split}
    \HHF = \Hzero + \sum_{m=1,2}\sum_{\langle{ij}\rangle_m} \Vgen_m \Big( & \abs{\xi_{ij}}^2 - \xi_{ij}\cdop_j\cop_i - \xi_{ij}^*\cdop_i\cop_j \\[-2mm] & + \bar{n}_i\nop_j+\bar{n}_j\nop_i -\bar{n}_i\bar{n}_j \Big).
\end{split}
\label{eq:hamiltonianmf}
\end{equation}
The Hartree-Fock values $\xi_{ij}$ and $\bar{n}_i$ are found by solving iteratively the resulting self-consistent quadratic Hamiltonian, as described in Appendix~\ref{sec:apphf}.
Figure~\ref{fig:01_intro}(b) shows the half-filling phase diagram of $\HHF$ zero temperature ~\cite{Sur2018,julia-farre_self-trapped_2020}.
In the limit of vanishing hopping $t \rightarrow 0$, the phase diagram hosts two insulating phases which spontaneously break the lattice translational symmetry, as can be seen in Fig.~\ref{fig:01_intro}(c)-(d).
The state resulting from the symmetry breaking is determined by the competition between $\Vgen_1$ and $\Vgen_2$.
Consequence of the repulsive density-density interaction is an energy cost of $\Vgen_1$ on pairs of particles occupying nearest-neighboring sites, and of $\Vgen_2$ for next-nearest-neighbors.
For dominant $\Vgen_1$, low-energy states are characterized by a minimal number of nearest-neighboring pairs, conjoined with a maximal density imbalance $\rho_n\equiv \bar{n}_\text{A}-\bar{n}_\text{B}$ between the two sub-lattices, thereby giving rise to the so-called site-nematic order.
By the same argument, for dominant $\Vgen_2$, the energy penalty of next-nearest-neighbor pairs favors states with stripe density order, characterized by a finite value of the density imbalance $\rho_s \equiv \bar{n}_{\rm{S_1}}-\bar{n}_{\rm{S_2}}$ between, e.g., the stripes $S_1$ and $S_2$ in Fig.~\ref{fig:01_intro}(d).
As shown in Fig.~\ref{fig:01_intro}(b), the transition between these two charge-ordered phases happens along the line $\Vgen_2=\Vgen_1/2$, when interactions dominate over the tunnelling amplitude.
However, when the kinetic energy becomes comparable to the interactions, quantum fluctuations lead to frustration between the two competing charge orders close to the phase transition.
In this scenario of charge homogeneity (translational symmetry), the ground state can still be insulating due to the appearance of a current loop order across nearest neighbors which spontaneously breaks time-reversal symmetry [see Fig.~\ref{fig:01_intro}(e)].
The local order parameter is defined as the staggered sum of currents in a closed loop of nearest-neighbors bonds,
\begin{equation}
    \xi_{\text{QAH}}\equiv \frac{1}{4}\sum_{<ij>\in \ \textrm{loop}}\epsilon_{ij}\, \text{Im}\, \xi_{ij},
    \label{eq:loop}
\end{equation} 
where $\epsilon_{ij}=+1$ if the bond $i\rightarrow j$ follows the red arrow convention of Fig.~\ref{fig:01_intro}(e), and $\epsilon_{ij}=-1$ otherwise.
This phase is known as topological Mott insulator or interaction-induced quantum anomalous Hall (QAH) phase, as each of its two  symmetry-breaking ground states with opposite current chiralities is characterized by a global topological invariant, the Chern number~\cite{thouless_quantized_1982},
\begin{equation}
    \nu = \frac{1}{2\pi \im}\int_{\rm{BZ}}\, d^2\mathbf{k}\left( \braket{\partial_{k_x}u^{0}_{\mathbf{k}}}{\partial_{k_y}u^{0}_{\mathbf{k}}}-\braket{\partial_{k_y}u^{0}_{\mathbf{k}}}{\partial_{k_x}u^{0}_{\mathbf{k}}}\right).
\end{equation}
Here $\ket{u^{0}_{\mathbf{k}}}$ is the lowest single-particle Hartree-Fock band of the Hamiltonian in Eq.~\eqref{eq:hamiltonianmf}, which includes the effect of interactions at the mean-field level.
The integral is performed over the first Brillouin zone (BZ) of the checkerboard lattice, assuming translational invariance of the two-site unit cell. The Chern number is quantized to integer values in systems with a band gap and is related to the Hall conductivity by $\sigma_H=\nu e^2/h$~\cite{thouless_quantized_1982,Haldane88}. For the topological Mott insulator it assumes one of the two non-trivial values $\nu =\pm 1$ corresponding to the two sectors of the spontaneous symmetry breaking. For the two other insulating phases present in the phase diagram, the topological invariant takes the value $\nu=0$, indicating that these phases are topologically trivial.
In the next sections, we show that this QAH phase remains present also in the scenario of the realistic long-range interaction potential that describes the interaction between laser-dressed Rydberg atom pairs.


\section{Quantum simulation using Rydberg atoms}
\label{sec:qs}

Numerical analyses aiming at unveiling the presence of a topological phase in quantum models are typically carried out in the thermodynamic limit in the mean-field approximation or, when including interactions, using exact or quasi-exact methods but considering systems of limited size.
In general, it is computationally hard to study the ground-state properties of interacting two-dimensional systems in the thermodynamic limit and observe its phenomena, such as a spontaneous symmetry breaking and the emergence of a quantized Chern number.
When direct observation in, e.g., quantum materials is not practicable, quantum simulation offers an alternative way to reveal theoretically predicted physical properties. In this context, ultracold gases trapped in optical lattices represent a pre-eminent platform for the quantum simulation of interacting Hubbard models~\cite{greiner_quantum_2002,gross_quantum_2017} such as the one given by Eq.~\eqref{eq:full_ham}. The platform enjoys a high level of experimental tunability, allowing for the control on tunnelling and on-site interaction~\cite{bloch_quantum_2012}. Furthermore, various detection methods are available for state inspection, from time-of-flight measurements to quantum gas microscopy~\cite{Gross2021}, or magnification~\cite{asteria_quantum_2021} techniques.

In this Section we discuss how cold Rydberg gases in suitable lattice geometries represent an ideal platform for the engineering of the interacting Hamiltonians that give rise to the discussed TMI phase. In particular, we will show how the phase can be realized in a checkerboard lattice with a $\pi$-flux and with the required long-range interaction terms. Remarkably, the demonstration of all essential elements of the Hamiltonian~\eqref{eq:full_ham} has been reported in currently available experimental setups for the parameters of our concern.

\subsection{Free Hamiltonian}

The TMI has been numerically identified in various lattices with a quadratic band touching, including the kagome ~\cite{PhysRevLett.117.066403,zhu_interaction-driven_2016} and the checkerboard lattice~\cite{zeng_tuning_2018,Sur2018,PhysRevLett.117.066403}. 
In this work we are interested in the latter case, where the checkerboard is obtained from a square lattice in which a sub-lattice-dependent $\pi$-flux on NNN plaquettes is introduced [Eq.~\eqref{eq:hamiltonianfree}].
Regarding the lattice, the design of a wide variety of optical lattice geometries, including the square, has been demonstrated experimentally~\cite{luhmann_quantum_2014, tarruell_creating_2012} and can be realized by properly adjusting the interference pattern of the standing laser beams.
The injection of an artificial flux on NNN plaquettes generates the checkerboard lattice with a quadratic band touching; this can be resolved via band mapping techniques, already used to certify the presence of Dirac points in a free Fermi gas on a tunable honeycomb lattice~\cite{tarruell_creating_2012}. The flux insertion has been demonstrated  experimentally in cold gases quantum simulators ~\cite{Jaksch2003,Goldman_2014,Cooper_2019} and requires control over the magnitude, sign and complex phase of the hopping amplitude $t$.
The dynamics of cold gases in lattices, a tight-binding system, occurs via hopping between nearest-neighbors and, marginally, next-nearest-neighbors.
Coherent control of the hopping amplitude can be attained with several methods.
Periodic perturbations of the optical lattice (Floquet techniques)~\cite{eckardt_superfluid-insulator_2005} make it possible to reduce, suppress and eventually change the sign of the tunnelling amplitude~\cite{lignier_dynamical_2007}.
Combining a strong lattice tilting with assisted tunnelling allows to exert selective control on hopping terms.
The tilting inhibits the tunnelling by introducing inter-site energy barriers larger than the hopping amplitude.
Then, the hopping can be activated again in a selective manner and with control on the hopping amplitude, using lattice amplitude modulation~\cite{ma_photon-assisted_2011}, or Raman-assisted tunnelling~\cite{Jaksch2003}.
In particular, Raman-assisted tunnelling allows one to engineer hopping terms with complex amplitudes $e^{i2\pi \phi_{jk}} \cdop_{k}\cop_{j} $, which can result in finite effective magnetic fluxes on closed paths~\cite{Goldman_2014}.
The engineering of artificial fluxes was a crucial step for the experimental simulation of static Abelian gauge fields~\cite{aidelsburger_realization_2013, miyake_realizing_2013}.
The method can be readily adapted to the $\pi$-flux case $\phi=1/2$ discussed in this paper, which induces the quadratic band touching present in the checkerboard lattice. It is worth stressing that the $\pi$-flux does not break explicitly the time-reversal symmetry, as opposed to generic finite fluxes $\phi \neq 1/2$.
As discussed, the symmetry breaking in TMIs occurs by effect of the interactions. 

\subsection{Interacting Hamiltonian}
\label{sec:int_ham}

\begin{figure}[h!]
\includegraphics[width=.9\columnwidth]{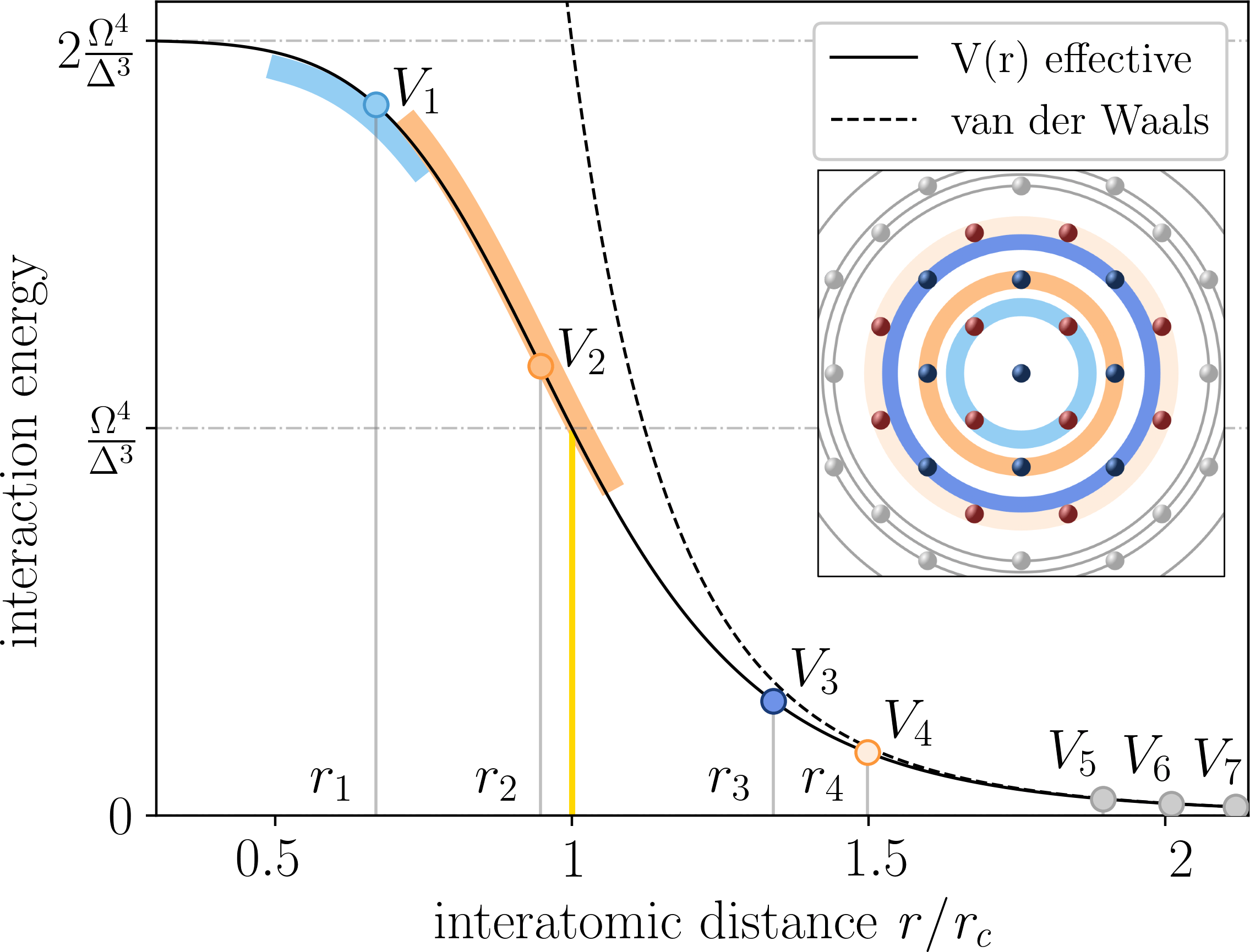}
\caption{
Amplitude of the interactions between Rydberg-dressed atom pairs, in a checkerboard lattice.
The continuous curve shows the effective interaction potential $\dVRyd$ from Eq.~\eqref{eq:vv4th} including only distance-dependent terms to fourth order in $\Omega/\Delta$ - see Sec.~\ref{sec:int_ham}.
At large inter-particle distances, $\dVRyd$ decays as $\beta^4 r^{-6}$, consistently with the repulsive van der Waals interaction in the doubly-excited component of the two-atom dressed state (dashed line).
Within the blockade radius, $r < r_c$, $\dVRyd$ converges to an energy plateau.
The colored slabs show the ranges of $\VRyd_1$ and $\VRyd_2$ for which a topological QAH phase emerges, as presented in Figs.~\ref{fig:05_pdrydberg} and~\ref{fig:10_DMRG}.
The inset shows an excerpt of the lattice, illustrating the succession of inter-site distances $r_i$: e.g., nearest-neighboring atoms sit on the cyan circle of radius $r_1$.
The proximity of $r_3$ and $r_4$ and the comparable magnitude of $\VRyd_3$ and $\VRyd_4$ require the inclusion of both terms in an analysis beyond $\VRyd_2$. Here, $r_1/r_c=0.67$.
}
\label{fig:02_rydpot}
\end{figure}

Let us next discuss how a density-density inter-particle interaction as in Eq.~\eqref{eq:hamiltonianint} can be engineered in cold gases experiments.
We consider a scheme based on effective interactions between Rydberg-dressed atoms.
Rydberg states are electronically excited atomic states with a large principal quantum number~\cite{gallagher_rydberg_1994,sibalic_rydberg_2018,adams_rydberg_2019, browaeys_many-body_2020}.
The laser-driven (single- or two-photon) transition that couples the electronic ground-state or low-lying (meta-)stable state $\ket{g}$ and an excited Rydberg state $\ket{r}$~\cite{adams_rydberg_2019}, in rotating-wave approximation and in the co-rotating frame, is described by the single-particle Hamiltonian $\Hop_{\mathrm{c}} = (\Omega \ket{r}\bra{g} + \hc) + \Delta \ket{r}\bra{r}$,
with effective Rabi frequency $\Omega$ and detuning $\Delta$.
In this work, we consider the repulsive two-body van der Waals interactions experienced between two atoms in the same Rydberg state $\ket{r}$, described by the van der Waals potential $\uvwr = C_6/r^6$, where $r$ is the inter-atomic distance and $C_6$ depends on the Rydberg state~\cite{sibalic_rydberg_2018,adams_rydberg_2019}.
The van der Waals interaction between Rydberg atoms is long-ranged and strong at short distances; as an example, $\uvw(r_1) = 90$ MHz for the $\ket{28P}$ Rydberg state of $\ce{^{6}Li}$ at $r_1 = 752$ nm (in this case attractive)~\cite{guardado-sanchez_quench_2021}.
A characteristic effect of the strong Rydberg potential at short distances is the dipole blockade: the laser excitation to the Rydberg state of multiple atoms within a certain exclusion volume is inhibited, as the strong interaction shifts the energy level of a state with multiple Rydberg atoms by more than the line width~\cite{sibalic_rydberg_2018}.
The Rydberg blockade has been observed in numerous experiments (see e.g.~\cite{Singer2004,Tong2004,Vogt2006,Heidemann2007,Raitzsch2008,urban_observation_2009, gaetan_observation_2009,isenhower_demonstration_2010}) and lies at the heart of Rydberg-based analog quantum simulation, e.g.,~of quantum spin models~\cite{browaeys_many-body_2020}.
Also, the implementation of entangling gates based on the blockade has been demonstrated~\cite{wilk_entanglement_2010,isenhower_demonstration_2010}, and represents the basis for potential applications in quantum computations~\cite{Lukin2001}, under rapid development in recent years~\cite{levine_high-fidelity_2018, omran_generation_2019,henriet_quantum_2020,cong_hardware-efficient_2021,Xu2021}.

The strong interaction within the blockade radius can also be advantageously used in a Rydberg-dressing scheme~\cite{dauphin_rydberg-atom_2012} for the quantum simulation of extended Hubbard-type models.
In the limit of far off-resonant laser coupling, i.e.,~for small values of the parameter $\beta=\Omega/\Delta \ll 1$, the transition between $\ket{g}$ and $\ket{r}$ are energetically suppressed.
The laser induces a weak hybridization of the electronic ground state with the Rydberg state; the strong van der Waals interaction occurring in the marginal Rydberg component of the admixture results in a finite and attenuated spatial-dependent soft-core potential~\cite{sibalic_rydberg_2018}.
One can understand the two-body interaction and derive the resulting effective potential surface by looking at a two-atom system.

The ground-state energy of a laser-dressed two-atom system can be obtained as a power series of the perturbation parameter $\beta$ using, e.g.,~van Vleck's perturbation theory~\cite{shavitt_quasidegenerate_1980,dauphin_rydberg-atom_2012}.
The spatial-dependent correction to the unperturbed electronic ground state energy up to fourth order in $\Omega/\Delta$ reads (see Appendix~\ref{sec:app03_vanvleck} for the derivation):
\begin{equation}
\label{eq:vv4th}
    \dVRyd = 2 \frac{\Omega^4}{\Delta^3} \left[ \frac{\uvwr}{2\Delta + \uvwr} \right].
\end{equation}
Here, we have not included the interaction-independent single-particle AC Stark shift $2(\Omega^2/\Delta)[1-(\Omega/\Delta)^2]$.
We can conveniently fix $E_p=2\Omega^4/\Delta^3$ as an energy scale for the effective interaction and define in the usual way the critical length, or blockade radius,
\begin{equation}\label{eq:critical_r}
    r_c^6=C_6/(2\Delta),
\end{equation}
at the full width at half maximum of $\dVRyd$. Also, we express the discrete inter-site distances in the lattice in units of the lattice spacing $r_1$, $r_i = \dis{i} r_1$, where $\dis{i} : \{ 1,\sqrt{2},2,\sqrt{5},2\sqrt{2} \}$
and $i$ is an index labelling neighbors radii (see Fig.~\ref{fig:02_rydpot}).
As a result, the inter-site density-density effective interaction reads
\begin{equation}
\label{eq:rydeffpot}
    \VRyd_i = E_p \left[ \frac{1}{1 + (\dis{i}r_1 / r_c)^6} \right].
\end{equation}
Figure~\ref{fig:02_rydpot} shows the plot of $\dVRyd$ as well as the discrete $\VRyd_i$ values. At large distances, the effective potential decays as
\begin{equation}
\VRyd(r\gg r_c)\sim\beta^4 \uvwr;
\end{equation}
this shows the suppression of the bare van der Waals interaction by the small prefactor $\beta^4$, that is the probability to find the dressed two-atom system in a doubly excited Rydberg state.
At short interatomic distances, the strong enhancement of the van der Waals interaction is accompanied by a vanishing population of the Rydberg-Rydberg component of the admixture, resulting in a soft-core potential,
\begin{equation}
\VRyd(r\ll r_c)\sim E_p.
\end{equation}
The energy plateau picture does not hold at very short distances, where the overlap of the electronic wave functions becomes more relevant and the van der Waals interaction ceases to correctly describe the interparticle interaction~\cite{gallagher_rydberg_1994}.

It should be emphasized at this point that the effective interaction potential $\VRyd_i$ is a particular case of the interaction Hamiltonian \eqref{eq:hamiltonianint}, i.e., $\VRyd_i$ is a constrained parametrization of the more general $\Vgen_m$, and depends on a range of controllable independent laser and atomic parameters: the Rabi frequency $\Omega$, the detuning $\Delta$, the lattice spacing $\alatt$, and the van der Waals interaction coefficient $C_6$.
\begin{figure}[t]
\centering
\includegraphics[width=0.95\columnwidth]{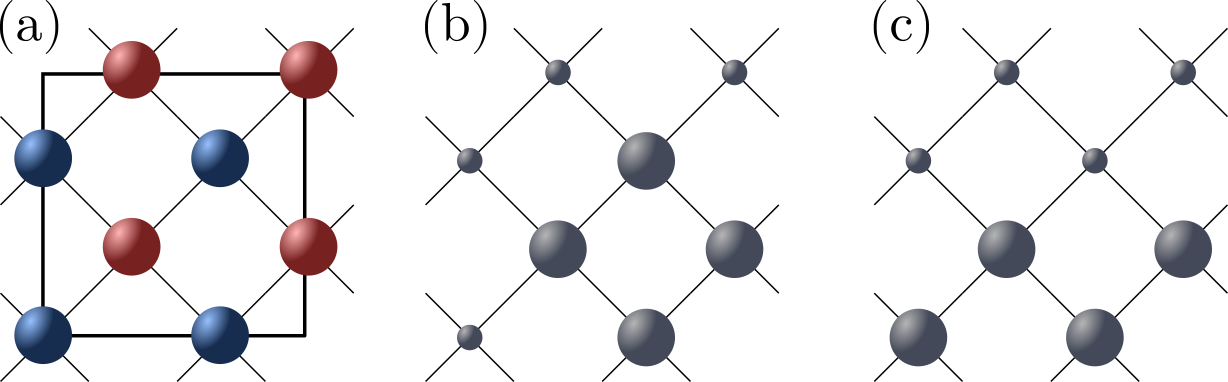}
\caption{Lattice unit cell adopted in the mean-field numerical analysis (a). The interaction between third neighbors $\Vgen_3$ favors two types of charge order: squares (b) and zig-zag (c).}
\label{fig:03_eightsites}
\end{figure}
Below, in Sec.~\ref{sec:exp_param}, we show how these parameters can be adequately tuned to adjust $\VRyd_i$ and access QAH states in a quantum simulation.
Here, we note that fixing $\VRyd_1$ and $\VRyd_2$, or any other pair of $\VRyd_i$, uniquely determines $\epl$ and $r_1/r_c$, and thereby determines the value of the remaining $\VRyd_i$.
We also remark that the ratio between two interaction amplitudes $\VRyd_i/\VRyd_j$, with $i<j$, is a monotonically increasing function of $r_1/r_c$, with a lower bound set by $\VRyd_i/\VRyd_j \to (d_i/d_j)^6$ in the limit $r_1 \gg r_c$. In particular, $\VRyd_2/\VRyd_1$ has a lower bound of $1/8$. The other limit $r_1 \ll r_c$ corresponds to an unphysical regime where Eq.~\eqref{eq:rydeffpot} is no longer valid, as all long-distance $\VRyd_i$ would be comparable in magnitude to $\VRyd_1$.

Since we are interested in studying how the physical properties of model \eqref{eq:full_ham} with $M=2$ change when we include a finite number of sub-leading long-distance interaction terms, we limit our investigation to a regime in which a truncation of the effective Rydberg potential to $\VRyd_4$ represents a meaningful approximation of the entire effective potential, including the tail.
To this aim, we chose to set the condition of $\VRyd_1$ always being at least an order of magnitude larger than the largest discarded interaction term, i.e., $\VRyd_1 > 10 \VRyd_5$, corresponding to $r_1/r_c \geq 0.51$.

Having comprehensively introduced the model Hamiltonian, we can now proceed to presenting the results of our numerical study of this model.


\section{Phase diagram with Rydberg interactions}
\label{sec:phase_diag}

In this Section, we present an extended numerical analysis of the Hamiltonian in Eq.~\eqref{eq:full_ham} in the presence of long-range interactions beyond next-nearest neighbors, motivated by the long-range character of the effective Rydberg potential, Eq.~\eqref{eq:rydeffpot}. After showing the effect of adding an arbitrary $\Vgen_3$ interaction in the ground-state phase diagram at half filling, we focus on the particular shape of interactions given by the effective Rydberg potential. For the latter, we study the presence of the QAH phase in the phase diagram with the mean-field Hartree-Fock method, and we also discuss the effects of incommensurate fillings on finite-sized systems.
Then, in the prospect of a quantum simulation, we examine the robustness of the QAH phase against thermal fluctuations with the finite-temperature Hartree-Fock method. Furthermore, we analyze the stability of the phase beyond the Hartree-Fock ansatz using the DMRG method at zero temperature, which accurately describes the ground states of gapped two-dimensional systems in cylinder geometries with finite widths~\cite{Stoudenmire2012}.

\subsection{Hartree-Fock phase diagram}

\subsubsection{Half filling}
\label{sec:mfpdrydT0}

To inspect the ground-state phase diagram, we perform a Hartree-Fock study in a large unit cell containing eight sites, illustrated in Figure~\ref{fig:03_eightsites}(a), which can host long-range correlators and capture charge orders with a large spatial periodicity.

\begin{figure}[t]
\centering
\includegraphics[width=0.95\columnwidth]{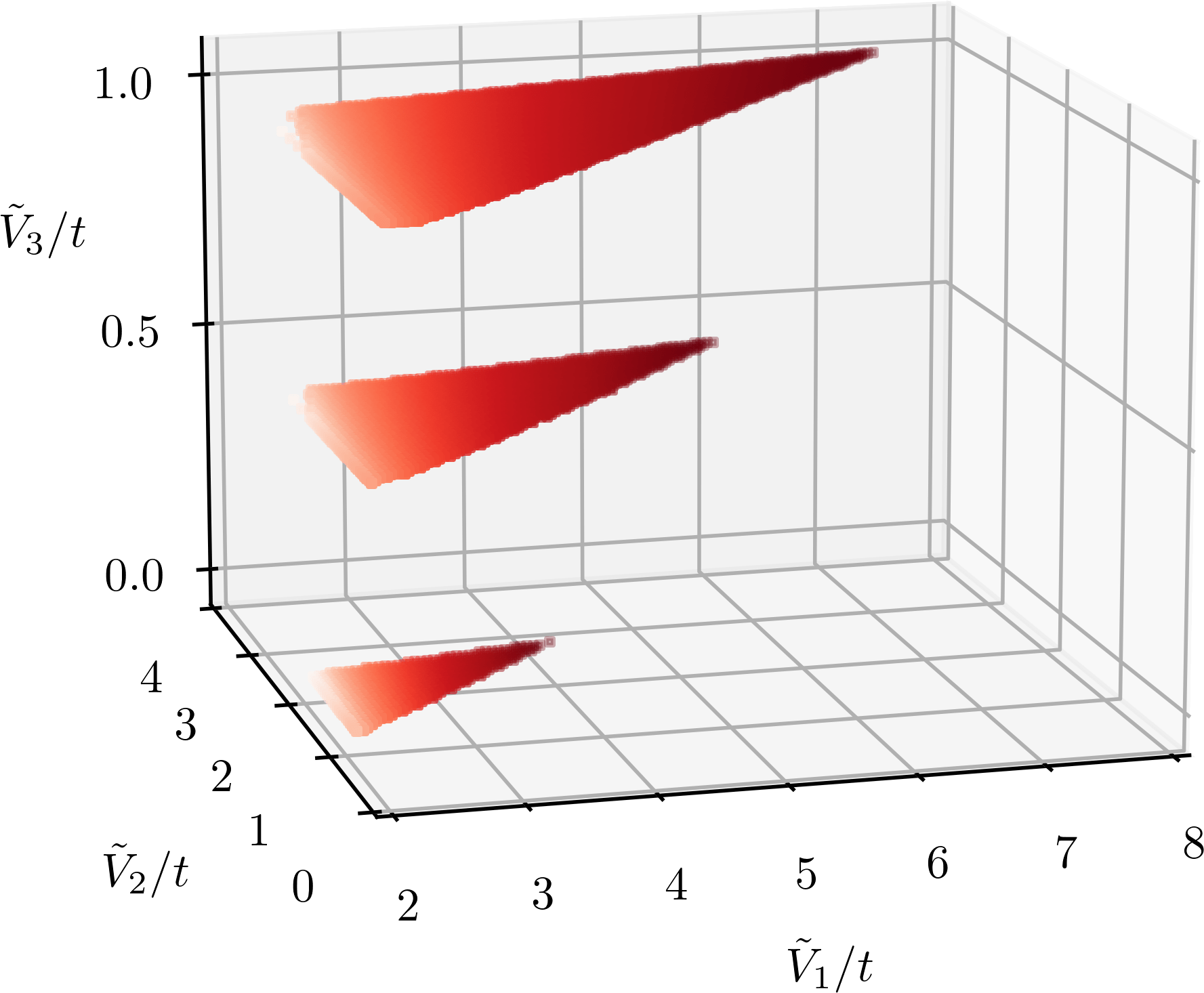}
\caption{Effect of $\Vgen_3/t$ on the topological phase.
Red areas represent the regions of QAH phase in the mean-field ground-state phase diagrams, at $T=0$ and for $\Vgen_3\ \in \ \{0,\ 0.5,\ 1\}$. The color gradient indicates the magnitude of the current loop order, larger for darker red. For increasing $\Vgen_3$, the topological phase is energetically favored over phases with trivial charge order (string and site-nematic) on larger regions of $\{ \Vgen_1,\ \Vgen_2 \}$. The $\Vgen_3=0$ layer corresponds to the QAH phase in Fig.~\ref{fig:01_intro}.
}
\label{fig:04_stackedQAH_V3}
\end{figure}

First, we survey the phase diagram for various, unconstrained $\Vgen_1$.
As seen in Sec.~\ref{sec:M2}, for $\Vgen_1$ ($\Vgen_2$) much larger than any other energy scale in the Hamiltonian, the system is in a gapped site-nematic (stripe) phase.
For dominant $\Vgen_3$ interactions, the density distribution presents two types of charge orders, depicted in Figs.~\ref{fig:03_eightsites}(b)-(c).
A first observation to make is that these two orders favoured by $\Vgen_3$ are incompatible with the density orders generated by $\Vgen_1$ and $\Vgen_2$.
The consequence of this is an enhanced competition between charge orders as $\Vgen_3$ becomes larger.
Figure~\ref{fig:04_stackedQAH_V3} shows the size of the QAH region in the $\Vgen_1-\Vgen_2$ plane for different values of $\Vgen_3$. 
One can observe that a finite $\Vgen_3$ augments the area of the QAH phase in parameter space.
The QAH phase benefits, indeed, from the frustration between competing charge orders: $\Vgen_3$ supports a different charge order than that of the site-nematic or the stripe phases, ultimately favoring the topologically ordered phase.

\begin{figure}[t]
\centering
\includegraphics[width=0.95\columnwidth]{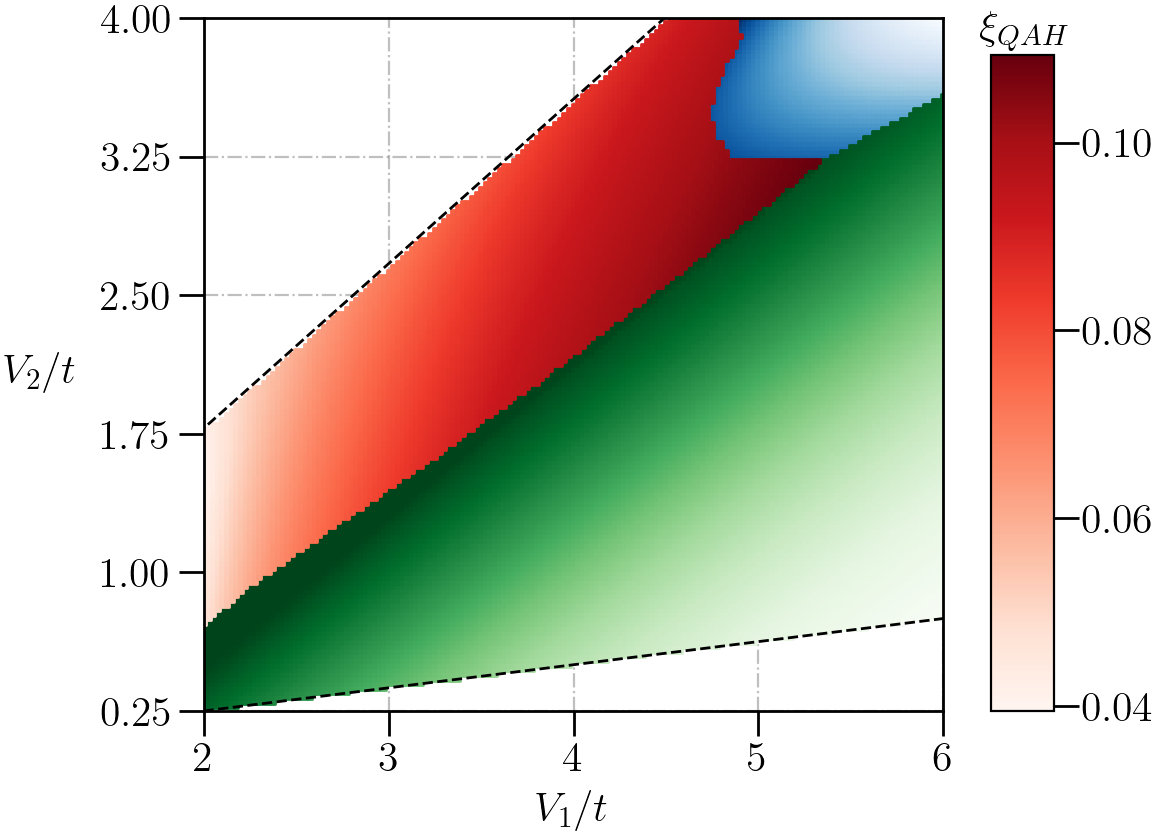}
\caption{Ground-state phase diagram for effective Rydberg interactions at $T=0$.
We retrieve the three insulating phases observed in the $M=2$ case and shown in Fig.~\ref{fig:01_intro}: site-nematic in green, stripe in blue and topological QAH phase in red. The QAH phase has a larger order parameter at large $\VRyd_1/t$ and $\VRyd_2/t$.
The diagram is bounded below by the largest ratio $\VRyd_1/\VRyd_2=8$ attainable within the Rydberg-dressed potential and above by the truncation condition  $\VRyd_1>10\,\VRyd_5$.}
\label{fig:05_pdrydberg}
\end{figure}

Let us now come to the Hamiltonian describing dressed Rydberg atoms,
\begin{equation}
\begin{split}
    \hat{H}_{\textrm{R}} = \Hzero + \sum_{m=1}^4\sum_{\langle{ij}\rangle_m} \VRyd_m \nop_i\nop_j,
\end{split}
\label{eq:hamiltonianM4}
\end{equation}
where we emphasize that $\VRyd_m$ is constrained by Eq.~\eqref{eq:rydeffpot} and by our truncation condition $\VRyd_1>10\VRyd_5$.
Notice that the Hamiltonian includes a finite $\VRyd_3$ term which promotes the stabilization of the QAH phase, as discussed above, but also a finite $\VRyd_4$ term, which favors the site-nematic order generated by $\VRyd_1$.
Notwithstanding, since $\VRyd_4$ is a subleading term, we expect the appearance of the QAH for $\hat{H}_{\textrm{R}}$ also.
This is indeed what we observe in the phase diagram of $\hat{H}_\textrm{R}$, shown in Fig.~\ref{fig:05_pdrydberg}.
The site-nematic phase prevails in a large part of the phase diagram, owing to the predominance of $\VRyd_1$ over the other interactions.
A QAH phase emerges as $\VRyd_2/\VRyd_1$ increases, and it can approximately be located in the window of $V_1/t \in \{2,6\}$ and $\VRyd_2/\VRyd_1 \in \{0.9,0.5\}$.
This latter corresponds to $r_1/r_c \in \{0.51,0.74\}$, as can be easily verified using Eq. \eqref{eq:rydeffpot}.
This interval of $r_1$ is indicated in Fig.~\ref{fig:02_rydpot} by a cyan slab; the orange slab shows the corresponding range of $r_2/r_c$.
Note that $r_2/r_1$ is determined by the lattice geometry; consequently, $V_2$ does not span the orange slab independently from $V_1$.
These slabs illustrate, for the checkerboard model and in the presence of dressed van der Waals interactions, where the QAH is to be found on the soft-core potential curve.
A first-neighbors distance $r_1$ too close to the critical distance $r_c$ leads the system into a deep site-nematic phase, because all ratios $V_1/V_i$ increase for increasing $r_1/r_c$; this determines the right limit of the cyan band, $r_1 = 0.74$.
On the opposite end, the slab is limited by the criterion of $V_1$ being an order of magnitude larger than $V_5$, which we impose to work with a potential truncated to $V_4$.
The current loop order parameter $\xi_\text{QAH}$ takes larger values at larger $\VRyd_1/t$ and $\VRyd_2/t$, as indicated by the darker red color.
For $\VRyd_1/t \leq 2$ (not shown) we find no presence of either orders, as the system enters a metallic phase.
The behavior of $\xi_\text{QAH}$ in both limits is congruent with what we observed for the $\tilde{V}_1-\tilde{V}_2$ model, in Fig.~\ref{fig:01_intro}, and reaffirms the emergence of the current loop order from an interplay between kinetic energy and interactions.
As compared to the $\tilde{V}_1-\tilde{V}_2$ model \eqref{eq:hamiltonianmf}, however, we can appreciate a considerably larger QAH region with the Rydberg dressing, as effect of the frustration introduced by the competition between multiple charge orders.

\subsubsection{Incommensurate fillings}

\begin{figure}[t]
\centering
\includegraphics[width=0.95\columnwidth]{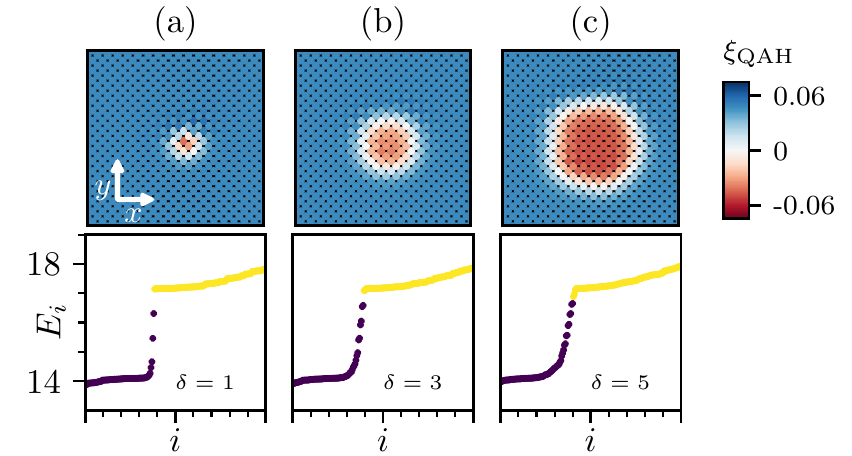}
\caption{Incommensurate solutions at zero temperature in a $24\times 24$ unit cells lattice. Here we choose the Rydberg potential such that $\VRyd_1=4t$, and $\VRyd_2=2.5t$, leading to $\VRyd_3=0.63t$, and $\VRyd_4=0.34t$ (corresponding to the full circles in Fig.~\ref{fig:02_rydpot}). The number of added particles is $\delta=1,\ 3,\ 5$ in (a),(b), and (c), respectively. (upper panels) Real-space profiles of $\xi_\text{QAH}$. (lower panels) Hartree-Fock single-particle spectra corresponding to the solutions shown in the upper panels. Yellow (dark purple) points represent empty (occupied) sites.}
\label{fig:06_doping}
\end{figure}
The interaction-induced QAH phase presents several differences in contrast to a non-interacting fermionic Chern insulator. On the one hand, it exhibits a twofold degeneracy of the ground state at half filling, corresponding to the two sectors of the spontaneous time-reversal symmetry breaking. On the other hand, the rigid band picture around half-filling breaks down due to the presence of correlations, and localized states can appear inside the topological gap. 
These properties lead to exotic solutions at incommensurate fillings, such as self-trapped polarons or domain walls interpolating between the two sectors of the spontaneous symmetry breaking~\cite{julia-farre_self-trapped_2020}.

We find these solutions also in the presence of the effective Rydberg potential, as shown in Fig.~\ref{fig:06_doping}, with the unrestricted Hartree-Fock method described in Appendix~\ref{sec:apphf}. The quantity $\delta$ counts the number of particles added to the half-filled state. In the case $\delta=1$, Fig.~\ref{fig:06_doping}(a) shows that the added particle does not populate the conducting band but instead occupies a midgap localized state induced by interactions, a self-trapped polaron. In this solution, the local current loop order $\xi_\textrm{QAH}$ changes its sign inside the polaron region, which can be understood as a collapsed domain wall. As $\delta$ increases, the number of mid-gap states and the polaron size increases [see Fig.~\ref{fig:06_doping}(b)]. Eventually, we observe the formation of a ring-shaped domain wall separating an inner and outer region with opposite current chiralities [see Fig.~\ref{fig:06_doping}(c)] which correspond to opposite Chern numbers inside and outside the ring.

\subsubsection{Finite temperature analysis}
\label{sec:temperature}

\begin{figure}[b]
\centering
\includegraphics[width=0.95\columnwidth]{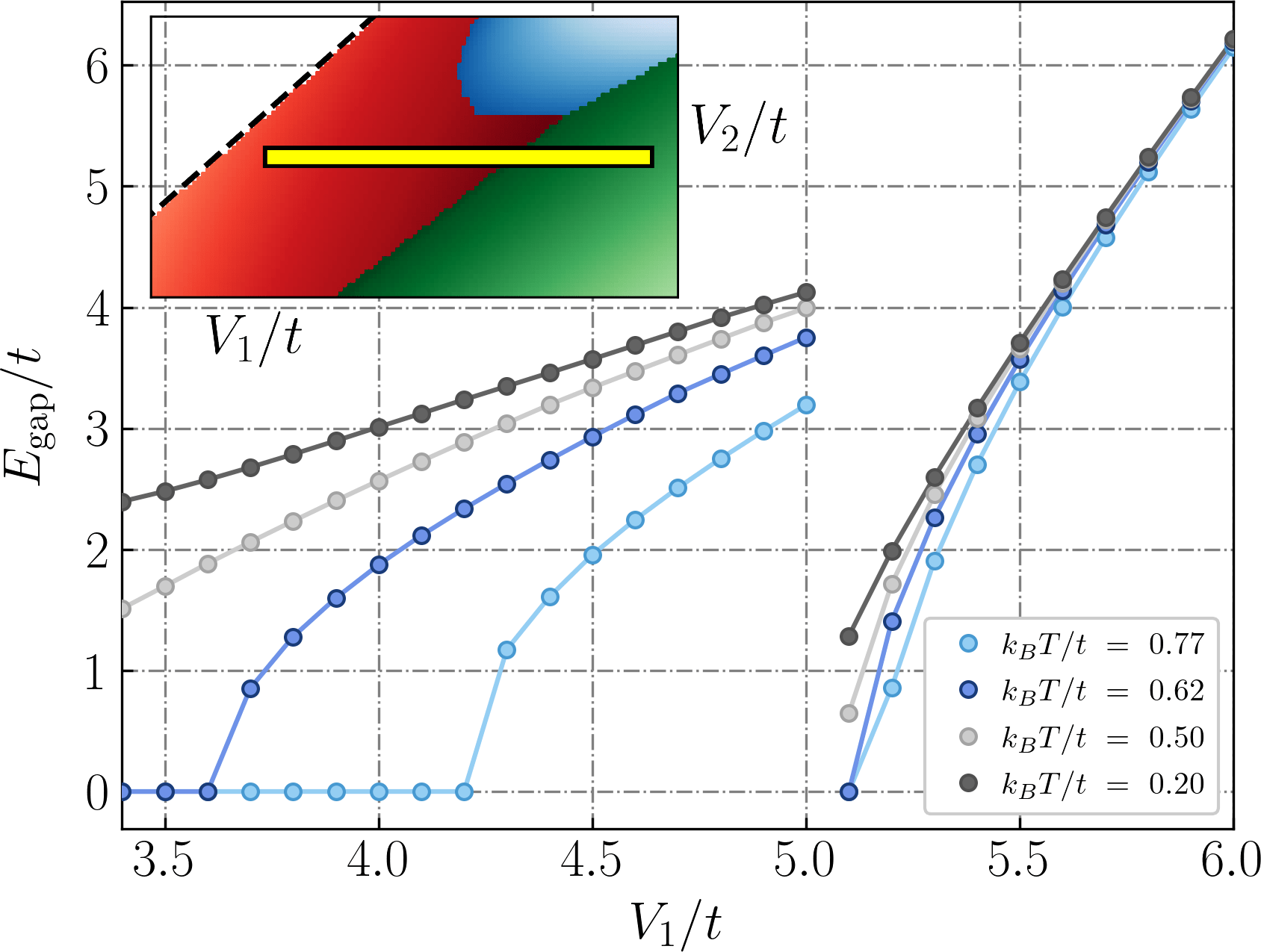}
\caption{Gap in the energy band structure at finite temperatures, for effective Rydberg interactions with $\VRyd_2/t=3$, corresponding to the yellow cut in the inset (clip of the phase diagram in Fig.~\ref{fig:05_pdrydberg}).
At $\VRyd_1/t \simeq 5$ we observe the transition between two insulating phases with a finite gap, from QAH to site-nematic.
No evident effects to the zero-temperature gap $E_{\mathrm{gap}}(T=0)$ (not shown) are observed for $k_B T/t \leq 0.2$.
At higher temperatures, the gap begins to close, affecting first and mostly the QAH phase.}
\label{fig:07_gap}
\end{figure}
We have seen above that, at zero temperature, the QAH phase appears within the Rydberg potential for a wide range of interactions. Let us now study the stability of the phase with respect to temperature by means of the finite-temperature Hartree-Fock method (see Appendix~\ref{sec:apphf}). Here, the occupations of the Hartree-Fock single-particle states with energies $E_i$ are given by the Fermi-Dirac distribution,
\begin{equation}
    f(E_i)=\frac{1}{1+e^{(E_i-\mu)/(k_BT)}}.
\end{equation}
 Moreover, we study the typical temperatures needed in order to resolve the spatial structures around half filling, shown in Fig.~\ref{fig:06_doping}.
\begin{figure}[t]
\centering
\includegraphics[width=0.95\columnwidth]{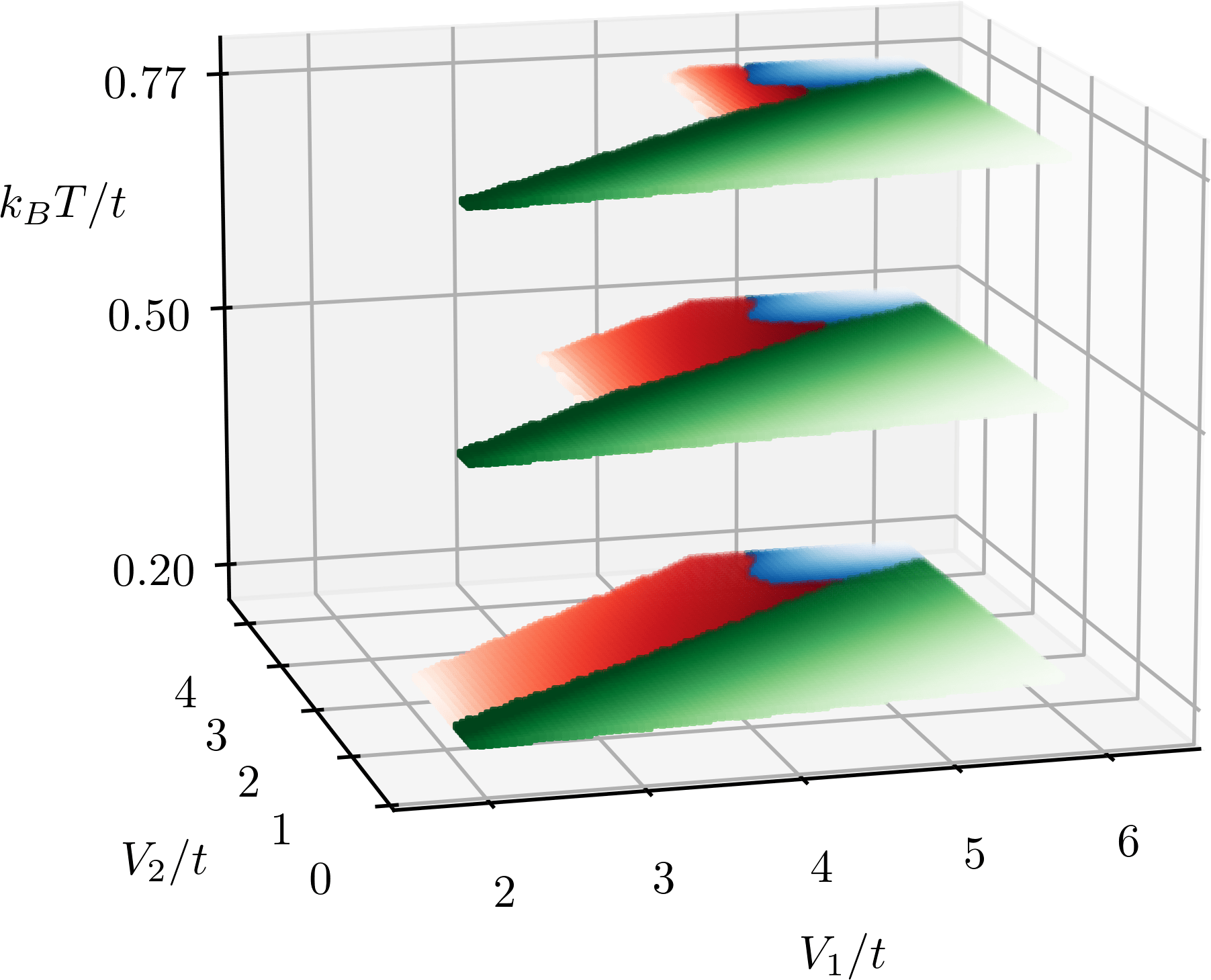}
\caption{Mean-field phase diagrams at finite temperature. Darker red indicates larger current loop order, as in Fig.~\ref{fig:05_pdrydberg}. All phases stand thermal excitations up to~ $k_BT/t\simeq0.2$. Higher temperatures progressively break the topological phase until it is no longer observed, above $k_BT/t \simeq 1$.
For a realistic value of the hopping $t/\hbar=1.7\ \textrm{kHz}$~\cite{guardado-sanchez_quench_2021}, the temperatures in the z-axis correspond to $T \in \{16,\ 40,\ 63\}$ nK.}
\label{fig:08_stackedQAH_temp}
\end{figure}

\textbf{Homogeneous phase at half filling.}
Figure~\ref{fig:07_gap} presents the energy band gap $E_{\mathrm{gap}}$ along a cut in the phase diagram, indicated by the yellow line in the inset. The gap refers to the Hartree-Fock single-particle band structure, at half-filling and for different temperatures. While the traditional notion of topology is typically defined at zero temperature, one can still use the notion of a topological invariant at finite temperature for the density matrix~\cite{Rivas2013,Viyuela2014,Huang2014,Budich2015,Bardyn2018}, provided that the thermal energy scale $k_BT$ is lower than this insulating gap $E_{\mathrm{gap}}$.

As a first remark, we note that the zero-temperature gap $E_{\mathrm{gap}}(T=0)$, not shown, is indistinguishable from the gap at $k_B T/t=0.2$.
The sharp discontinuity at $\VRyd_1 \simeq 5t$ pins the phase transition between the QAH and the site-nematic phase. This jump in the value of the gap, in agreement with the first-order nature of the transition, can be understood from the fact that this quantity is correlated with the value of the order parameter of the respective phase: when approaching the transition from the QAH side, both the current loop order as well as the gap are enhanced, whereas when approaching it from the site-nematic phase both the charge order and the gap vanish.  
With regards to the QAH phase, the gap $E_{\mathrm{gap}}$ is of the order of the hopping rate $t$ in most of the QAH region, taking the maximum value of about $4t$ around $\VRyd_1=5t,\ \VRyd_2=3t$.
From the zero temperature gap analysis, one would estimate that the topological QAH phase is robust for temperatures up to a few $t/k_B$.
However, given the interacting nature of the Hartree-Fock band structure, a finite temperature calculation of the gap is required in order to establish the critical temperature of the QAH phase.
As shown in Fig.~\ref{fig:07_gap}, the gap decreases non-linearly with increasing temperature, affecting most rapidly states with smaller gap at $T=0$.
Ultimately, we can roughly estimate a critical temperature for the appearance of the QAH phase of about $T_c = t/k_B$, well below the temperature $T \simeq 4t/k_B$ suggested by the gap structure at $T=0$.
Using $t=1.7\ \textrm{kHz}$, a value on the scale of current experimental realizations~\cite{guardado-sanchez_quench_2021}, we obtain a critical temperature of $T_c=82\ \mathrm{nK}$ around $V_1=5t$ and $V_2=3t$. 

It is also worth to visualize the effect of the temperature without restrictions to a point or a line of the phase diagram. To this end, in Figure~\ref{fig:08_stackedQAH_temp}, we show the whole phase diagram of $\Hop_\textrm{R}$ at three different finite temperatures.
Up to a temperature of $k_BT/t = 0.2$ no appreciable alteration to the phase diagram is observed. 
The QAH phase emerges from the competition between kinetic energy and the frustrated charge order driven by the interaction.
As such, it results to be most fragile against thermal fluctuations.
As the temperature increases, the valence-conduction gap of all insulating phases progressively reduces, with a major impact on the QAH phase.
The QAH gap closes first in the region of lower $\VRyd_1$ and $\VRyd_2$, where the zero-temperature current loop is smaller, leading to a gapless semi-metallic phase.
\begin{figure}[b]
\centering
\includegraphics[width=0.95\columnwidth]{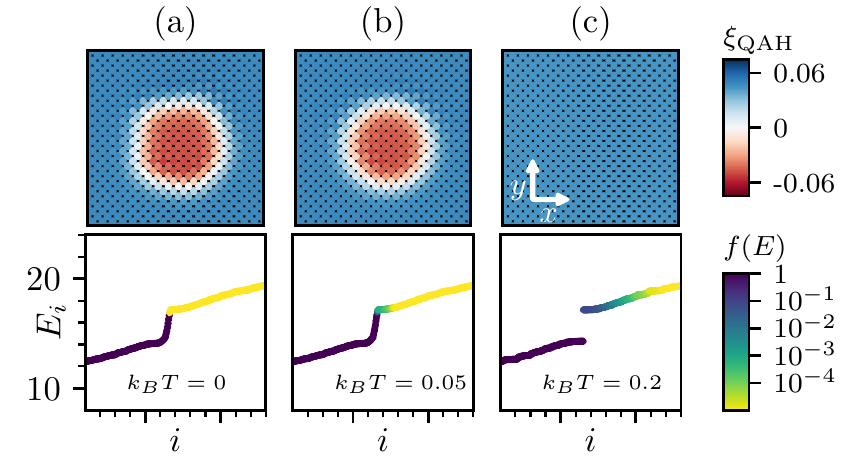}
\caption{Temperature behavior of the ring solution obtained for $\delta=5$. Here we consider the same lattice size and interactions used in Fig.~\ref{fig:06_doping}. The temperature in (a),(b), and (c) is $k_BT/t=0,\ 0.05,\ 0.2$, respectively. (upper panels) Real-space profile of $\xi_\text{QAH}$ for different temperatures.  (lower panels) Hartree-Fock single-particle spectrum corresponding to the upper panels. The color code indicates the Fermi occupation $f(E)$ of each state.}
\label{fig:09_temp_defects}
\end{figure}

\textbf{Defects around half filling.} Along with a closing gap, at rising temperatures, the mid-gap states progressively disappear, as they mix with the lower bulk band. As an example of such behavior, we study the effect of finite temperature for the case $\delta=5$, which at zero temperature corresponds to a ring-shaped domain wall [Fig.~\ref{fig:09_temp_defects}(a)]. In Fig.~\ref{fig:09_temp_defects}(b) we show the results for a finite temperature $k_BT/t=0.05$; while there is no appreciable difference of the order parameter in real space compared to the zero temperature case, the conducting band starts to be populated. For an even higher temperature $k_BT/t=0.2$ the spatially homogeneous QAH phase without mid-gap states is recovered, as seen in Fig.~\ref{fig:09_temp_defects}(c). However, notice that in this homogeneous solution the excess particles are distributed in the upper band, destroying the gap insulating nature of the phase. 

\subsection{DMRG phase diagram}

In order to corroborate the stability of the QAH phase beyond the mean-field approach used in the previous section, in the case of Rydberg interactions, we perform a density-matrix-renormalization group (DMRG) study in its matrix-product-state (MPS) formulation~\cite{schollwock_density-matrix_2011,tenpy,Stoudenmire2012}. We consider a cylindrical geometry of the checkerboard lattice with an infinite size along the longitudinal direction (iDMRG). Due to the one-dimensional nature of the DMRG algorithm, the cylinder is mapped to a one-dimensional chain in a snake-like folding along the radial direction, at the cost of introducing effective long-range couplings. The latter limits us to cylinder widths up to $L_y=6$ unit cells (12 physical sites). By using a maximum bond dimension $\chi_{\text{max}}=3000$, we get truncation errors of the infinite MPS of the order $10^{-5}$ at most.

\begin{figure}[t]
\centering
\includegraphics[width=0.95\columnwidth]{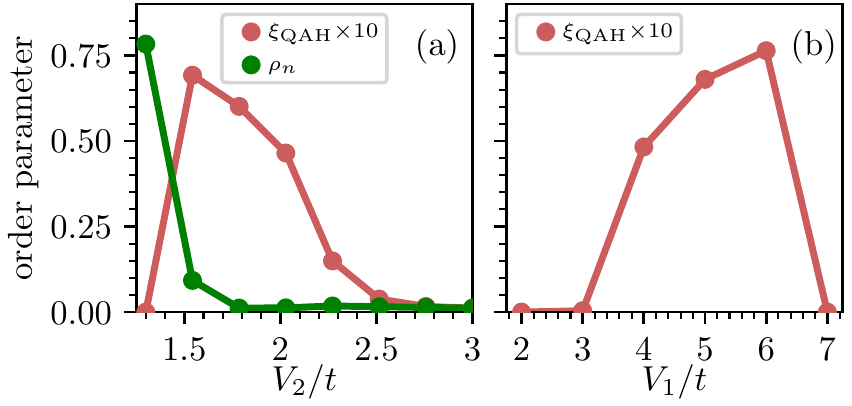}
\caption{iDMRG study of the QAH phase considering the Rydberg potential. (a) QAH and site-nematic order parameters as a function of $\VRyd_2$ for a fixed $\VRyd_1/t=4$.  (b) QAH order parameter as a function of $\VRyd_1/t$ along the line $\VRyd_2=\VRyd_1/2$.}
\label{fig:10_DMRG}
\end{figure}
As shown in Fig.~\ref{fig:10_DMRG}(a), the DMRG calculation confirms the site-nematic to QAH phase transition when varying $\VRyd_2/t$ for a fixed $\VRyd_1/t=4$, which determines the value of $\VRyd_3/t$ and $\VRyd_4/t$ according to the dressed Rydberg potential of Eq.~\eqref{eq:rydeffpot}. 
One observes a shift of the QAH boundary compared to the Hartree-Fock phase diagram of Fig.~\ref{fig:05_pdrydberg}, which is expected since mean-field methods are known to be less accurate in the vicinity of a quantum phase transition. Specifically, at $\VRyd_1/t=4$ with DMRG the QAH phase appears at $\VRyd_2/t\simeq 1.5$ and disappears for $\VRyd_2/t>2.5$. In contrast, with a Hartree-Fock ansatz the QAH phase is not present until $\VRyd_2/t>2$ and disappears for $\VRyd_2/t>3.5$.
The shift points in a direction advantageous from an experimental point of view; the realization of coherent systems with strong long-range interactions is difficult, therefore the possible appearance of the QAH phase at lower $\VRyd_2$, already smaller than $\VRyd_1$, is a favorable sign.
We also establish the existence of the QAH phase for a wide range of $\VRyd_1/t$ ratios, as shown in Fig.~\ref{fig:10_DMRG}(b). As hinted by the previous Hartree-Fock calculations, the Rydberg potential stabilizes the phase in a larger window of interaction strengths compared to the simplified $\Vgen_1$-$\Vgen_2$ model~\cite{Sur2018}.


\section{Experimental parameters analysis}
\label{sec:exp_param}
In this section we study the relevant experimental parameter regimes to simulate the model of Eq.~\eqref{eq:hamiltonianM4} with Rydberg dressed atoms~\cite{Santos2000,*PhysRevLett.88.139904,Pupillo-2010,Nath-2010,Johnson2010} in an optical lattice.
Notice that this dressing technique has been already widely used in  several recent experiments~\cite{Jau2016,Zeiher2016,Zeiher2017,Borish2020,Hollerith2021,guardado-sanchez_quench_2021}, which have succeeded in engineering tunable long-range interactions in two-dimensional systems.
Of particular interest for the simulation of the TMI phase is the experiment of Ref.~\cite{guardado-sanchez_quench_2021}, which has allowed for the observation of a long coherence time in a two-dimensional Fermi lattice gas, in the presence of tunnelling and inducing strong non-local interactions~\cite{guardado-sanchez_quench_2021}.
Such system paves the way towards the quantum simulation of other Fermi-Hubbard Hamiltonians with long-range interactions, including models with topological properties.

Motivated by this prospect, we investigated and identified parametric regions for which a similar experimental system would well approximate the model of~\eqref{eq:hamiltonianM4} and allow one to reach the interaction-induced topological phase.
To begin the analysis, we consider the coherence time as a crucial figure of merit in the context of many-body quantum simulations.
In a Rydberg-dressed cold gas, coherence is affected by spontaneous decay of atoms from the Rydberg state, which limits the time scale up to which the effective Hamiltonian~\eqref{eq:hamiltonianM4} faithfully describes the system; in a simple single-particle picture, the time scale is given by the effective Rydberg decay rate 
\begin{equation}\label{eq:rydberg_decay}
    \geff = \left(\frac{\Omega}{\Delta}\right)^2 \Gamma_0, 
\end{equation}
where $\Gamma_0$ is the bare decay rate of the Rydberg state. Thus, $\geff$ sets a lower limit for both the hopping and the interaction rates, namely that $t, \VRyd_i > \geff$.

We now show that the experimental conditions allowing one to engineer and observe the topological phase can be attained by a suitable choice of atomic and laser system parameters.
The choice of the atomic species and the target Rydberg state determines the van der Waals interaction strength $C_6$, the bare decay rate $\Gamma_0$, the lattice spacing $a_{\textrm{latt}}$ at which the species can be trapped, and the order of magnitude of the tunneling amplitude $t$~\footnote{The tunneling amplitude depends on the atomic mass but can be further fine tuned by changing the lattice depth.}.
The remaining free parameters are the detuning $\Delta$ and the Rabi frequency $\Omega$ of the dressing laser fields.

We envision the quantum simulation being based on a light atomic species in order to favour the itinerant character of the TMI phase.
An example would be the fermionic isotope $\ce{^{6}Li}$ of Lithium.
This species can be trapped in optical lattices with a lattice spacing $\alatt=752\ \textrm{nm}$ \cite{guardado-sanchez_quench_2021}, and is accompanied by a fast tunneling which we assume to be $t\simeq 3$ kHz in the following.
A repulsive and isotropic van der Waals interaction can be found between atoms in $\ket{nS}$ Rydberg states, for which the series of $C_6$ values can be calculated using, e.g., the library from Ref.~ \cite{Weber2017}.
Here we consider a Rydberg coefficient $C_6\simeq 100\  \textrm{MHz} \ \alatt^6$, which can be attained for principal quantum numbers roughly above $n=33$.
For these states, the radiative lifetime is estimated to be approximately $1/\Gamma_0\simeq 30\ \mathrm{\mu s}$ \cite{Beterov2009}.
We fix the parameters mentioned so far and leave out $\Omega$ and $\Delta$ as independent variables.

\begin{figure}[t]
\centering
\includegraphics[width=0.95\columnwidth]{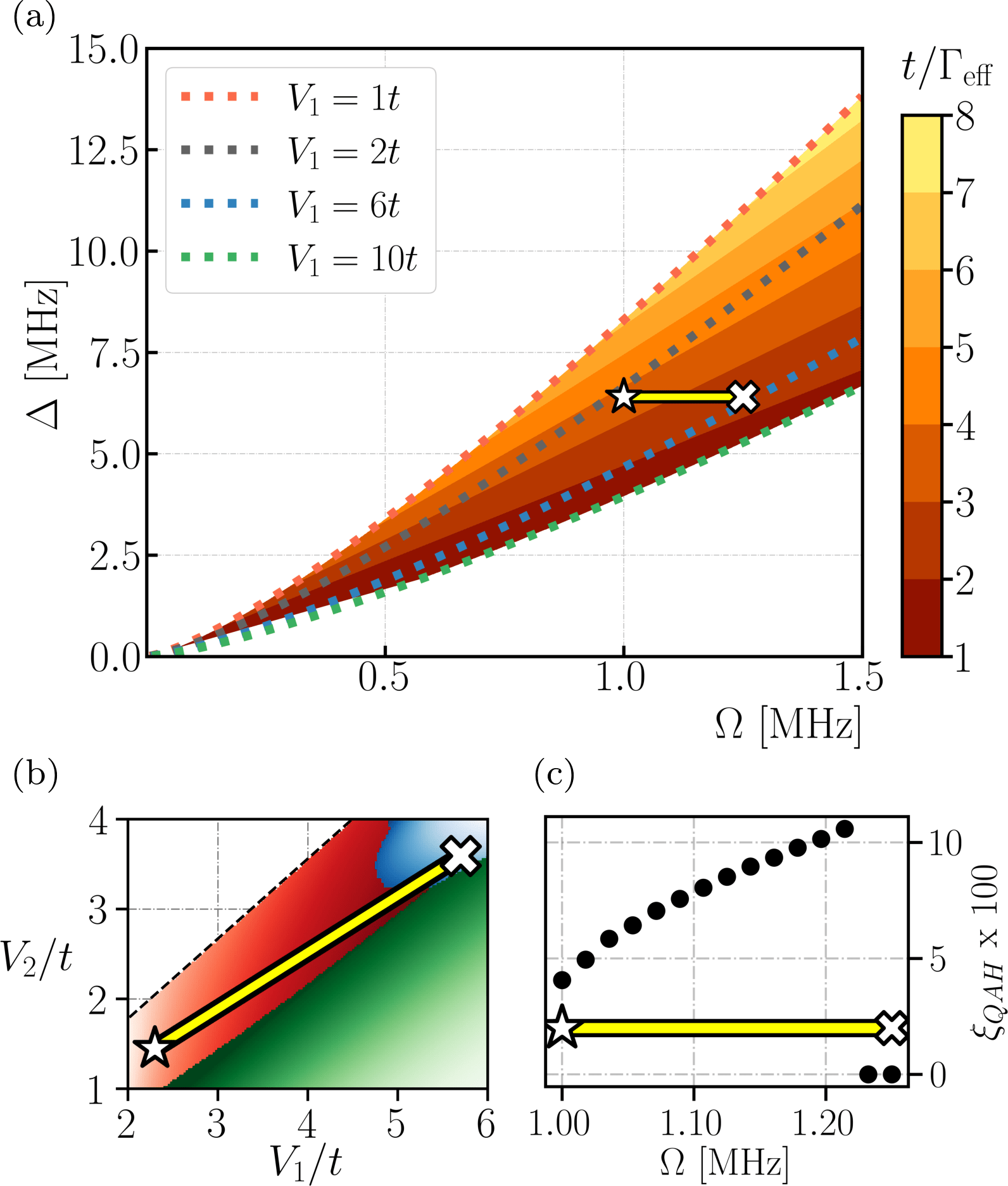}
\caption{QAH phase in laser parameters space.
Relevant parameters are fixed to plausible values for this type of experiments: $\tau_0=30 \ \mathrm{\mu s}$, $C_6 = 100 \ \mathrm{MHz} \ \alatt^6$, $t = 3 \ \textrm{kHz}$.
The color map in (a) refers to the hopping rate in units of the effective decay rate, $t/\geff$.
The QAH phase emerges for $V_1/t \in \{2, 6\}$ and $\VRyd_2/\VRyd_1 \in \{0.5, 0.9\}$, the latter corresponding to $\Delta \in \{8, 1\}$ (see Sec.~\ref{sec:mfpdrydT0}).
In this region, the hopping is always larger than $\geff$.
The yellow line indicates a possible trajectory to cross the QAH phase: $\Delta=6\ \textrm{MHz}$, $\VRyd_2/\VRyd_1 = 0.62$, $r_1/r_c = 0.68$.
Subplots (b), a clip of Fig.~\ref{fig:05_pdrydberg}, and (c) present the corresponding path in the phase diagram and the amplitude of the QAH order parameter along the oriented line.}
\label{fig:11_region_experiment}
\end{figure}

Figure~\ref{fig:11_region_experiment}(a) shows
the figure of merit  $t/\geff$, i.e., the ratio between the tunneling rate and the effective Rydberg decay rate, as a function of the free parameters $\Omega$ and $\Delta$, as given by Eq.~\eqref{eq:rydberg_decay}. An important remark here is that, for each point of this Fig.~\ref{fig:11_region_experiment}(a), the values of the interactions $V_i$ are determined by the rest of the fixed experimental parameters discussed above. In particular, we only show $t/\geff$ in the region comprised between the equipotential lines $\VRyd_1=t$ and $\VRyd_1=10t$ given by the condition
\begin{equation}
\frac{V_1}{t}=\frac{2\Omega^4}{t\Delta^3}\left(1+2a_{\textrm{latt}}^6 \Delta/C_6\right).
\label{eq:equi_V}
\end{equation}
Notice that, for this range of interaction strengths, relevant for the simulation of the TMI phase, the hopping rate is always larger than the decay rate, i.e., $t/\geff > 1$.

We now discuss a specific region of $\Delta$ and $\Omega$ that allows one to reach the TMI phase. In particular, we consider the yellow line at a fixed $\Delta=6\ \textrm{MHz}$ in Fig.~\ref{fig:11_region_experiment}(a), which corresponds to fixing the ratio $V_1/V_2=0.62$, as can be inferred from Eqs.~\eqref{eq:critical_r}-\eqref{eq:rydeffpot}. The interaction parameters spanned by this yellow line are mapped into the phase diagram of Rydberg dressed atoms in Fig.~\ref{fig:11_region_experiment}(b). The current loop order parameter $\xi_\text{QAH}$ along this cut is also shown in Fig.~\ref{fig:11_region_experiment}(c): within this parametric span one can access the topological phase.

As an example, we can choose to pin $\VRyd_1=4t$, which corresponds to $\VRyd_2 = 2.5t$, $\VRyd_3 = 0.63t, \VRyd_4 = 0.34t$; this specific set of $\VRyd_i$ is illustrated by the circles shown in Fig.~\ref{fig:02_rydpot}.
For this example, we obtain a large current loop order $\xi_\textrm{QAH}\simeq 0.08$, a moderate decay rate leading to $t/\geff \simeq 4$, and the ground state is relatively robust to thermal fluctuations with a critical temperature of $k_\textrm{B}T\simeq 50\ \textrm{nK}$, obtained from the finite-temperature gap analysis of Fig.~\ref{fig:07_gap}.

As a conclusive note, we remark that the observation of collective phenomena in the quantum simulation set-up considered here hinges on the validity of Eq.~\eqref{eq:rydberg_decay} to describe the effective decay rate for Rydberg-dressed atoms. Previous experiments have observed a rather large decay rate, scaling with the number of particles.
This phenomenon is modelled as a blackbody-driven collective resonant decay from the dressed state, an avalanche mechanism triggered by the first individual atom decay that drives the entire system out of its simulation task.
The suppression of this avalanche mechanism will be crucial in order to scale the system size and was to a good extent achieved by the authors of~\cite{guardado-sanchez_quench_2021}, thereby opening the door to quantum simulating the extended Fermi-Hubbard model and topological Mott insulating phase discussed in our present work. 


\section{Conclusions and outlook}

In this work, we investigated a topological Mott insulating model and addressed the question regarding its quantum simulation with state-of-the art experimental methods.
Precisely, we considered a Fermi-Hubbard Hamiltonian on a checkerboard lattice with inter-site density-density interactions, a model which is known in the literature to host a quantum anomalous Hall phase in the $V_1-V_2$ case, i.e.,~when featuring interactions up to next-nearest neighbors~\cite{zeng_tuning_2018,PhysRevLett.117.066403,Sur2018,julia-farre_self-trapped_2020}.
Here, we studied the impact of longer-range interactions on the $V_1-V_2$ model and observed that the interaction between third-neighbors stabilize and enlarge the QAH region in the ground-state phase diagram.
This result eases the requirements for an experimental realization, given that realistic long-range interaction potential profiles generally comprise a non-vanishing coupling beyond second-neighbors.
We focused on the Hamiltonian modelling a lattice gas of Rydberg-dressed Fermi atoms, exhibiting a long-range two-body effective interaction potential, and we studied the physical properties of this model.
The choice of the effective Rydberg potential is motivated by the technological progress reached in the field of optically trapped cold Rydberg atoms~\cite{browaeys_many-body_2020}, as highlighted by recent experiments~\cite{guardado-sanchez_quench_2021}.

Interestingly, we observed that the Rydberg-dressed model Hamiltonian hosts a larger QAH phase in the ground-state phase diagram, as compared to the $V_1-V_2$ model.
This constitutes an encouraging result for the purpose of quantum simulations, since realistic long-range interactions generically present finite beyond-second-neighbors coupling.
We substantiated our analysis addressing real-system effects that can arise under ordinary conditions in a laboratory: we studied the fate of self-trapped polarons and interaction-induced domain walls at incommensurate fillings, we analyzed the stability of the QAH phase with respect to temperature with a mean-field approach and with respect to quantum fluctuations with DMRG.
Furthermore, we discussed realistic ranges of the experimental parameters in a cold gases setup, which allow to access a TMI state of matter.

This work provides a clear route towards the experimental realization of an interaction-induced topological phase in cold-atom quantum simulators. 
In this context, it is important to better understand how to bring the quantum state into the interaction-induced topological phase.
The task requires finding paths in the phase diagram to cross from, e.g.,~an initial state in the charge ordered phase, which can be prepared in experiments, to the QAH phase via a continuous phase transition. One route could be an adiabatic state preparation through a ramping protocol~\cite{Motruk2017,He2017}. Another interesting question concerns the detection of the QAH phase. For this, detection schemes developed for non-interacting topological phases in cold atom quantum simulators exist~\cite{PhysRevLett.107.235301,dauphin2013,Aidelsburger_2014,Tran2017,Goldman2013b,Asteria_2018,Goldman_2016} which require to be generalized and adapted to interacting systems.
Finally, as the system also exhibits topological defects, it would be interesting to study the dynamics of the formation of defects~\cite{Zurek_2005,DelCampo_2014,KibbleZurekRydberg} when changing the speed of a ramping protocol for state preparation, and by means of that characterize the topological nature of the interaction-induced QAH phase.   \\

\textbf{Code availability.} The codes for the unrestricted and restricted Hartree-Fock studies of this work are openly available in the repository~\cite{repo}.

\textbf{Acknowledgments}. DMRG calculations were performed using the TeNPy Library~\cite{tenpy}.
ICFO group acknowledges support from: ERC AdG NOQIA; Agencia Estatal de Investigación ({R\&D} project CEX2019-000910-S, funded by MCIN/ AEI/10.13039/501100011033, Plan National FIDEUA PID2019-106901GB-I00, FPI, QUANTERA MAQS PCI2019-111828-2, Proyectos de I+D+I “Retos Colaboración” QUSPIN RTC2019-007196-7); Fundació Cellex; Fundació Mir-Puig; Generalitat de Catalunya through the European Social Fund FEDER and CERCA program (AGAUR Grant No. 2017 SGR 134, QuantumCAT \ U16-011424, co-funded by ERDF Operational Program of Catalonia 2014-2020); EU Horizon 2020 FET-OPEN OPTOlogic (Grant No 899794); National Science Centre, Poland (Symfonia Grant No. 2016/20/W/ST4/00314); European Union’s Horizon 2020 research and innovation programme under the Marie-Sk\l odowska-Curie grant agreement No 101029393 (STREDCH) and No 847648 (“La Caixa” Junior Leaders fellowships ID100010434: LCF/BQ/PI19/11690013, LCF/BQ/PI20/11760031, LCF/BQ/PR20/11770012, LCF/BQ/PR21/11840013). S. J.-F. acknowledges financial support from MCIN/AEI/10.13039/501100011033 and FSE “El FSE invierte en tu futuro” (reference code BES-2017-082118). AD further acknowledges the financial support from a fellowship granted by la Caixa Foundation (ID 100010434, fellowship code LCF/BQ/PR20/11770012).
The RWTH and FZJ group acknowledges support by the ERC Starting Grant QNets Grant Number 804247,
the EU H2020-FETFLAG-2018-03 under Grant Agreement number 820495,
by the Germany ministry of science and education (BMBF) via the VDI within the project IQuAn,
by the Deutsche Forschungsgemeinschaft through Grant No. 449905436,
and by US A.R.O. through Grant No. W911NF-21-1-0007,
and by the Office of the Director of National Intelligence (ODNI), Intelligence Advanced Research Projects Activity (IARPA), via US ARO Grant number W911NF-16-1-0070.
All statements of fact, opinions or conclusions contained herein are those of the authors and should not be construed as representing the official views or policies of ODNI, the IARPA, or the US Government.
\bibliography{Bibliography}

\begin{thebibliography}{128}%
\makeatletter
\providecommand \@ifxundefined [1]{%
 \@ifx{#1\undefined}
}%
\providecommand \@ifnum [1]{%
 \ifnum #1\expandafter \@firstoftwo
 \else \expandafter \@secondoftwo
 \fi
}%
\providecommand \@ifx [1]{%
 \ifx #1\expandafter \@firstoftwo
 \else \expandafter \@secondoftwo
 \fi
}%
\providecommand \natexlab [1]{#1}%
\providecommand \enquote  [1]{``#1''}%
\providecommand \bibnamefont  [1]{#1}%
\providecommand \bibfnamefont [1]{#1}%
\providecommand \citenamefont [1]{#1}%
\providecommand \href@noop [0]{\@secondoftwo}%
\providecommand \href [0]{\begingroup \@sanitize@url \@href}%
\providecommand \@href[1]{\@@startlink{#1}\@@href}%
\providecommand \@@href[1]{\endgroup#1\@@endlink}%
\providecommand \@sanitize@url [0]{\catcode `\\12\catcode `\$12\catcode
  `\&12\catcode `\#12\catcode `\^12\catcode `\_12\catcode `\%12\relax}%
\providecommand \@@startlink[1]{}%
\providecommand \@@endlink[0]{}%
\providecommand \url  [0]{\begingroup\@sanitize@url \@url }%
\providecommand \@url [1]{\endgroup\@href {#1}{\urlprefix }}%
\providecommand \urlprefix  [0]{URL }%
\providecommand \Eprint [0]{\href }%
\providecommand \doibase [0]{https://doi.org/}%
\providecommand \selectlanguage [0]{\@gobble}%
\providecommand \bibinfo  [0]{\@secondoftwo}%
\providecommand \bibfield  [0]{\@secondoftwo}%
\providecommand \translation [1]{[#1]}%
\providecommand \BibitemOpen [0]{}%
\providecommand \bibitemStop [0]{}%
\providecommand \bibitemNoStop [0]{.\EOS\space}%
\providecommand \EOS [0]{\spacefactor3000\relax}%
\providecommand \BibitemShut  [1]{\csname bibitem#1\endcsname}%
\let\auto@bib@innerbib\@empty
\bibitem [{\citenamefont {Feynman}(1982)}]{Feynman_1982}%
  \BibitemOpen
  \bibfield  {author} {\bibinfo {author} {\bibfnamefont {R.~P.}\ \bibnamefont
  {Feynman}},\ }\bibfield  {title} {\bibinfo {title} {Simulating physics with
  computers},\ }\href {https://doi.org/10.1007/bf02650179} {\bibfield
  {journal} {\bibinfo  {journal} {Int. J. Theor. Phys.}\ }\textbf {\bibinfo
  {volume} {21}},\ \bibinfo {pages} {467} (\bibinfo {year} {1982})}\BibitemShut
  {NoStop}%
\bibitem [{\citenamefont {Trabesinger}(2012)}]{Trabesinger2012}%
  \BibitemOpen
  \bibfield  {author} {\bibinfo {author} {\bibfnamefont {A.}~\bibnamefont
  {Trabesinger}},\ }\bibfield  {title} {\bibinfo {title} {Quantum simulation},\
  }\href {https://doi.org/10.1038/nphys2258} {\bibfield  {journal} {\bibinfo
  {journal} {Nature Physics 2012 8:4}\ }\textbf {\bibinfo {volume} {8}},\
  \bibinfo {pages} {263} (\bibinfo {year} {2012})}\BibitemShut {NoStop}%
\bibitem [{\citenamefont {Lewenstein}\ \emph {et~al.}(2017)\citenamefont
  {Lewenstein}, \citenamefont {Sanpera},\ and\ \citenamefont
  {Ahufinger}}]{lewenstein2017}%
  \BibitemOpen
  \bibfield  {author} {\bibinfo {author} {\bibfnamefont {M.}~\bibnamefont
  {Lewenstein}}, \bibinfo {author} {\bibfnamefont {A.}~\bibnamefont
  {Sanpera}},\ and\ \bibinfo {author} {\bibfnamefont {V.}~\bibnamefont
  {Ahufinger}},\ }\href@noop {} {\emph {\bibinfo {title} {Ultracold Atoms in
  Optical Lattices: Simulating Quantum Many Body Systems}}},\ \bibinfo
  {edition} {2nd}\ ed.\ (\bibinfo  {publisher} {Oxford University Press,
  Oxford},\ \bibinfo {year} {2017})\BibitemShut {NoStop}%
\bibitem [{\citenamefont {Brierley}\ and\ \citenamefont
  {Yun}(2021)}]{Brierley2021}%
  \BibitemOpen
  \bibfield  {author} {\bibinfo {author} {\bibfnamefont {R.}~\bibnamefont
  {Brierley}}\ and\ \bibinfo {author} {\bibfnamefont {L.}~\bibnamefont {Yun}},\
  }\bibfield  {title} {\bibinfo {title} {Ultracold quantum technologies},\
  }\href {https://doi.org/10.1038/s41567-021-01461-3} {\bibfield  {journal}
  {\bibinfo  {journal} {Nature Physics 2021 17:12}\ }\textbf {\bibinfo {volume}
  {17}},\ \bibinfo {pages} {1293} (\bibinfo {year} {2021})}\BibitemShut
  {NoStop}%
\bibitem [{\citenamefont {Santos}\ \emph {et~al.}(2000)\citenamefont {Santos},
  \citenamefont {Shlyapnikov}, \citenamefont {Zoller},\ and\ \citenamefont
  {Lewenstein}}]{Santos2000}%
  \BibitemOpen
  \bibfield  {author} {\bibinfo {author} {\bibfnamefont {L.}~\bibnamefont
  {Santos}}, \bibinfo {author} {\bibfnamefont {G.~V.}\ \bibnamefont
  {Shlyapnikov}}, \bibinfo {author} {\bibfnamefont {P.}~\bibnamefont
  {Zoller}},\ and\ \bibinfo {author} {\bibfnamefont {M.}~\bibnamefont
  {Lewenstein}},\ }\bibfield  {title} {\bibinfo {title} {Bose-{E}instein
  {C}ondensation in {T}rapped {D}ipolar {G}ases},\ }\href
  {https://doi.org/10.1103/PhysRevLett.85.1791} {\bibfield  {journal} {\bibinfo
   {journal} {Phys. Rev. Lett.}\ }\textbf {\bibinfo {volume} {85}},\ \bibinfo
  {pages} {1791} (\bibinfo {year} {2000})}\BibitemShut {NoStop}%
\bibitem [{\citenamefont {Santos}\ \emph {et~al.}(2002)\citenamefont {Santos},
  \citenamefont {Shlyapnikov}, \citenamefont {Zoller},\ and\ \citenamefont
  {Lewenstein}}]{PhysRevLett.88.139904}%
  \BibitemOpen
  \bibfield  {author} {\bibinfo {author} {\bibfnamefont {L.}~\bibnamefont
  {Santos}}, \bibinfo {author} {\bibfnamefont {G.~V.}\ \bibnamefont
  {Shlyapnikov}}, \bibinfo {author} {\bibfnamefont {P.}~\bibnamefont
  {Zoller}},\ and\ \bibinfo {author} {\bibfnamefont {M.}~\bibnamefont
  {Lewenstein}},\ }\bibfield  {title} {\bibinfo {title} {\emph{erratum}},\
  }\href {https://doi.org/10.1103/PhysRevLett.88.139904} {\bibfield  {journal}
  {\bibinfo  {journal} {Phys. Rev. Lett.}\ }\textbf {\bibinfo {volume} {88}},\
  \bibinfo {pages} {139904} (\bibinfo {year} {2002})}\BibitemShut {NoStop}%
\bibitem [{\citenamefont {Saffman}\ \emph {et~al.}(2010)\citenamefont
  {Saffman}, \citenamefont {Walker},\ and\ \citenamefont
  {M\o{}lmer}}]{saffman-rmp-82-2313}%
  \BibitemOpen
  \bibfield  {author} {\bibinfo {author} {\bibfnamefont {M.}~\bibnamefont
  {Saffman}}, \bibinfo {author} {\bibfnamefont {T.~G.}\ \bibnamefont
  {Walker}},\ and\ \bibinfo {author} {\bibfnamefont {K.}~\bibnamefont
  {M\o{}lmer}},\ }\bibfield  {title} {\bibinfo {title} {Quantum information
  with {R}ydberg atoms},\ }\href {https://doi.org/10.1103/RevModPhys.82.2313}
  {\bibfield  {journal} {\bibinfo  {journal} {Rev. Mod. Phys.}\ }\textbf
  {\bibinfo {volume} {82}},\ \bibinfo {pages} {2313} (\bibinfo {year}
  {2010})}\BibitemShut {NoStop}%
\bibitem [{\citenamefont {Heidemann}\ \emph
  {et~al.}(2007{\natexlab{a}})\citenamefont {Heidemann}, \citenamefont
  {Raitzsch}, \citenamefont {Bendkowsky}, \citenamefont {Butscher},
  \citenamefont {L{\"o}w}, \citenamefont {Santos},\ and\ \citenamefont
  {Pfau}}]{heidemann_evidence_2007}%
  \BibitemOpen
  \bibfield  {author} {\bibinfo {author} {\bibfnamefont {R.}~\bibnamefont
  {Heidemann}}, \bibinfo {author} {\bibfnamefont {U.}~\bibnamefont {Raitzsch}},
  \bibinfo {author} {\bibfnamefont {V.}~\bibnamefont {Bendkowsky}}, \bibinfo
  {author} {\bibfnamefont {B.}~\bibnamefont {Butscher}}, \bibinfo {author}
  {\bibfnamefont {R.}~\bibnamefont {L{\"o}w}}, \bibinfo {author} {\bibfnamefont
  {L.}~\bibnamefont {Santos}},\ and\ \bibinfo {author} {\bibfnamefont
  {T.}~\bibnamefont {Pfau}},\ }\bibfield  {title} {\bibinfo {title} {{Evidence
  for {C}oherent {C}ollective {R}ydberg {E}xcitation in the {S}trong {B}lockade
  {R}egime}},\ }\href {https://doi.org/10.1103/PhysRevLett.99.163601}
  {\bibfield  {journal} {\bibinfo  {journal} {Phys. Rev. Lett.}\ }\textbf
  {\bibinfo {volume} {99}},\ \bibinfo {pages} {163601} (\bibinfo {year}
  {2007}{\natexlab{a}})}\BibitemShut {NoStop}%
\bibitem [{\citenamefont {Browaeys}\ and\ \citenamefont
  {Lahaye}(2020)}]{browaeys_many-body_2020}%
  \BibitemOpen
  \bibfield  {author} {\bibinfo {author} {\bibfnamefont {A.}~\bibnamefont
  {Browaeys}}\ and\ \bibinfo {author} {\bibfnamefont {T.}~\bibnamefont
  {Lahaye}},\ }\bibfield  {title} {\bibinfo {title} {Many-body physics with
  individually controlled {R}ydberg atoms},\ }\href
  {https://doi.org/10.1038/s41567-019-0733-z} {\bibfield  {journal} {\bibinfo
  {journal} {Nat. Phys.}\ }\textbf {\bibinfo {volume} {16}},\ \bibinfo {pages}
  {132} (\bibinfo {year} {2020})}\BibitemShut {NoStop}%
\bibitem [{\citenamefont {Isenhower}\ \emph {et~al.}(2010)\citenamefont
  {Isenhower}, \citenamefont {Urban}, \citenamefont {Zhang}, \citenamefont
  {Gill}, \citenamefont {Henage}, \citenamefont {Johnson}, \citenamefont
  {Walker},\ and\ \citenamefont {Saffman}}]{isenhower_demonstration_2010}%
  \BibitemOpen
  \bibfield  {author} {\bibinfo {author} {\bibfnamefont {L.}~\bibnamefont
  {Isenhower}}, \bibinfo {author} {\bibfnamefont {E.}~\bibnamefont {Urban}},
  \bibinfo {author} {\bibfnamefont {X.~L.}\ \bibnamefont {Zhang}}, \bibinfo
  {author} {\bibfnamefont {A.~T.}\ \bibnamefont {Gill}}, \bibinfo {author}
  {\bibfnamefont {T.}~\bibnamefont {Henage}}, \bibinfo {author} {\bibfnamefont
  {T.~A.}\ \bibnamefont {Johnson}}, \bibinfo {author} {\bibfnamefont {T.~G.}\
  \bibnamefont {Walker}},\ and\ \bibinfo {author} {\bibfnamefont
  {M.}~\bibnamefont {Saffman}},\ }\bibfield  {title} {\bibinfo {title}
  {Demonstration of a {N}eutral {A}tom {C}ontrolled-{NOT} {Q}uantum {G}ate},\
  }\href {https://doi.org/10.1103/PhysRevLett.104.010503} {\bibfield  {journal}
  {\bibinfo  {journal} {Phys. Rev. Lett.}\ }\textbf {\bibinfo {volume} {104}},\
  \bibinfo {pages} {010503} (\bibinfo {year} {2010})}\BibitemShut {NoStop}%
\bibitem [{\citenamefont {Wilk}\ \emph {et~al.}(2010)\citenamefont {Wilk},
  \citenamefont {Ga{\"e}tan}, \citenamefont {Evellin}, \citenamefont {Wolters},
  \citenamefont {Miroshnychenko}, \citenamefont {Grangier},\ and\ \citenamefont
  {Browaeys}}]{wilk_entanglement_2010}%
  \BibitemOpen
  \bibfield  {author} {\bibinfo {author} {\bibfnamefont {T.}~\bibnamefont
  {Wilk}}, \bibinfo {author} {\bibfnamefont {A.}~\bibnamefont {Ga{\"e}tan}},
  \bibinfo {author} {\bibfnamefont {C.}~\bibnamefont {Evellin}}, \bibinfo
  {author} {\bibfnamefont {J.}~\bibnamefont {Wolters}}, \bibinfo {author}
  {\bibfnamefont {Y.}~\bibnamefont {Miroshnychenko}}, \bibinfo {author}
  {\bibfnamefont {P.}~\bibnamefont {Grangier}},\ and\ \bibinfo {author}
  {\bibfnamefont {A.}~\bibnamefont {Browaeys}},\ }\bibfield  {title} {\bibinfo
  {title} {Entanglement of {T}wo {I}ndividual {N}eutral {A}toms {U}sing
  {R}ydberg {B}lockade},\ }\href
  {https://doi.org/10.1103/PhysRevLett.104.010502} {\bibfield  {journal}
  {\bibinfo  {journal} {Phys. Rev. Lett.}\ }\textbf {\bibinfo {volume} {104}},\
  \bibinfo {pages} {010502} (\bibinfo {year} {2010})}\BibitemShut {NoStop}%
\bibitem [{\citenamefont {Levine}\ \emph {et~al.}(2018)\citenamefont {Levine},
  \citenamefont {Keesling}, \citenamefont {Omran}, \citenamefont {Bernien},
  \citenamefont {Schwartz}, \citenamefont {Zibrov}, \citenamefont {Endres},
  \citenamefont {Greiner}, \citenamefont {Vuleti{\'c}},\ and\ \citenamefont
  {Lukin}}]{levine_high-fidelity_2018}%
  \BibitemOpen
  \bibfield  {author} {\bibinfo {author} {\bibfnamefont {H.}~\bibnamefont
  {Levine}}, \bibinfo {author} {\bibfnamefont {A.}~\bibnamefont {Keesling}},
  \bibinfo {author} {\bibfnamefont {A.}~\bibnamefont {Omran}}, \bibinfo
  {author} {\bibfnamefont {H.}~\bibnamefont {Bernien}}, \bibinfo {author}
  {\bibfnamefont {S.}~\bibnamefont {Schwartz}}, \bibinfo {author}
  {\bibfnamefont {A.~S.}\ \bibnamefont {Zibrov}}, \bibinfo {author}
  {\bibfnamefont {M.}~\bibnamefont {Endres}}, \bibinfo {author} {\bibfnamefont
  {M.}~\bibnamefont {Greiner}}, \bibinfo {author} {\bibfnamefont
  {V.}~\bibnamefont {Vuleti{\'c}}},\ and\ \bibinfo {author} {\bibfnamefont
  {M.~D.}\ \bibnamefont {Lukin}},\ }\bibfield  {title} {\bibinfo {title}
  {High-{F}idelity {C}ontrol and {E}ntanglement of {R}ydberg-{A}tom {Q}ubits},\
  }\href {https://doi.org/10.1103/PhysRevLett.121.123603} {\bibfield  {journal}
  {\bibinfo  {journal} {Phys. Rev. Lett.}\ }\textbf {\bibinfo {volume} {121}},\
  \bibinfo {pages} {123603} (\bibinfo {year} {2018})}\BibitemShut {NoStop}%
\bibitem [{\citenamefont {Omran}\ \emph {et~al.}(2019)\citenamefont {Omran},
  \citenamefont {Levine}, \citenamefont {Keesling}, \citenamefont {Semeghini},
  \citenamefont {Wang}, \citenamefont {Ebadi}, \citenamefont {Bernien},
  \citenamefont {Zibrov}, \citenamefont {Pichler}, \citenamefont {Choi},
  \citenamefont {Cui}, \citenamefont {Rossignolo}, \citenamefont {Rembold},
  \citenamefont {Montangero}, \citenamefont {Calarco}, \citenamefont {Endres},
  \citenamefont {Greiner}, \citenamefont {Vuleti{\'c}},\ and\ \citenamefont
  {Lukin}}]{omran_generation_2019}%
  \BibitemOpen
  \bibfield  {author} {\bibinfo {author} {\bibfnamefont {A.}~\bibnamefont
  {Omran}}, \bibinfo {author} {\bibfnamefont {H.}~\bibnamefont {Levine}},
  \bibinfo {author} {\bibfnamefont {A.}~\bibnamefont {Keesling}}, \bibinfo
  {author} {\bibfnamefont {G.}~\bibnamefont {Semeghini}}, \bibinfo {author}
  {\bibfnamefont {T.~T.}\ \bibnamefont {Wang}}, \bibinfo {author}
  {\bibfnamefont {S.}~\bibnamefont {Ebadi}}, \bibinfo {author} {\bibfnamefont
  {H.}~\bibnamefont {Bernien}}, \bibinfo {author} {\bibfnamefont {A.~S.}\
  \bibnamefont {Zibrov}}, \bibinfo {author} {\bibfnamefont {H.}~\bibnamefont
  {Pichler}}, \bibinfo {author} {\bibfnamefont {S.}~\bibnamefont {Choi}},
  \bibinfo {author} {\bibfnamefont {J.}~\bibnamefont {Cui}}, \bibinfo {author}
  {\bibfnamefont {M.}~\bibnamefont {Rossignolo}}, \bibinfo {author}
  {\bibfnamefont {P.}~\bibnamefont {Rembold}}, \bibinfo {author} {\bibfnamefont
  {S.}~\bibnamefont {Montangero}}, \bibinfo {author} {\bibfnamefont
  {T.}~\bibnamefont {Calarco}}, \bibinfo {author} {\bibfnamefont
  {M.}~\bibnamefont {Endres}}, \bibinfo {author} {\bibfnamefont
  {M.}~\bibnamefont {Greiner}}, \bibinfo {author} {\bibfnamefont
  {V.}~\bibnamefont {Vuleti{\'c}}},\ and\ \bibinfo {author} {\bibfnamefont
  {M.~D.}\ \bibnamefont {Lukin}},\ }\bibfield  {title} {\bibinfo {title}
  {Generation and manipulation of {S}chr{\"o}dinger cat states in {R}ydberg
  atom arrays},\ }\href {https://doi.org/10.1126/science.aax9743} {\bibfield
  {journal} {\bibinfo  {journal} {Science}\ }\textbf {\bibinfo {volume}
  {365}},\ \bibinfo {pages} {570} (\bibinfo {year} {2019})}\BibitemShut
  {NoStop}%
\bibitem [{\citenamefont {Henriet}\ \emph {et~al.}(2020)\citenamefont
  {Henriet}, \citenamefont {Beguin}, \citenamefont {Signoles}, \citenamefont
  {Lahaye}, \citenamefont {Browaeys}, \citenamefont {Reymond},\ and\
  \citenamefont {Jurczak}}]{henriet_quantum_2020}%
  \BibitemOpen
  \bibfield  {author} {\bibinfo {author} {\bibfnamefont {L.}~\bibnamefont
  {Henriet}}, \bibinfo {author} {\bibfnamefont {L.}~\bibnamefont {Beguin}},
  \bibinfo {author} {\bibfnamefont {A.}~\bibnamefont {Signoles}}, \bibinfo
  {author} {\bibfnamefont {T.}~\bibnamefont {Lahaye}}, \bibinfo {author}
  {\bibfnamefont {A.}~\bibnamefont {Browaeys}}, \bibinfo {author}
  {\bibfnamefont {G.-O.}\ \bibnamefont {Reymond}},\ and\ \bibinfo {author}
  {\bibfnamefont {C.}~\bibnamefont {Jurczak}},\ }\bibfield  {title} {\bibinfo
  {title} {Quantum computing with neutral atoms},\ }\href
  {https://doi.org/10.22331/q-2020-09-21-327} {\bibfield  {journal} {\bibinfo
  {journal} {Quantum}\ }\textbf {\bibinfo {volume} {4}},\ \bibinfo {pages}
  {327} (\bibinfo {year} {2020})}\BibitemShut {NoStop}%
\bibitem [{\citenamefont {Cong}\ \emph {et~al.}(2021)\citenamefont {Cong},
  \citenamefont {Wang}, \citenamefont {Levine}, \citenamefont {Keesling},\ and\
  \citenamefont {Lukin}}]{cong_hardware-efficient_2021}%
  \BibitemOpen
  \bibfield  {author} {\bibinfo {author} {\bibfnamefont {I.}~\bibnamefont
  {Cong}}, \bibinfo {author} {\bibfnamefont {S.-T.}\ \bibnamefont {Wang}},
  \bibinfo {author} {\bibfnamefont {H.}~\bibnamefont {Levine}}, \bibinfo
  {author} {\bibfnamefont {A.}~\bibnamefont {Keesling}},\ and\ \bibinfo
  {author} {\bibfnamefont {M.~D.}\ \bibnamefont {Lukin}},\ }\href@noop {}
  {\bibinfo {title} {Hardware-{E}fficient, {F}ault-{T}olerant {Q}uantum
  {C}omputation with {R}ydberg {A}toms}} (\bibinfo {year} {2021}),\ \Eprint
  {https://arxiv.org/abs/2105.13501} {arXiv:2105.13501 [quant-ph]} \BibitemShut
  {NoStop}%
\bibitem [{\citenamefont {Endres}\ \emph {et~al.}(2016)\citenamefont {Endres},
  \citenamefont {Bernien}, \citenamefont {Keesling}, \citenamefont {Levine},
  \citenamefont {Anschuetz}, \citenamefont {Krajenbrink}, \citenamefont
  {Senko}, \citenamefont {Vuletic}, \citenamefont {Greiner},\ and\
  \citenamefont {Lukin}}]{endres_atom-by-atom_2016}%
  \BibitemOpen
  \bibfield  {author} {\bibinfo {author} {\bibfnamefont {M.}~\bibnamefont
  {Endres}}, \bibinfo {author} {\bibfnamefont {H.}~\bibnamefont {Bernien}},
  \bibinfo {author} {\bibfnamefont {A.}~\bibnamefont {Keesling}}, \bibinfo
  {author} {\bibfnamefont {H.}~\bibnamefont {Levine}}, \bibinfo {author}
  {\bibfnamefont {E.~R.}\ \bibnamefont {Anschuetz}}, \bibinfo {author}
  {\bibfnamefont {A.}~\bibnamefont {Krajenbrink}}, \bibinfo {author}
  {\bibfnamefont {C.}~\bibnamefont {Senko}}, \bibinfo {author} {\bibfnamefont
  {V.}~\bibnamefont {Vuletic}}, \bibinfo {author} {\bibfnamefont
  {M.}~\bibnamefont {Greiner}},\ and\ \bibinfo {author} {\bibfnamefont {M.~D.}\
  \bibnamefont {Lukin}},\ }\bibfield  {title} {\bibinfo {title} {Atom-by-atom
  assembly of defect-free one-dimensional cold atom arrays},\ }\href
  {https://doi.org/10.1126/science.aah3752} {\bibfield  {journal} {\bibinfo
  {journal} {Science}\ }\textbf {\bibinfo {volume} {354}},\ \bibinfo {pages}
  {1024} (\bibinfo {year} {2016})}\BibitemShut {NoStop}%
\bibitem [{\citenamefont {de~L{\'e}s{\'e}leuc}\ \emph
  {et~al.}(2019)\citenamefont {de~L{\'e}s{\'e}leuc}, \citenamefont {Lienhard},
  \citenamefont {Scholl}, \citenamefont {Barredo}, \citenamefont {Weber},
  \citenamefont {Lang}, \citenamefont {B{\"u}chler}, \citenamefont {Lahaye},\
  and\ \citenamefont {Browaeys}}]{de_leseleuc_observation_2019}%
  \BibitemOpen
  \bibfield  {author} {\bibinfo {author} {\bibfnamefont {S.}~\bibnamefont
  {de~L{\'e}s{\'e}leuc}}, \bibinfo {author} {\bibfnamefont {V.}~\bibnamefont
  {Lienhard}}, \bibinfo {author} {\bibfnamefont {P.}~\bibnamefont {Scholl}},
  \bibinfo {author} {\bibfnamefont {D.}~\bibnamefont {Barredo}}, \bibinfo
  {author} {\bibfnamefont {S.}~\bibnamefont {Weber}}, \bibinfo {author}
  {\bibfnamefont {N.}~\bibnamefont {Lang}}, \bibinfo {author} {\bibfnamefont
  {H.~P.}\ \bibnamefont {B{\"u}chler}}, \bibinfo {author} {\bibfnamefont
  {T.}~\bibnamefont {Lahaye}},\ and\ \bibinfo {author} {\bibfnamefont
  {A.}~\bibnamefont {Browaeys}},\ }\bibfield  {title} {\bibinfo {title}
  {Observation of a symmetry-protected topological phase of interacting bosons
  with {R}ydberg atoms},\ }\href {https://doi.org/10.1126/science.aav9105}
  {\bibfield  {journal} {\bibinfo  {journal} {Science}\ }\textbf {\bibinfo
  {volume} {365}},\ \bibinfo {pages} {775} (\bibinfo {year}
  {2019})}\BibitemShut {NoStop}%
\bibitem [{\citenamefont {Bernien}\ \emph {et~al.}(2017)\citenamefont
  {Bernien}, \citenamefont {Schwartz}, \citenamefont {Keesling}, \citenamefont
  {Levine}, \citenamefont {Omran}, \citenamefont {Pichler}, \citenamefont
  {Choi}, \citenamefont {Zibrov}, \citenamefont {Endres}, \citenamefont
  {Greiner}, \citenamefont {Vuleti{\'c}},\ and\ \citenamefont
  {Lukin}}]{bernien_probing_2017}%
  \BibitemOpen
  \bibfield  {author} {\bibinfo {author} {\bibfnamefont {H.}~\bibnamefont
  {Bernien}}, \bibinfo {author} {\bibfnamefont {S.}~\bibnamefont {Schwartz}},
  \bibinfo {author} {\bibfnamefont {A.}~\bibnamefont {Keesling}}, \bibinfo
  {author} {\bibfnamefont {H.}~\bibnamefont {Levine}}, \bibinfo {author}
  {\bibfnamefont {A.}~\bibnamefont {Omran}}, \bibinfo {author} {\bibfnamefont
  {H.}~\bibnamefont {Pichler}}, \bibinfo {author} {\bibfnamefont
  {S.}~\bibnamefont {Choi}}, \bibinfo {author} {\bibfnamefont {A.~S.}\
  \bibnamefont {Zibrov}}, \bibinfo {author} {\bibfnamefont {M.}~\bibnamefont
  {Endres}}, \bibinfo {author} {\bibfnamefont {M.}~\bibnamefont {Greiner}},
  \bibinfo {author} {\bibfnamefont {V.}~\bibnamefont {Vuleti{\'c}}},\ and\
  \bibinfo {author} {\bibfnamefont {M.~D.}\ \bibnamefont {Lukin}},\ }\bibfield
  {title} {\bibinfo {title} {Probing many-body dynamics on a 51-atom quantum
  simulator},\ }\href {https://doi.org/10.1038/nature24622} {\bibfield
  {journal} {\bibinfo  {journal} {Nature}\ }\textbf {\bibinfo {volume} {551}},\
  \bibinfo {pages} {579} (\bibinfo {year} {2017})}\BibitemShut {NoStop}%
\bibitem [{\citenamefont {Keesling}\ \emph
  {et~al.}(2019{\natexlab{a}})\citenamefont {Keesling}, \citenamefont {Omran},
  \citenamefont {Levine}, \citenamefont {Bernien}, \citenamefont {Pichler},
  \citenamefont {Choi}, \citenamefont {Samajdar}, \citenamefont {Schwartz},
  \citenamefont {Silvi}, \citenamefont {Sachdev}, \citenamefont {Zoller},
  \citenamefont {Endres}, \citenamefont {Greiner}, \citenamefont
  {Vuleti{\'c}},\ and\ \citenamefont {Lukin}}]{keesling_quantum_2019}%
  \BibitemOpen
  \bibfield  {author} {\bibinfo {author} {\bibfnamefont {A.}~\bibnamefont
  {Keesling}}, \bibinfo {author} {\bibfnamefont {A.}~\bibnamefont {Omran}},
  \bibinfo {author} {\bibfnamefont {H.}~\bibnamefont {Levine}}, \bibinfo
  {author} {\bibfnamefont {H.}~\bibnamefont {Bernien}}, \bibinfo {author}
  {\bibfnamefont {H.}~\bibnamefont {Pichler}}, \bibinfo {author} {\bibfnamefont
  {S.}~\bibnamefont {Choi}}, \bibinfo {author} {\bibfnamefont {R.}~\bibnamefont
  {Samajdar}}, \bibinfo {author} {\bibfnamefont {S.}~\bibnamefont {Schwartz}},
  \bibinfo {author} {\bibfnamefont {P.}~\bibnamefont {Silvi}}, \bibinfo
  {author} {\bibfnamefont {S.}~\bibnamefont {Sachdev}}, \bibinfo {author}
  {\bibfnamefont {P.}~\bibnamefont {Zoller}}, \bibinfo {author} {\bibfnamefont
  {M.}~\bibnamefont {Endres}}, \bibinfo {author} {\bibfnamefont
  {M.}~\bibnamefont {Greiner}}, \bibinfo {author} {\bibfnamefont
  {V.}~\bibnamefont {Vuleti{\'c}}},\ and\ \bibinfo {author} {\bibfnamefont
  {M.~D.}\ \bibnamefont {Lukin}},\ }\bibfield  {title} {\bibinfo {title}
  {Quantum {K}ibble--{Z}urek mechanism and critical dynamics on a programmable
  {R}ydberg simulator},\ }\href {https://doi.org/10.1038/s41586-019-1070-1}
  {\bibfield  {journal} {\bibinfo  {journal} {Nature}\ }\textbf {\bibinfo
  {volume} {568}},\ \bibinfo {pages} {207} (\bibinfo {year}
  {2019}{\natexlab{a}})}\BibitemShut {NoStop}%
\bibitem [{\citenamefont {Labuhn}\ \emph {et~al.}(2016)\citenamefont {Labuhn},
  \citenamefont {Barredo}, \citenamefont {Ravets}, \citenamefont
  {de~L{\'e}s{\'e}leuc}, \citenamefont {Macr{\`\i}}, \citenamefont {Lahaye},\
  and\ \citenamefont {Browaeys}}]{labuhn_tunable_2016}%
  \BibitemOpen
  \bibfield  {author} {\bibinfo {author} {\bibfnamefont {H.}~\bibnamefont
  {Labuhn}}, \bibinfo {author} {\bibfnamefont {D.}~\bibnamefont {Barredo}},
  \bibinfo {author} {\bibfnamefont {S.}~\bibnamefont {Ravets}}, \bibinfo
  {author} {\bibfnamefont {S.}~\bibnamefont {de~L{\'e}s{\'e}leuc}}, \bibinfo
  {author} {\bibfnamefont {T.}~\bibnamefont {Macr{\`\i}}}, \bibinfo {author}
  {\bibfnamefont {T.}~\bibnamefont {Lahaye}},\ and\ \bibinfo {author}
  {\bibfnamefont {A.}~\bibnamefont {Browaeys}},\ }\bibfield  {title} {\bibinfo
  {title} {Tunable two-dimensional arrays of single {R}ydberg atoms for
  realizing quantum {I}sing models},\ }\href
  {https://doi.org/10.1038/nature18274} {\bibfield  {journal} {\bibinfo
  {journal} {Nature}\ }\textbf {\bibinfo {volume} {534}},\ \bibinfo {pages}
  {667} (\bibinfo {year} {2016})}\BibitemShut {NoStop}%
\bibitem [{\citenamefont {Scholl}\ \emph {et~al.}(2021)\citenamefont {Scholl},
  \citenamefont {Schuler}, \citenamefont {Williams}, \citenamefont
  {Eberharter}, \citenamefont {Barredo}, \citenamefont {Schymik}, \citenamefont
  {Lienhard}, \citenamefont {Henry}, \citenamefont {Lang}, \citenamefont
  {Lahaye}, \citenamefont {L{\"a}uchli},\ and\ \citenamefont
  {Browaeys}}]{scholl_quantum_2021}%
  \BibitemOpen
  \bibfield  {author} {\bibinfo {author} {\bibfnamefont {P.}~\bibnamefont
  {Scholl}}, \bibinfo {author} {\bibfnamefont {M.}~\bibnamefont {Schuler}},
  \bibinfo {author} {\bibfnamefont {H.~J.}\ \bibnamefont {Williams}}, \bibinfo
  {author} {\bibfnamefont {A.~A.}\ \bibnamefont {Eberharter}}, \bibinfo
  {author} {\bibfnamefont {D.}~\bibnamefont {Barredo}}, \bibinfo {author}
  {\bibfnamefont {K.-N.}\ \bibnamefont {Schymik}}, \bibinfo {author}
  {\bibfnamefont {V.}~\bibnamefont {Lienhard}}, \bibinfo {author}
  {\bibfnamefont {L.-P.}\ \bibnamefont {Henry}}, \bibinfo {author}
  {\bibfnamefont {T.~C.}\ \bibnamefont {Lang}}, \bibinfo {author}
  {\bibfnamefont {T.}~\bibnamefont {Lahaye}}, \bibinfo {author} {\bibfnamefont
  {A.~M.}\ \bibnamefont {L{\"a}uchli}},\ and\ \bibinfo {author} {\bibfnamefont
  {A.}~\bibnamefont {Browaeys}},\ }\bibfield  {title} {\bibinfo {title}
  {Quantum simulation of 2{D} antiferromagnets with hundreds of {R}ydberg
  atoms},\ }\href {https://doi.org/10.1038/s41586-021-03585-1} {\bibfield
  {journal} {\bibinfo  {journal} {Nature}\ }\textbf {\bibinfo {volume} {595}},\
  \bibinfo {pages} {233} (\bibinfo {year} {2021})}\BibitemShut {NoStop}%
\bibitem [{\citenamefont {Ebadi}\ \emph {et~al.}(2021)\citenamefont {Ebadi},
  \citenamefont {Wang}, \citenamefont {Levine}, \citenamefont {Keesling},
  \citenamefont {Semeghini}, \citenamefont {Omran}, \citenamefont {Bluvstein},
  \citenamefont {Samajdar}, \citenamefont {Pichler}, \citenamefont {Ho},
  \citenamefont {Choi}, \citenamefont {Sachdev}, \citenamefont {Greiner},
  \citenamefont {Vuleti{\'c}},\ and\ \citenamefont
  {Lukin}}]{ebadi_quantum_2021}%
  \BibitemOpen
  \bibfield  {author} {\bibinfo {author} {\bibfnamefont {S.}~\bibnamefont
  {Ebadi}}, \bibinfo {author} {\bibfnamefont {T.~T.}\ \bibnamefont {Wang}},
  \bibinfo {author} {\bibfnamefont {H.}~\bibnamefont {Levine}}, \bibinfo
  {author} {\bibfnamefont {A.}~\bibnamefont {Keesling}}, \bibinfo {author}
  {\bibfnamefont {G.}~\bibnamefont {Semeghini}}, \bibinfo {author}
  {\bibfnamefont {A.}~\bibnamefont {Omran}}, \bibinfo {author} {\bibfnamefont
  {D.}~\bibnamefont {Bluvstein}}, \bibinfo {author} {\bibfnamefont
  {R.}~\bibnamefont {Samajdar}}, \bibinfo {author} {\bibfnamefont
  {H.}~\bibnamefont {Pichler}}, \bibinfo {author} {\bibfnamefont {W.~W.}\
  \bibnamefont {Ho}}, \bibinfo {author} {\bibfnamefont {S.}~\bibnamefont
  {Choi}}, \bibinfo {author} {\bibfnamefont {S.}~\bibnamefont {Sachdev}},
  \bibinfo {author} {\bibfnamefont {M.}~\bibnamefont {Greiner}}, \bibinfo
  {author} {\bibfnamefont {V.}~\bibnamefont {Vuleti{\'c}}},\ and\ \bibinfo
  {author} {\bibfnamefont {M.~D.}\ \bibnamefont {Lukin}},\ }\bibfield  {title}
  {\bibinfo {title} {Quantum phases of matter on a 256-atom programmable
  quantum simulator},\ }\href {https://doi.org/10.1038/s41586-021-03582-4}
  {\bibfield  {journal} {\bibinfo  {journal} {Nature}\ }\textbf {\bibinfo
  {volume} {595}},\ \bibinfo {pages} {227} (\bibinfo {year}
  {2021})}\BibitemShut {NoStop}%
\bibitem [{\citenamefont {Semeghini}\ \emph {et~al.}(2021)\citenamefont
  {Semeghini}, \citenamefont {Levine}, \citenamefont {Keesling}, \citenamefont
  {Ebadi}, \citenamefont {Wang}, \citenamefont {Bluvstein}, \citenamefont
  {Verresen}, \citenamefont {Pichler}, \citenamefont {Kalinowski},
  \citenamefont {Samajdar}, \citenamefont {Omran}, \citenamefont {Sachdev},
  \citenamefont {Vishwanath}, \citenamefont {Greiner}, \citenamefont
  {Vuleti{\'c}},\ and\ \citenamefont {Lukin}}]{semeghini_probing_2021}%
  \BibitemOpen
  \bibfield  {author} {\bibinfo {author} {\bibfnamefont {G.}~\bibnamefont
  {Semeghini}}, \bibinfo {author} {\bibfnamefont {H.}~\bibnamefont {Levine}},
  \bibinfo {author} {\bibfnamefont {A.}~\bibnamefont {Keesling}}, \bibinfo
  {author} {\bibfnamefont {S.}~\bibnamefont {Ebadi}}, \bibinfo {author}
  {\bibfnamefont {T.~T.}\ \bibnamefont {Wang}}, \bibinfo {author}
  {\bibfnamefont {D.}~\bibnamefont {Bluvstein}}, \bibinfo {author}
  {\bibfnamefont {R.}~\bibnamefont {Verresen}}, \bibinfo {author}
  {\bibfnamefont {H.}~\bibnamefont {Pichler}}, \bibinfo {author} {\bibfnamefont
  {M.}~\bibnamefont {Kalinowski}}, \bibinfo {author} {\bibfnamefont
  {R.}~\bibnamefont {Samajdar}}, \bibinfo {author} {\bibfnamefont
  {A.}~\bibnamefont {Omran}}, \bibinfo {author} {\bibfnamefont
  {S.}~\bibnamefont {Sachdev}}, \bibinfo {author} {\bibfnamefont
  {A.}~\bibnamefont {Vishwanath}}, \bibinfo {author} {\bibfnamefont
  {M.}~\bibnamefont {Greiner}}, \bibinfo {author} {\bibfnamefont
  {V.}~\bibnamefont {Vuleti{\'c}}},\ and\ \bibinfo {author} {\bibfnamefont
  {M.~D.}\ \bibnamefont {Lukin}},\ }\bibfield  {title} {\bibinfo {title}
  {Probing topological spin liquids on a programmable quantum simulator},\
  }\href {https://doi.org/10.1126/science.abi8794} {\bibfield  {journal}
  {\bibinfo  {journal} {Science}\ }\textbf {\bibinfo {volume} {374}},\ \bibinfo
  {pages} {1242} (\bibinfo {year} {2021})}\BibitemShut {NoStop}%
\bibitem [{\citenamefont {Chomaz}\ \emph {et~al.}(2016)\citenamefont {Chomaz},
  \citenamefont {Baier}, \citenamefont {Petter}, \citenamefont {Mark},
  \citenamefont {W{\"a}chtler}, \citenamefont {Santos},\ and\ \citenamefont
  {Ferlaino}}]{Chomaz2016}%
  \BibitemOpen
  \bibfield  {author} {\bibinfo {author} {\bibfnamefont {L.}~\bibnamefont
  {Chomaz}}, \bibinfo {author} {\bibfnamefont {S.}~\bibnamefont {Baier}},
  \bibinfo {author} {\bibfnamefont {D.}~\bibnamefont {Petter}}, \bibinfo
  {author} {\bibfnamefont {M.~J.}\ \bibnamefont {Mark}}, \bibinfo {author}
  {\bibfnamefont {F.}~\bibnamefont {W{\"a}chtler}}, \bibinfo {author}
  {\bibfnamefont {L.}~\bibnamefont {Santos}},\ and\ \bibinfo {author}
  {\bibfnamefont {F.}~\bibnamefont {Ferlaino}},\ }\bibfield  {title} {\bibinfo
  {title} {Quantum-{F}luctuation-{D}riven {C}rossover from a {D}ilute
  {B}ose-{E}instein {C}ondensate to a {M}acrodroplet in a {D}ipolar {Q}uantum
  {F}luid},\ }\href {https://doi.org/10.1103/PhysRevX.6.041039} {\bibfield
  {journal} {\bibinfo  {journal} {Phys. Rev. X}\ }\textbf {\bibinfo {volume}
  {6}},\ \bibinfo {pages} {041039} (\bibinfo {year} {2016})}\BibitemShut
  {NoStop}%
\bibitem [{\citenamefont {Schmitt}\ \emph {et~al.}(2016)\citenamefont
  {Schmitt}, \citenamefont {Wenzel}, \citenamefont {B{\"o}ttcher},
  \citenamefont {Ferrier-Barbut},\ and\ \citenamefont {Pfau}}]{Schmitt2016}%
  \BibitemOpen
  \bibfield  {author} {\bibinfo {author} {\bibfnamefont {M.}~\bibnamefont
  {Schmitt}}, \bibinfo {author} {\bibfnamefont {M.}~\bibnamefont {Wenzel}},
  \bibinfo {author} {\bibfnamefont {F.}~\bibnamefont {B{\"o}ttcher}}, \bibinfo
  {author} {\bibfnamefont {I.}~\bibnamefont {Ferrier-Barbut}},\ and\ \bibinfo
  {author} {\bibfnamefont {T.}~\bibnamefont {Pfau}},\ }\bibfield  {title}
  {\bibinfo {title} {Self-bound droplets of a dilute magnetic quantum liquid},\
  }\href {https://doi.org/https://doi.org/10.1038/nature20126} {\bibfield
  {journal} {\bibinfo  {journal} {Nature}\ }\textbf {\bibinfo {volume} {539}},\
  \bibinfo {pages} {259} (\bibinfo {year} {2016})}\BibitemShut {NoStop}%
\bibitem [{\citenamefont {B\"ottcher}\ \emph {et~al.}(2019)\citenamefont
  {B\"ottcher}, \citenamefont {Schmidt}, \citenamefont {Wenzel}, \citenamefont
  {Hertkorn}, \citenamefont {Guo}, \citenamefont {Langen},\ and\ \citenamefont
  {Pfau}}]{Boettcher2019}%
  \BibitemOpen
  \bibfield  {author} {\bibinfo {author} {\bibfnamefont {F.}~\bibnamefont
  {B\"ottcher}}, \bibinfo {author} {\bibfnamefont {J.-N.}\ \bibnamefont
  {Schmidt}}, \bibinfo {author} {\bibfnamefont {M.}~\bibnamefont {Wenzel}},
  \bibinfo {author} {\bibfnamefont {J.}~\bibnamefont {Hertkorn}}, \bibinfo
  {author} {\bibfnamefont {M.}~\bibnamefont {Guo}}, \bibinfo {author}
  {\bibfnamefont {T.}~\bibnamefont {Langen}},\ and\ \bibinfo {author}
  {\bibfnamefont {T.}~\bibnamefont {Pfau}},\ }\bibfield  {title} {\bibinfo
  {title} {Transient {S}upersolid {P}roperties in an {A}rray of {D}ipolar
  {Q}uantum {D}roplets},\ }\href {https://doi.org/10.1103/PhysRevX.9.011051}
  {\bibfield  {journal} {\bibinfo  {journal} {Phys. Rev. X}\ }\textbf {\bibinfo
  {volume} {9}},\ \bibinfo {pages} {011051} (\bibinfo {year}
  {2019})}\BibitemShut {NoStop}%
\bibitem [{\citenamefont {Tanzi}\ \emph {et~al.}(2019)\citenamefont {Tanzi},
  \citenamefont {Lucioni}, \citenamefont {Fam\`a}, \citenamefont {Catani},
  \citenamefont {Fioretti}, \citenamefont {Gabbanini}, \citenamefont {Bisset},
  \citenamefont {Santos},\ and\ \citenamefont {Modugno}}]{Tanzi2019a}%
  \BibitemOpen
  \bibfield  {author} {\bibinfo {author} {\bibfnamefont {L.}~\bibnamefont
  {Tanzi}}, \bibinfo {author} {\bibfnamefont {E.}~\bibnamefont {Lucioni}},
  \bibinfo {author} {\bibfnamefont {F.}~\bibnamefont {Fam\`a}}, \bibinfo
  {author} {\bibfnamefont {J.}~\bibnamefont {Catani}}, \bibinfo {author}
  {\bibfnamefont {A.}~\bibnamefont {Fioretti}}, \bibinfo {author}
  {\bibfnamefont {C.}~\bibnamefont {Gabbanini}}, \bibinfo {author}
  {\bibfnamefont {R.~N.}\ \bibnamefont {Bisset}}, \bibinfo {author}
  {\bibfnamefont {L.}~\bibnamefont {Santos}},\ and\ \bibinfo {author}
  {\bibfnamefont {G.}~\bibnamefont {Modugno}},\ }\bibfield  {title} {\bibinfo
  {title} {Observation of a {D}ipolar {Q}uantum {G}as with {M}etastable
  {S}upersolid {P}roperties},\ }\href
  {https://doi.org/10.1103/PhysRevLett.122.130405} {\bibfield  {journal}
  {\bibinfo  {journal} {Phys. Rev. Lett.}\ }\textbf {\bibinfo {volume} {122}},\
  \bibinfo {pages} {130405} (\bibinfo {year} {2019})}\BibitemShut {NoStop}%
\bibitem [{\citenamefont {Chomaz}\ \emph {et~al.}(2019)\citenamefont {Chomaz},
  \citenamefont {Petter}, \citenamefont {Ilzh{\"o}fer}, \citenamefont {Natale},
  \citenamefont {Trautmann}, \citenamefont {Politi}, \citenamefont
  {Durastante}, \citenamefont {van Bijnen}, \citenamefont {Patscheider},
  \citenamefont {Sohmen}, \citenamefont {Mark},\ and\ \citenamefont
  {Ferlaino}}]{Chomaz2019}%
  \BibitemOpen
  \bibfield  {author} {\bibinfo {author} {\bibfnamefont {L.}~\bibnamefont
  {Chomaz}}, \bibinfo {author} {\bibfnamefont {D.}~\bibnamefont {Petter}},
  \bibinfo {author} {\bibfnamefont {P.}~\bibnamefont {Ilzh{\"o}fer}}, \bibinfo
  {author} {\bibfnamefont {G.}~\bibnamefont {Natale}}, \bibinfo {author}
  {\bibfnamefont {A.}~\bibnamefont {Trautmann}}, \bibinfo {author}
  {\bibfnamefont {C.}~\bibnamefont {Politi}}, \bibinfo {author} {\bibfnamefont
  {G.}~\bibnamefont {Durastante}}, \bibinfo {author} {\bibfnamefont {R.~M.~W.}\
  \bibnamefont {van Bijnen}}, \bibinfo {author} {\bibfnamefont
  {A.}~\bibnamefont {Patscheider}}, \bibinfo {author} {\bibfnamefont
  {M.}~\bibnamefont {Sohmen}}, \bibinfo {author} {\bibfnamefont {M.~J.}\
  \bibnamefont {Mark}},\ and\ \bibinfo {author} {\bibfnamefont
  {F.}~\bibnamefont {Ferlaino}},\ }\bibfield  {title} {\bibinfo {title}
  {Long-{L}ived and {T}ransient {S}upersolid {B}ehaviors in {D}ipolar {Q}uantum
  {G}ases},\ }\href {https://doi.org/10.1103/PhysRevX.9.021012} {\bibfield
  {journal} {\bibinfo  {journal} {Phys. Rev. X}\ }\textbf {\bibinfo {volume}
  {9}},\ \bibinfo {pages} {021012} (\bibinfo {year} {2019})}\BibitemShut
  {NoStop}%
\bibitem [{\citenamefont {Norcia}\ \emph {et~al.}(2021)\citenamefont {Norcia},
  \citenamefont {Politi}, \citenamefont {Klaus}, \citenamefont {Poli},
  \citenamefont {Sohmen}, \citenamefont {Mark}, \citenamefont {Bisset},
  \citenamefont {Santos},\ and\ \citenamefont {Ferlaino}}]{Norcia2021}%
  \BibitemOpen
  \bibfield  {author} {\bibinfo {author} {\bibfnamefont {M.~A.}\ \bibnamefont
  {Norcia}}, \bibinfo {author} {\bibfnamefont {C.}~\bibnamefont {Politi}},
  \bibinfo {author} {\bibfnamefont {L.}~\bibnamefont {Klaus}}, \bibinfo
  {author} {\bibfnamefont {E.}~\bibnamefont {Poli}}, \bibinfo {author}
  {\bibfnamefont {M.}~\bibnamefont {Sohmen}}, \bibinfo {author} {\bibfnamefont
  {M.~J.}\ \bibnamefont {Mark}}, \bibinfo {author} {\bibfnamefont {R.~N.}\
  \bibnamefont {Bisset}}, \bibinfo {author} {\bibfnamefont {L.}~\bibnamefont
  {Santos}},\ and\ \bibinfo {author} {\bibfnamefont {F.}~\bibnamefont
  {Ferlaino}},\ }\bibfield  {title} {\bibinfo {title} {Two-dimensional
  supersolidity in a dipolar quantum gas},\ }\href
  {https://doi.org/https://doi.org/10.1038/s41586-021-03725-7} {\bibfield
  {journal} {\bibinfo  {journal} {Nature}\ }\textbf {\bibinfo {volume} {596}},\
  \bibinfo {pages} {357} (\bibinfo {year} {2021})}\BibitemShut {NoStop}%
\bibitem [{\citenamefont {Marco}\ \emph {et~al.}(2019)\citenamefont {Marco},
  \citenamefont {Valtolina}, \citenamefont {Matsuda}, \citenamefont {Tobias},
  \citenamefont {Covey},\ and\ \citenamefont {Ye}}]{deMarco2019}%
  \BibitemOpen
  \bibfield  {author} {\bibinfo {author} {\bibfnamefont {L.~D.}\ \bibnamefont
  {Marco}}, \bibinfo {author} {\bibfnamefont {G.}~\bibnamefont {Valtolina}},
  \bibinfo {author} {\bibfnamefont {K.}~\bibnamefont {Matsuda}}, \bibinfo
  {author} {\bibfnamefont {W.~G.}\ \bibnamefont {Tobias}}, \bibinfo {author}
  {\bibfnamefont {J.~P.}\ \bibnamefont {Covey}},\ and\ \bibinfo {author}
  {\bibfnamefont {J.}~\bibnamefont {Ye}},\ }\bibfield  {title} {\bibinfo
  {title} {A degenerate {F}ermi gas of polar molecules},\ }\href
  {https://doi.org/10.1126/science.aau7230} {\bibfield  {journal} {\bibinfo
  {journal} {Science}\ }\textbf {\bibinfo {volume} {363}},\ \bibinfo {pages}
  {853} (\bibinfo {year} {2019})}\BibitemShut {NoStop}%
\bibitem [{\citenamefont {Bohn}\ \emph {et~al.}(2017)\citenamefont {Bohn},
  \citenamefont {Rey},\ and\ \citenamefont {Ye}}]{Bohn2017}%
  \BibitemOpen
  \bibfield  {author} {\bibinfo {author} {\bibfnamefont {J.~L.}\ \bibnamefont
  {Bohn}}, \bibinfo {author} {\bibfnamefont {A.~M.}\ \bibnamefont {Rey}},\ and\
  \bibinfo {author} {\bibfnamefont {J.}~\bibnamefont {Ye}},\ }\bibfield
  {title} {\bibinfo {title} {Cold molecules: {P}rogress in quantum engineering
  of chemistry and quantum matter},\ }\href
  {https://doi.org/10.1126/science.aam6299} {\bibfield  {journal} {\bibinfo
  {journal} {Science}\ }\textbf {\bibinfo {volume} {357}},\ \bibinfo {pages}
  {1002} (\bibinfo {year} {2017})}\BibitemShut {NoStop}%
\bibitem [{\citenamefont {Trefzger}\ \emph {et~al.}(2011)\citenamefont
  {Trefzger}, \citenamefont {Menotti}, \citenamefont {Capogrosso-Sansone},\
  and\ \citenamefont {Lewenstein}}]{Trefzger2011}%
  \BibitemOpen
  \bibfield  {author} {\bibinfo {author} {\bibfnamefont {C.}~\bibnamefont
  {Trefzger}}, \bibinfo {author} {\bibfnamefont {C.}~\bibnamefont {Menotti}},
  \bibinfo {author} {\bibfnamefont {B.}~\bibnamefont {Capogrosso-Sansone}},\
  and\ \bibinfo {author} {\bibfnamefont {M.}~\bibnamefont {Lewenstein}},\
  }\bibfield  {title} {\bibinfo {title} {Ultracold dipolar gases in optical
  lattices},\ }\href {https://doi.org/10.1088/0953-4075/44/19/193001}
  {\bibfield  {journal} {\bibinfo  {journal} {J. Phys. B}\ }\textbf {\bibinfo
  {volume} {44}},\ \bibinfo {pages} {193001} (\bibinfo {year}
  {2011})}\BibitemShut {NoStop}%
\bibitem [{\citenamefont {de~Paz}\ \emph {et~al.}(2013)\citenamefont {de~Paz},
  \citenamefont {Sharma}, \citenamefont {Chotia}, \citenamefont {Mar\'echal},
  \citenamefont {Huckans}, \citenamefont {Pedri}, \citenamefont {Santos},
  \citenamefont {Gorceix}, \citenamefont {Vernac},\ and\ \citenamefont
  {Laburthe-Tolra}}]{dePaz2013}%
  \BibitemOpen
  \bibfield  {author} {\bibinfo {author} {\bibfnamefont {A.}~\bibnamefont
  {de~Paz}}, \bibinfo {author} {\bibfnamefont {A.}~\bibnamefont {Sharma}},
  \bibinfo {author} {\bibfnamefont {A.}~\bibnamefont {Chotia}}, \bibinfo
  {author} {\bibfnamefont {E.}~\bibnamefont {Mar\'echal}}, \bibinfo {author}
  {\bibfnamefont {J.~H.}\ \bibnamefont {Huckans}}, \bibinfo {author}
  {\bibfnamefont {P.}~\bibnamefont {Pedri}}, \bibinfo {author} {\bibfnamefont
  {L.}~\bibnamefont {Santos}}, \bibinfo {author} {\bibfnamefont
  {O.}~\bibnamefont {Gorceix}}, \bibinfo {author} {\bibfnamefont
  {L.}~\bibnamefont {Vernac}},\ and\ \bibinfo {author} {\bibfnamefont
  {B.}~\bibnamefont {Laburthe-Tolra}},\ }\bibfield  {title} {\bibinfo {title}
  {Nonequilibrium {Q}uantum {M}agnetism in a {D}ipolar {L}attice {G}as},\
  }\href {https://doi.org/10.1103/PhysRevLett.111.185305} {\bibfield  {journal}
  {\bibinfo  {journal} {Phys. Rev. Lett.}\ }\textbf {\bibinfo {volume} {111}},\
  \bibinfo {pages} {185305} (\bibinfo {year} {2013})}\BibitemShut {NoStop}%
\bibitem [{\citenamefont {Dutta}\ \emph {et~al.}(2015)\citenamefont {Dutta},
  \citenamefont {Gajda}, \citenamefont {Hauke}, \citenamefont {Lewenstein},
  \citenamefont {L{\"u}hmann}, \citenamefont {Malomed}, \citenamefont
  {Sowi{\'{n}}ski},\ and\ \citenamefont {Zakrzewski}}]{Dutta2015}%
  \BibitemOpen
  \bibfield  {author} {\bibinfo {author} {\bibfnamefont {O.}~\bibnamefont
  {Dutta}}, \bibinfo {author} {\bibfnamefont {M.}~\bibnamefont {Gajda}},
  \bibinfo {author} {\bibfnamefont {P.}~\bibnamefont {Hauke}}, \bibinfo
  {author} {\bibfnamefont {M.}~\bibnamefont {Lewenstein}}, \bibinfo {author}
  {\bibfnamefont {D.-S.}\ \bibnamefont {L{\"u}hmann}}, \bibinfo {author}
  {\bibfnamefont {B.~A.}\ \bibnamefont {Malomed}}, \bibinfo {author}
  {\bibfnamefont {T.}~\bibnamefont {Sowi{\'{n}}ski}},\ and\ \bibinfo {author}
  {\bibfnamefont {J.}~\bibnamefont {Zakrzewski}},\ }\bibfield  {title}
  {\bibinfo {title} {Non-standard {H}ubbard models in optical lattices: a
  review},\ }\href {https://doi.org/10.1088/0034-4885/78/6/066001} {\bibfield
  {journal} {\bibinfo  {journal} {Rep. Prog. Phys.}\ }\textbf {\bibinfo
  {volume} {78}},\ \bibinfo {pages} {066001} (\bibinfo {year}
  {2015})}\BibitemShut {NoStop}%
\bibitem [{\citenamefont {Baier}\ \emph {et~al.}(2016)\citenamefont {Baier},
  \citenamefont {Mark}, \citenamefont {Petter}, \citenamefont {Aikawa},
  \citenamefont {Chomaz}, \citenamefont {Cai}, \citenamefont {Baranov},
  \citenamefont {Zoller},\ and\ \citenamefont {Ferlaino}}]{Baier2016}%
  \BibitemOpen
  \bibfield  {author} {\bibinfo {author} {\bibfnamefont {S.}~\bibnamefont
  {Baier}}, \bibinfo {author} {\bibfnamefont {M.~J.}\ \bibnamefont {Mark}},
  \bibinfo {author} {\bibfnamefont {D.}~\bibnamefont {Petter}}, \bibinfo
  {author} {\bibfnamefont {K.}~\bibnamefont {Aikawa}}, \bibinfo {author}
  {\bibfnamefont {L.}~\bibnamefont {Chomaz}}, \bibinfo {author} {\bibfnamefont
  {Z.}~\bibnamefont {Cai}}, \bibinfo {author} {\bibfnamefont {M.}~\bibnamefont
  {Baranov}}, \bibinfo {author} {\bibfnamefont {P.}~\bibnamefont {Zoller}},\
  and\ \bibinfo {author} {\bibfnamefont {F.}~\bibnamefont {Ferlaino}},\
  }\bibfield  {title} {\bibinfo {title} {Extended {B}ose-{H}ubbard models with
  ultracold magnetic atoms},\ }\href {https://doi.org/10.1126/science.aac9812}
  {\bibfield  {journal} {\bibinfo  {journal} {Science}\ }\textbf {\bibinfo
  {volume} {352}},\ \bibinfo {pages} {201} (\bibinfo {year}
  {2016})}\BibitemShut {NoStop}%
\bibitem [{\citenamefont {Lepoutre}\ \emph {et~al.}(2019)\citenamefont
  {Lepoutre}, \citenamefont {Schachenmayer}, \citenamefont {Gabardos},
  \citenamefont {Zhu}, \citenamefont {Naylor}, \citenamefont {Mar{\'e}chal},
  \citenamefont {Gorceix}, \citenamefont {Rey}, \citenamefont {Vernac},\ and\
  \citenamefont {Laburthe-Tolra}}]{Lepoutre2019}%
  \BibitemOpen
  \bibfield  {author} {\bibinfo {author} {\bibfnamefont {S.}~\bibnamefont
  {Lepoutre}}, \bibinfo {author} {\bibfnamefont {J.}~\bibnamefont
  {Schachenmayer}}, \bibinfo {author} {\bibfnamefont {L.}~\bibnamefont
  {Gabardos}}, \bibinfo {author} {\bibfnamefont {B.}~\bibnamefont {Zhu}},
  \bibinfo {author} {\bibfnamefont {B.}~\bibnamefont {Naylor}}, \bibinfo
  {author} {\bibfnamefont {E.}~\bibnamefont {Mar{\'e}chal}}, \bibinfo {author}
  {\bibfnamefont {O.}~\bibnamefont {Gorceix}}, \bibinfo {author} {\bibfnamefont
  {A.~M.}\ \bibnamefont {Rey}}, \bibinfo {author} {\bibfnamefont
  {L.}~\bibnamefont {Vernac}},\ and\ \bibinfo {author} {\bibfnamefont
  {B.}~\bibnamefont {Laburthe-Tolra}},\ }\bibfield  {title} {\bibinfo {title}
  {Out-of-equilibrium quantum magnetism and thermalization in a spin-3
  many-body dipolar lattice system},\ }\href
  {https://doi.org/10.1038/s41467-019-09699-5} {\bibfield  {journal} {\bibinfo
  {journal} {Nat. Commun.}\ }\textbf {\bibinfo {volume} {10}},\ \bibinfo
  {pages} {1714} (\bibinfo {year} {2019})}\BibitemShut {NoStop}%
\bibitem [{\citenamefont {Patscheider}\ \emph {et~al.}(2020)\citenamefont
  {Patscheider}, \citenamefont {Zhu}, \citenamefont {Chomaz}, \citenamefont
  {Petter}, \citenamefont {Baier}, \citenamefont {Rey}, \citenamefont
  {Ferlaino},\ and\ \citenamefont {Mark}}]{Patscheider2020}%
  \BibitemOpen
  \bibfield  {author} {\bibinfo {author} {\bibfnamefont {A.}~\bibnamefont
  {Patscheider}}, \bibinfo {author} {\bibfnamefont {B.}~\bibnamefont {Zhu}},
  \bibinfo {author} {\bibfnamefont {L.}~\bibnamefont {Chomaz}}, \bibinfo
  {author} {\bibfnamefont {D.}~\bibnamefont {Petter}}, \bibinfo {author}
  {\bibfnamefont {S.}~\bibnamefont {Baier}}, \bibinfo {author} {\bibfnamefont
  {A.-M.}\ \bibnamefont {Rey}}, \bibinfo {author} {\bibfnamefont
  {F.}~\bibnamefont {Ferlaino}},\ and\ \bibinfo {author} {\bibfnamefont
  {M.~J.}\ \bibnamefont {Mark}},\ }\bibfield  {title} {\bibinfo {title}
  {Controlling dipolar exchange interactions in a dense three-dimensional array
  of large-spin fermions},\ }\href
  {https://doi.org/10.1103/PhysRevResearch.2.023050} {\bibfield  {journal}
  {\bibinfo  {journal} {Phys. Rev. Res.}\ }\textbf {\bibinfo {volume} {2}},\
  \bibinfo {pages} {023050} (\bibinfo {year} {2020})}\BibitemShut {NoStop}%
\bibitem [{\citenamefont {Pupillo}\ \emph {et~al.}(2010)\citenamefont
  {Pupillo}, \citenamefont {Micheli}, \citenamefont {Boninsegni}, \citenamefont
  {Lesanovsky},\ and\ \citenamefont {Zoller}}]{Pupillo-2010}%
  \BibitemOpen
  \bibfield  {author} {\bibinfo {author} {\bibfnamefont {G.}~\bibnamefont
  {Pupillo}}, \bibinfo {author} {\bibfnamefont {A.}~\bibnamefont {Micheli}},
  \bibinfo {author} {\bibfnamefont {M.}~\bibnamefont {Boninsegni}}, \bibinfo
  {author} {\bibfnamefont {I.}~\bibnamefont {Lesanovsky}},\ and\ \bibinfo
  {author} {\bibfnamefont {P.}~\bibnamefont {Zoller}},\ }\bibfield  {title}
  {\bibinfo {title} {Strongly {C}orrelated {G}ases of {R}ydberg-{D}ressed
  {A}toms: {Q}uantum and {C}lassical {D}ynamics},\ }\href
  {https://doi.org/10.1103/PhysRevLett.104.223002} {\bibfield  {journal}
  {\bibinfo  {journal} {Phys. Rev. Lett.}\ }\textbf {\bibinfo {volume} {104}},\
  \bibinfo {pages} {223002} (\bibinfo {year} {2010})}\BibitemShut {NoStop}%
\bibitem [{\citenamefont {Henkel}\ \emph {et~al.}(2010)\citenamefont {Henkel},
  \citenamefont {Nath},\ and\ \citenamefont {Pohl}}]{Nath-2010}%
  \BibitemOpen
  \bibfield  {author} {\bibinfo {author} {\bibfnamefont {N.}~\bibnamefont
  {Henkel}}, \bibinfo {author} {\bibfnamefont {R.}~\bibnamefont {Nath}},\ and\
  \bibinfo {author} {\bibfnamefont {T.}~\bibnamefont {Pohl}},\ }\bibfield
  {title} {\bibinfo {title} {Three-{D}imensional {R}oton {E}xcitations and
  {S}upersolid {F}ormation in {R}ydberg-{E}xcited {B}ose-{E}instein
  {C}ondensates},\ }\href {https://doi.org/10.1103/PhysRevLett.104.195302}
  {\bibfield  {journal} {\bibinfo  {journal} {Phys. Rev. Lett.}\ }\textbf
  {\bibinfo {volume} {104}},\ \bibinfo {pages} {195302} (\bibinfo {year}
  {2010})}\BibitemShut {NoStop}%
\bibitem [{\citenamefont {Johnson}\ and\ \citenamefont
  {Rolston}(2010)}]{Johnson2010}%
  \BibitemOpen
  \bibfield  {author} {\bibinfo {author} {\bibfnamefont {J.~E.}\ \bibnamefont
  {Johnson}}\ and\ \bibinfo {author} {\bibfnamefont {S.~L.}\ \bibnamefont
  {Rolston}},\ }\bibfield  {title} {\bibinfo {title} {Interactions between
  {R}ydberg-dressed atoms},\ }\href
  {https://doi.org/10.1103/PhysRevA.82.033412} {\bibfield  {journal} {\bibinfo
  {journal} {Phys. Rev. A}\ }\textbf {\bibinfo {volume} {82}},\ \bibinfo
  {pages} {033412} (\bibinfo {year} {2010})}\BibitemShut {NoStop}%
\bibitem [{\citenamefont {Jau}\ \emph {et~al.}(2016)\citenamefont {Jau},
  \citenamefont {Hankin}, \citenamefont {Keating}, \citenamefont {Deutsch},\
  and\ \citenamefont {Biedermann}}]{Jau2016}%
  \BibitemOpen
  \bibfield  {author} {\bibinfo {author} {\bibfnamefont {Y.~Y.}\ \bibnamefont
  {Jau}}, \bibinfo {author} {\bibfnamefont {A.~M.}\ \bibnamefont {Hankin}},
  \bibinfo {author} {\bibfnamefont {T.}~\bibnamefont {Keating}}, \bibinfo
  {author} {\bibfnamefont {I.~H.}\ \bibnamefont {Deutsch}},\ and\ \bibinfo
  {author} {\bibfnamefont {G.~W.}\ \bibnamefont {Biedermann}},\ }\bibfield
  {title} {\bibinfo {title} {{Entangling atomic spins with a Rydberg-dressed
  spin-flip blockade}},\ }\href
  {https://doi.org/https://doi.org/10.1038/nphys3487} {\bibfield  {journal}
  {\bibinfo  {journal} {Nat. Phys.}\ }\textbf {\bibinfo {volume} {12}},\
  \bibinfo {pages} {71} (\bibinfo {year} {2016})}\BibitemShut {NoStop}%
\bibitem [{\citenamefont {Zeiher}\ \emph {et~al.}(2016)\citenamefont {Zeiher},
  \citenamefont {van Bijnen}, \citenamefont {Schau{\ss}}, \citenamefont {Hild},
  \citenamefont {Choi}, \citenamefont {Pohl}, \citenamefont {Bloch},\ and\
  \citenamefont {Gross}}]{Zeiher2016}%
  \BibitemOpen
  \bibfield  {author} {\bibinfo {author} {\bibfnamefont {J.}~\bibnamefont
  {Zeiher}}, \bibinfo {author} {\bibfnamefont {R.}~\bibnamefont {van Bijnen}},
  \bibinfo {author} {\bibfnamefont {P.}~\bibnamefont {Schau{\ss}}}, \bibinfo
  {author} {\bibfnamefont {S.}~\bibnamefont {Hild}}, \bibinfo {author}
  {\bibfnamefont {J.-y.}\ \bibnamefont {Choi}}, \bibinfo {author}
  {\bibfnamefont {T.}~\bibnamefont {Pohl}}, \bibinfo {author} {\bibfnamefont
  {I.}~\bibnamefont {Bloch}},\ and\ \bibinfo {author} {\bibfnamefont
  {C.}~\bibnamefont {Gross}},\ }\bibfield  {title} {\bibinfo {title}
  {{Many-body interferometry of a Rydberg-dressed spin lattice}},\ }\href
  {https://doi.org/https://doi.org/10.1038/nphys3835} {\bibfield  {journal}
  {\bibinfo  {journal} {Nat. Phys.}\ }\textbf {\bibinfo {volume} {12}},\
  \bibinfo {pages} {1095} (\bibinfo {year} {2016})}\BibitemShut {NoStop}%
\bibitem [{\citenamefont {Zeiher}\ \emph {et~al.}(2017)\citenamefont {Zeiher},
  \citenamefont {Choi}, \citenamefont {Rubio-Abadal}, \citenamefont {Pohl},
  \citenamefont {van Bijnen}, \citenamefont {Bloch},\ and\ \citenamefont
  {Gross}}]{Zeiher2017}%
  \BibitemOpen
  \bibfield  {author} {\bibinfo {author} {\bibfnamefont {J.}~\bibnamefont
  {Zeiher}}, \bibinfo {author} {\bibfnamefont {J.-Y.}\ \bibnamefont {Choi}},
  \bibinfo {author} {\bibfnamefont {A.}~\bibnamefont {Rubio-Abadal}}, \bibinfo
  {author} {\bibfnamefont {T.}~\bibnamefont {Pohl}}, \bibinfo {author}
  {\bibfnamefont {R.}~\bibnamefont {van Bijnen}}, \bibinfo {author}
  {\bibfnamefont {I.}~\bibnamefont {Bloch}},\ and\ \bibinfo {author}
  {\bibfnamefont {C.}~\bibnamefont {Gross}},\ }\bibfield  {title} {\bibinfo
  {title} {{Coherent Many-Body Spin Dynamics in a Long-Range Interacting Ising
  Chain}},\ }\href {https://doi.org/10.1103/PhysRevX.7.041063} {\bibfield
  {journal} {\bibinfo  {journal} {Phys. Rev. X}\ }\textbf {\bibinfo {volume}
  {7}},\ \bibinfo {pages} {041063} (\bibinfo {year} {2017})}\BibitemShut
  {NoStop}%
\bibitem [{\citenamefont {Borish}\ \emph {et~al.}(2020)\citenamefont {Borish},
  \citenamefont {Markovi\ifmmode~\acute{c}\else \'{c}\fi{}}, \citenamefont
  {Hines}, \citenamefont {Rajagopal},\ and\ \citenamefont
  {Schleier-Smith}}]{Borish2020}%
  \BibitemOpen
  \bibfield  {author} {\bibinfo {author} {\bibfnamefont {V.}~\bibnamefont
  {Borish}}, \bibinfo {author} {\bibfnamefont {O.}~\bibnamefont
  {Markovi\ifmmode~\acute{c}\else \'{c}\fi{}}}, \bibinfo {author}
  {\bibfnamefont {J.~A.}\ \bibnamefont {Hines}}, \bibinfo {author}
  {\bibfnamefont {S.~V.}\ \bibnamefont {Rajagopal}},\ and\ \bibinfo {author}
  {\bibfnamefont {M.}~\bibnamefont {Schleier-Smith}},\ }\bibfield  {title}
  {\bibinfo {title} {{Transverse-Field Ising Dynamics in a Rydberg-Dressed
  Atomic Gas}},\ }\href {https://doi.org/10.1103/PhysRevLett.124.063601}
  {\bibfield  {journal} {\bibinfo  {journal} {Phys. Rev. Lett.}\ }\textbf
  {\bibinfo {volume} {124}},\ \bibinfo {pages} {063601} (\bibinfo {year}
  {2020})}\BibitemShut {NoStop}%
\bibitem [{\citenamefont {Hollerith}\ \emph {et~al.}(2021)\citenamefont
  {Hollerith}, \citenamefont {Srakaew}, \citenamefont {Wei}, \citenamefont
  {Rubio-Abadal}, \citenamefont {Adler}, \citenamefont {Weckesser},
  \citenamefont {Kruckenhauser}, \citenamefont {Walther}, \citenamefont {van
  Bijnen}, \citenamefont {Rui}, \citenamefont {Gross}, \citenamefont {Bloch},\
  and\ \citenamefont {Zeiher}}]{Hollerith2021}%
  \BibitemOpen
  \bibfield  {author} {\bibinfo {author} {\bibfnamefont {S.}~\bibnamefont
  {Hollerith}}, \bibinfo {author} {\bibfnamefont {K.}~\bibnamefont {Srakaew}},
  \bibinfo {author} {\bibfnamefont {D.}~\bibnamefont {Wei}}, \bibinfo {author}
  {\bibfnamefont {A.}~\bibnamefont {Rubio-Abadal}}, \bibinfo {author}
  {\bibfnamefont {D.}~\bibnamefont {Adler}}, \bibinfo {author} {\bibfnamefont
  {P.}~\bibnamefont {Weckesser}}, \bibinfo {author} {\bibfnamefont
  {A.}~\bibnamefont {Kruckenhauser}}, \bibinfo {author} {\bibfnamefont
  {V.}~\bibnamefont {Walther}}, \bibinfo {author} {\bibfnamefont
  {R.}~\bibnamefont {van Bijnen}}, \bibinfo {author} {\bibfnamefont
  {J.}~\bibnamefont {Rui}}, \bibinfo {author} {\bibfnamefont {C.}~\bibnamefont
  {Gross}}, \bibinfo {author} {\bibfnamefont {I.}~\bibnamefont {Bloch}},\ and\
  \bibinfo {author} {\bibfnamefont {J.}~\bibnamefont {Zeiher}},\ }\bibfield
  {title} {\bibinfo {title} {Realizing distance-selective interactions in a
  rydberg-dressed atom array},\ }\href
  {https://doi.org/10.1103/PHYSREVLETT.128.113602/FIGURES/4/MEDIUM} {\bibfield
  {journal} {\bibinfo  {journal} {Phys. Rev. Lett.}\ }\textbf {\bibinfo
  {volume} {128}},\ \bibinfo {pages} {113602} (\bibinfo {year}
  {2021})}\BibitemShut {NoStop}%
\bibitem [{\citenamefont {Guardado-Sanchez}\ \emph {et~al.}(2021)\citenamefont
  {Guardado-Sanchez}, \citenamefont {Spar}, \citenamefont {Schauss},
  \citenamefont {Belyansky}, \citenamefont {Young}, \citenamefont {Bienias},
  \citenamefont {Gorshkov}, \citenamefont {Iadecola},\ and\ \citenamefont
  {Bakr}}]{guardado-sanchez_quench_2021}%
  \BibitemOpen
  \bibfield  {author} {\bibinfo {author} {\bibfnamefont {E.}~\bibnamefont
  {Guardado-Sanchez}}, \bibinfo {author} {\bibfnamefont {B.~M.}\ \bibnamefont
  {Spar}}, \bibinfo {author} {\bibfnamefont {P.}~\bibnamefont {Schauss}},
  \bibinfo {author} {\bibfnamefont {R.}~\bibnamefont {Belyansky}}, \bibinfo
  {author} {\bibfnamefont {J.~T.}\ \bibnamefont {Young}}, \bibinfo {author}
  {\bibfnamefont {P.}~\bibnamefont {Bienias}}, \bibinfo {author} {\bibfnamefont
  {A.~V.}\ \bibnamefont {Gorshkov}}, \bibinfo {author} {\bibfnamefont
  {T.}~\bibnamefont {Iadecola}},\ and\ \bibinfo {author} {\bibfnamefont
  {W.~S.}\ \bibnamefont {Bakr}},\ }\bibfield  {title} {\bibinfo {title} {Quench
  {D}ynamics of a {F}ermi {G}as with {S}trong {N}onlocal {I}nteractions},\
  }\href {https://doi.org/10.1103/PhysRevX.11.021036} {\bibfield  {journal}
  {\bibinfo  {journal} {Phys. Rev. X}\ }\textbf {\bibinfo {volume} {11}},\
  \bibinfo {pages} {021036} (\bibinfo {year} {2021})}\BibitemShut {NoStop}%
\bibitem [{\citenamefont {Rachel}(2018)}]{Rachel_2018}%
  \BibitemOpen
  \bibfield  {author} {\bibinfo {author} {\bibfnamefont {S.}~\bibnamefont
  {Rachel}},\ }\bibfield  {title} {\bibinfo {title} {Interacting topological
  insulators: a review},\ }\href {https://doi.org/10.1088/1361-6633/aad6a6}
  {\bibfield  {journal} {\bibinfo  {journal} {Rep. Prog. Phys.}\ }\textbf
  {\bibinfo {volume} {81}},\ \bibinfo {pages} {116501} (\bibinfo {year}
  {2018})}\BibitemShut {NoStop}%
\bibitem [{\citenamefont {Hasan}\ and\ \citenamefont
  {Kane}(2010)}]{RevModPhys.82.3045}%
  \BibitemOpen
  \bibfield  {author} {\bibinfo {author} {\bibfnamefont {M.~Z.}\ \bibnamefont
  {Hasan}}\ and\ \bibinfo {author} {\bibfnamefont {C.~L.}\ \bibnamefont
  {Kane}},\ }\bibfield  {title} {\bibinfo {title} {Colloquium: {T}opological
  insulators},\ }\href {https://doi.org/10.1103/RevModPhys.82.3045} {\bibfield
  {journal} {\bibinfo  {journal} {Rev. Mod. Phys.}\ }\textbf {\bibinfo {volume}
  {82}},\ \bibinfo {pages} {3045} (\bibinfo {year} {2010})}\BibitemShut
  {NoStop}%
\bibitem [{\citenamefont {Qi}\ and\ \citenamefont {Zhang}(2011)}]{Qi2011}%
  \BibitemOpen
  \bibfield  {author} {\bibinfo {author} {\bibfnamefont {X.-L.}\ \bibnamefont
  {Qi}}\ and\ \bibinfo {author} {\bibfnamefont {S.-C.}\ \bibnamefont {Zhang}},\
  }\bibfield  {title} {\bibinfo {title} {Topological insulators and
  superconductors},\ }\href {https://doi.org/10.1103/RevModPhys.83.1057}
  {\bibfield  {journal} {\bibinfo  {journal} {Rev. Mod. Phys.}\ }\textbf
  {\bibinfo {volume} {83}},\ \bibinfo {pages} {1057} (\bibinfo {year}
  {2011})}\BibitemShut {NoStop}%
\bibitem [{\citenamefont {Aidelsburger}\ \emph {et~al.}(2013)\citenamefont
  {Aidelsburger}, \citenamefont {Atala}, \citenamefont {Lohse}, \citenamefont
  {Barreiro}, \citenamefont {Paredes},\ and\ \citenamefont
  {Bloch}}]{aidelsburger_realization_2013}%
  \BibitemOpen
  \bibfield  {author} {\bibinfo {author} {\bibfnamefont {M.}~\bibnamefont
  {Aidelsburger}}, \bibinfo {author} {\bibfnamefont {M.}~\bibnamefont {Atala}},
  \bibinfo {author} {\bibfnamefont {M.}~\bibnamefont {Lohse}}, \bibinfo
  {author} {\bibfnamefont {J.~T.}\ \bibnamefont {Barreiro}}, \bibinfo {author}
  {\bibfnamefont {B.}~\bibnamefont {Paredes}},\ and\ \bibinfo {author}
  {\bibfnamefont {I.}~\bibnamefont {Bloch}},\ }\bibfield  {title} {\bibinfo
  {title} {Realization of the {H}ofstadter {H}amiltonian with {U}ltracold
  {A}toms in {O}ptical {L}attices},\ }\href
  {https://doi.org/10.1103/PhysRevLett.111.185301} {\bibfield  {journal}
  {\bibinfo  {journal} {Phys. Rev. Lett.}\ }\textbf {\bibinfo {volume} {111}},\
  \bibinfo {pages} {185301} (\bibinfo {year} {2013})}\BibitemShut {NoStop}%
\bibitem [{\citenamefont {Aidelsburger}\ \emph {et~al.}(2015)\citenamefont
  {Aidelsburger}, \citenamefont {Lohse}, \citenamefont {Schweizer},
  \citenamefont {Atala}, \citenamefont {Barreiro}, \citenamefont
  {Nascimb{\`e}ne}, \citenamefont {Cooper}, \citenamefont {Bloch},\ and\
  \citenamefont {Goldman}}]{Aidelsburger_2014}%
  \BibitemOpen
  \bibfield  {author} {\bibinfo {author} {\bibfnamefont {M.}~\bibnamefont
  {Aidelsburger}}, \bibinfo {author} {\bibfnamefont {M.}~\bibnamefont {Lohse}},
  \bibinfo {author} {\bibfnamefont {C.}~\bibnamefont {Schweizer}}, \bibinfo
  {author} {\bibfnamefont {M.}~\bibnamefont {Atala}}, \bibinfo {author}
  {\bibfnamefont {J.~T.}\ \bibnamefont {Barreiro}}, \bibinfo {author}
  {\bibfnamefont {S.}~\bibnamefont {Nascimb{\`e}ne}}, \bibinfo {author}
  {\bibfnamefont {N.~R.}\ \bibnamefont {Cooper}}, \bibinfo {author}
  {\bibfnamefont {I.}~\bibnamefont {Bloch}},\ and\ \bibinfo {author}
  {\bibfnamefont {N.}~\bibnamefont {Goldman}},\ }\bibfield  {title} {\bibinfo
  {title} {Measuring the {C}hern number of {H}ofstadter bands with ultracold
  bosonic atoms},\ }\href {https://doi.org/10.1038/nphys3171} {\bibfield
  {journal} {\bibinfo  {journal} {Nat. Phys.}\ }\textbf {\bibinfo {volume}
  {11}},\ \bibinfo {pages} {162} (\bibinfo {year} {2015})}\BibitemShut
  {NoStop}%
\bibitem [{\citenamefont {Jotzu}\ \emph {et~al.}(2014)\citenamefont {Jotzu},
  \citenamefont {Messer}, \citenamefont {Desbuquois}, \citenamefont {Lebrat},
  \citenamefont {Uehlinger}, \citenamefont {Greif},\ and\ \citenamefont
  {Esslinger}}]{Jotzu_2014}%
  \BibitemOpen
  \bibfield  {author} {\bibinfo {author} {\bibfnamefont {G.}~\bibnamefont
  {Jotzu}}, \bibinfo {author} {\bibfnamefont {M.}~\bibnamefont {Messer}},
  \bibinfo {author} {\bibfnamefont {R.}~\bibnamefont {Desbuquois}}, \bibinfo
  {author} {\bibfnamefont {M.}~\bibnamefont {Lebrat}}, \bibinfo {author}
  {\bibfnamefont {T.}~\bibnamefont {Uehlinger}}, \bibinfo {author}
  {\bibfnamefont {D.}~\bibnamefont {Greif}},\ and\ \bibinfo {author}
  {\bibfnamefont {T.}~\bibnamefont {Esslinger}},\ }\bibfield  {title} {\bibinfo
  {title} {Experimental realization of the topological {H}aldane model with
  ultracold fermions},\ }\href {https://doi.org/10.1038/nature13915} {\bibfield
   {journal} {\bibinfo  {journal} {Nature}\ }\textbf {\bibinfo {volume}
  {515}},\ \bibinfo {pages} {237} (\bibinfo {year} {2014})}\BibitemShut
  {NoStop}%
\bibitem [{\citenamefont {Mancini}\ \emph
  {et~al.}(2015{\natexlab{a}})\citenamefont {Mancini}, \citenamefont {Pagano},
  \citenamefont {Cappellini}, \citenamefont {Livi}, \citenamefont {Rider},
  \citenamefont {Catani}, \citenamefont {Sias}, \citenamefont {Zoller},
  \citenamefont {Inguscio}, \citenamefont {Dalmonte},\ and\ \citenamefont
  {Fallani}}]{Mancini2015}%
  \BibitemOpen
  \bibfield  {author} {\bibinfo {author} {\bibfnamefont {M.}~\bibnamefont
  {Mancini}}, \bibinfo {author} {\bibfnamefont {G.}~\bibnamefont {Pagano}},
  \bibinfo {author} {\bibfnamefont {G.}~\bibnamefont {Cappellini}}, \bibinfo
  {author} {\bibfnamefont {L.}~\bibnamefont {Livi}}, \bibinfo {author}
  {\bibfnamefont {M.}~\bibnamefont {Rider}}, \bibinfo {author} {\bibfnamefont
  {J.}~\bibnamefont {Catani}}, \bibinfo {author} {\bibfnamefont
  {C.}~\bibnamefont {Sias}}, \bibinfo {author} {\bibfnamefont {P.}~\bibnamefont
  {Zoller}}, \bibinfo {author} {\bibfnamefont {M.}~\bibnamefont {Inguscio}},
  \bibinfo {author} {\bibfnamefont {M.}~\bibnamefont {Dalmonte}},\ and\
  \bibinfo {author} {\bibfnamefont {L.}~\bibnamefont {Fallani}},\ }\bibfield
  {title} {\bibinfo {title} {Observation of chiral edge states with neutral
  fermions in synthetic {H}all ribbons},\ }\href
  {https://doi.org/10.1126/science.aaa8736} {\bibfield  {journal} {\bibinfo
  {journal} {Science}\ }\textbf {\bibinfo {volume} {349}},\ \bibinfo {pages}
  {1510} (\bibinfo {year} {2015}{\natexlab{a}})}\BibitemShut {NoStop}%
\bibitem [{\citenamefont {Asteria}\ \emph
  {et~al.}(2019{\natexlab{a}})\citenamefont {Asteria}, \citenamefont {Tran},
  \citenamefont {Ozawa}, \citenamefont {Tarnowski}, \citenamefont {Rem},
  \citenamefont {Fl{\"a}schner}, \citenamefont {Sengstock}, \citenamefont
  {Goldman},\ and\ \citenamefont {Weitenberg}}]{Asteria_2019}%
  \BibitemOpen
  \bibfield  {author} {\bibinfo {author} {\bibfnamefont {L.}~\bibnamefont
  {Asteria}}, \bibinfo {author} {\bibfnamefont {D.~T.}\ \bibnamefont {Tran}},
  \bibinfo {author} {\bibfnamefont {T.}~\bibnamefont {Ozawa}}, \bibinfo
  {author} {\bibfnamefont {M.}~\bibnamefont {Tarnowski}}, \bibinfo {author}
  {\bibfnamefont {B.~S.}\ \bibnamefont {Rem}}, \bibinfo {author} {\bibfnamefont
  {N.}~\bibnamefont {Fl{\"a}schner}}, \bibinfo {author} {\bibfnamefont
  {K.}~\bibnamefont {Sengstock}}, \bibinfo {author} {\bibfnamefont
  {N.}~\bibnamefont {Goldman}},\ and\ \bibinfo {author} {\bibfnamefont
  {C.}~\bibnamefont {Weitenberg}},\ }\bibfield  {title} {\bibinfo {title}
  {Measuring quantized circular dichroism in ultracold topological matter},\
  }\href {https://doi.org/10.1038/s41567-019-0417-8} {\bibfield  {journal}
  {\bibinfo  {journal} {Nat. Phys.}\ }\textbf {\bibinfo {volume} {15}},\
  \bibinfo {pages} {449} (\bibinfo {year} {2019}{\natexlab{a}})}\BibitemShut
  {NoStop}%
\bibitem [{\citenamefont {Mancini}\ \emph
  {et~al.}(2015{\natexlab{b}})\citenamefont {Mancini}, \citenamefont {Pagano},
  \citenamefont {Cappellini}, \citenamefont {Livi}, \citenamefont {Rider},
  \citenamefont {Catani}, \citenamefont {Sias}, \citenamefont {Zoller},
  \citenamefont {Inguscio}, \citenamefont {Dalmonte},\ and\ \citenamefont
  {Fallani}}]{mancini_2015}%
  \BibitemOpen
  \bibfield  {author} {\bibinfo {author} {\bibfnamefont {M.}~\bibnamefont
  {Mancini}}, \bibinfo {author} {\bibfnamefont {G.}~\bibnamefont {Pagano}},
  \bibinfo {author} {\bibfnamefont {G.}~\bibnamefont {Cappellini}}, \bibinfo
  {author} {\bibfnamefont {L.}~\bibnamefont {Livi}}, \bibinfo {author}
  {\bibfnamefont {M.}~\bibnamefont {Rider}}, \bibinfo {author} {\bibfnamefont
  {J.}~\bibnamefont {Catani}}, \bibinfo {author} {\bibfnamefont
  {C.}~\bibnamefont {Sias}}, \bibinfo {author} {\bibfnamefont {P.}~\bibnamefont
  {Zoller}}, \bibinfo {author} {\bibfnamefont {M.}~\bibnamefont {Inguscio}},
  \bibinfo {author} {\bibfnamefont {M.}~\bibnamefont {Dalmonte}},\ and\
  \bibinfo {author} {\bibfnamefont {L.}~\bibnamefont {Fallani}},\ }\bibfield
  {title} {\bibinfo {title} {Observation of chiral edge states with neutral
  fermions in synthetic {H}all ribbons},\ }\href
  {https://doi.org/10.1126/science.aaa8736} {\bibfield  {journal} {\bibinfo
  {journal} {Science}\ }\textbf {\bibinfo {volume} {349}},\ \bibinfo {pages}
  {1510} (\bibinfo {year} {2015}{\natexlab{b}})}\BibitemShut {NoStop}%
\bibitem [{\citenamefont {Stuhl}\ \emph {et~al.}(2015)\citenamefont {Stuhl},
  \citenamefont {Lu}, \citenamefont {Aycock}, \citenamefont {Genkina},\ and\
  \citenamefont {Spielman}}]{Stuhl_2015}%
  \BibitemOpen
  \bibfield  {author} {\bibinfo {author} {\bibfnamefont {B.~K.}\ \bibnamefont
  {Stuhl}}, \bibinfo {author} {\bibfnamefont {H.-I.}\ \bibnamefont {Lu}},
  \bibinfo {author} {\bibfnamefont {L.~M.}\ \bibnamefont {Aycock}}, \bibinfo
  {author} {\bibfnamefont {D.}~\bibnamefont {Genkina}},\ and\ \bibinfo {author}
  {\bibfnamefont {I.~B.}\ \bibnamefont {Spielman}},\ }\bibfield  {title}
  {\bibinfo {title} {Visualizing edge states with an atomic {B}ose gas in the
  quantum {H}all regime},\ }\href {https://doi.org/10.1126/science.aaa8515}
  {\bibfield  {journal} {\bibinfo  {journal} {Science}\ }\textbf {\bibinfo
  {volume} {349}},\ \bibinfo {pages} {1514} (\bibinfo {year}
  {2015})}\BibitemShut {NoStop}%
\bibitem [{\citenamefont {Goldman}\ \emph {et~al.}(2014)\citenamefont
  {Goldman}, \citenamefont {Juzeli{\={u}}nas}, \citenamefont {{\"O}hberg},\
  and\ \citenamefont {Spielman}}]{Goldman_2014}%
  \BibitemOpen
  \bibfield  {author} {\bibinfo {author} {\bibfnamefont {N.}~\bibnamefont
  {Goldman}}, \bibinfo {author} {\bibfnamefont {G.}~\bibnamefont
  {Juzeli{\={u}}nas}}, \bibinfo {author} {\bibfnamefont {P.}~\bibnamefont
  {{\"O}hberg}},\ and\ \bibinfo {author} {\bibfnamefont {I.~B.}\ \bibnamefont
  {Spielman}},\ }\bibfield  {title} {\bibinfo {title} {Light-induced gauge
  fields for ultracold atoms},\ }\href
  {https://doi.org/10.1088/0034-4885/77/12/126401} {\bibfield  {journal}
  {\bibinfo  {journal} {Rep. Prog. Phys.}\ }\textbf {\bibinfo {volume} {77}},\
  \bibinfo {pages} {126401} (\bibinfo {year} {2014})}\BibitemShut {NoStop}%
\bibitem [{\citenamefont {Cooper}\ \emph {et~al.}(2019)\citenamefont {Cooper},
  \citenamefont {Dalibard},\ and\ \citenamefont {Spielman}}]{Cooper_2019}%
  \BibitemOpen
  \bibfield  {author} {\bibinfo {author} {\bibfnamefont {N.~R.}\ \bibnamefont
  {Cooper}}, \bibinfo {author} {\bibfnamefont {J.}~\bibnamefont {Dalibard}},\
  and\ \bibinfo {author} {\bibfnamefont {I.~B.}\ \bibnamefont {Spielman}},\
  }\bibfield  {title} {\bibinfo {title} {Topological bands for ultracold
  atoms},\ }\href {https://doi.org/10.1103/RevModPhys.91.015005} {\bibfield
  {journal} {\bibinfo  {journal} {Rev. Mod. Phys.}\ }\textbf {\bibinfo {volume}
  {91}},\ \bibinfo {pages} {015005} (\bibinfo {year} {2019})}\BibitemShut
  {NoStop}%
\bibitem [{\citenamefont {Raghu}\ \emph {et~al.}(2008)\citenamefont {Raghu},
  \citenamefont {Qi}, \citenamefont {Honerkamp},\ and\ \citenamefont
  {Zhang}}]{raghu_topological_2008}%
  \BibitemOpen
  \bibfield  {author} {\bibinfo {author} {\bibfnamefont {S.}~\bibnamefont
  {Raghu}}, \bibinfo {author} {\bibfnamefont {X.-L.}\ \bibnamefont {Qi}},
  \bibinfo {author} {\bibfnamefont {C.}~\bibnamefont {Honerkamp}},\ and\
  \bibinfo {author} {\bibfnamefont {S.-C.}\ \bibnamefont {Zhang}},\ }\bibfield
  {title} {\bibinfo {title} {Topological {M}ott {I}nsulators},\ }\href
  {https://doi.org/10.1103/PhysRevLett.100.156401} {\bibfield  {journal}
  {\bibinfo  {journal} {Phys. Rev. Lett.}\ }\textbf {\bibinfo {volume} {100}},\
  \bibinfo {pages} {156401} (\bibinfo {year} {2008})}\BibitemShut {NoStop}%
\bibitem [{\citenamefont {Sun}\ \emph {et~al.}(2009)\citenamefont {Sun},
  \citenamefont {Yao}, \citenamefont {Fradkin},\ and\ \citenamefont
  {Kivelson}}]{Sun2009}%
  \BibitemOpen
  \bibfield  {author} {\bibinfo {author} {\bibfnamefont {K.}~\bibnamefont
  {Sun}}, \bibinfo {author} {\bibfnamefont {H.}~\bibnamefont {Yao}}, \bibinfo
  {author} {\bibfnamefont {E.}~\bibnamefont {Fradkin}},\ and\ \bibinfo {author}
  {\bibfnamefont {S.~A.}\ \bibnamefont {Kivelson}},\ }\bibfield  {title}
  {\bibinfo {title} {Topological {I}nsulators and {N}ematic {P}hases from
  {S}pontaneous {S}ymmetry {B}reaking in 2{D} {F}ermi {S}ystems with a
  {Q}uadratic {B}and {C}rossing},\ }\href
  {https://doi.org/10.1103/PhysRevLett.103.046811} {\bibfield  {journal}
  {\bibinfo  {journal} {Phys. Rev. Lett.}\ }\textbf {\bibinfo {volume} {103}},\
  \bibinfo {pages} {046811} (\bibinfo {year} {2009})}\BibitemShut {NoStop}%
\bibitem [{\citenamefont {Zhu}\ \emph {et~al.}(2016)\citenamefont {Zhu},
  \citenamefont {Gong}, \citenamefont {Zeng}, \citenamefont {Fu},\ and\
  \citenamefont {Sheng}}]{zhu_interaction-driven_2016}%
  \BibitemOpen
  \bibfield  {author} {\bibinfo {author} {\bibfnamefont {W.}~\bibnamefont
  {Zhu}}, \bibinfo {author} {\bibfnamefont {S.-S.}\ \bibnamefont {Gong}},
  \bibinfo {author} {\bibfnamefont {T.-S.}\ \bibnamefont {Zeng}}, \bibinfo
  {author} {\bibfnamefont {L.}~\bibnamefont {Fu}},\ and\ \bibinfo {author}
  {\bibfnamefont {D.}~\bibnamefont {Sheng}},\ }\bibfield  {title} {\bibinfo
  {title} {Interaction-{D}riven {S}pontaneous {Q}uantum {H}all {E}ffect on a
  {K}agome {L}attice},\ }\href {https://doi.org/10.1103/PhysRevLett.117.096402}
  {\bibfield  {journal} {\bibinfo  {journal} {Phys. Rev. Lett.}\ }\textbf
  {\bibinfo {volume} {117}},\ \bibinfo {pages} {096402} (\bibinfo {year}
  {2016})}\BibitemShut {NoStop}%
\bibitem [{\citenamefont {Dauphin}\ \emph {et~al.}(2012)\citenamefont
  {Dauphin}, \citenamefont {M{\"u}ller},\ and\ \citenamefont
  {Martin-Delgado}}]{dauphin_rydberg-atom_2012}%
  \BibitemOpen
  \bibfield  {author} {\bibinfo {author} {\bibfnamefont {A.}~\bibnamefont
  {Dauphin}}, \bibinfo {author} {\bibfnamefont {M.}~\bibnamefont
  {M{\"u}ller}},\ and\ \bibinfo {author} {\bibfnamefont {M.~A.}\ \bibnamefont
  {Martin-Delgado}},\ }\bibfield  {title} {\bibinfo {title} {{Rydberg-atom
  quantum simulation and Chern-number characterization of a topological Mott
  insulator}},\ }\href {https://doi.org/10.1103/PhysRevA.86.053618} {\bibfield
  {journal} {\bibinfo  {journal} {Phys. Rev. A}\ }\textbf {\bibinfo {volume}
  {86}},\ \bibinfo {pages} {053618} (\bibinfo {year} {2012})}\BibitemShut
  {NoStop}%
\bibitem [{\citenamefont {Zeng}\ \emph {et~al.}(2018)\citenamefont {Zeng},
  \citenamefont {Zhu},\ and\ \citenamefont {Sheng}}]{zeng_tuning_2018}%
  \BibitemOpen
  \bibfield  {author} {\bibinfo {author} {\bibfnamefont {T.-S.}\ \bibnamefont
  {Zeng}}, \bibinfo {author} {\bibfnamefont {W.}~\bibnamefont {Zhu}},\ and\
  \bibinfo {author} {\bibfnamefont {D.}~\bibnamefont {Sheng}},\ }\bibfield
  {title} {\bibinfo {title} {Tuning topological phase and quantum anomalous
  {H}all effect by interaction in quadratic band touching systems},\ }\href
  {https://doi.org/10.1038/s41535-018-0120-5} {\bibfield  {journal} {\bibinfo
  {journal} {npj Quant. Mater.}\ }\textbf {\bibinfo {volume} {3}},\ \bibinfo
  {pages} {1} (\bibinfo {year} {2018})}\BibitemShut {NoStop}%
\bibitem [{\citenamefont {Sur}\ \emph {et~al.}(2018)\citenamefont {Sur},
  \citenamefont {Gong}, \citenamefont {Yang},\ and\ \citenamefont
  {Vafek}}]{Sur2018}%
  \BibitemOpen
  \bibfield  {author} {\bibinfo {author} {\bibfnamefont {S.}~\bibnamefont
  {Sur}}, \bibinfo {author} {\bibfnamefont {S.-S.}\ \bibnamefont {Gong}},
  \bibinfo {author} {\bibfnamefont {K.}~\bibnamefont {Yang}},\ and\ \bibinfo
  {author} {\bibfnamefont {O.}~\bibnamefont {Vafek}},\ }\bibfield  {title}
  {\bibinfo {title} {Quantum anomalous {H}all insulator stabilized by competing
  interactions},\ }\href {https://doi.org/10.1103/PhysRevB.98.125144}
  {\bibfield  {journal} {\bibinfo  {journal} {Phys. Rev. B}\ }\textbf {\bibinfo
  {volume} {98}},\ \bibinfo {pages} {125144} (\bibinfo {year}
  {2018})}\BibitemShut {NoStop}%
\bibitem [{\citenamefont {Wu}\ \emph {et~al.}(2016)\citenamefont {Wu},
  \citenamefont {He}, \citenamefont {Fang}, \citenamefont {Meng},\ and\
  \citenamefont {Lu}}]{PhysRevLett.117.066403}%
  \BibitemOpen
  \bibfield  {author} {\bibinfo {author} {\bibfnamefont {H.-Q.}\ \bibnamefont
  {Wu}}, \bibinfo {author} {\bibfnamefont {Y.-Y.}\ \bibnamefont {He}}, \bibinfo
  {author} {\bibfnamefont {C.}~\bibnamefont {Fang}}, \bibinfo {author}
  {\bibfnamefont {Z.~Y.}\ \bibnamefont {Meng}},\ and\ \bibinfo {author}
  {\bibfnamefont {Z.-Y.}\ \bibnamefont {Lu}},\ }\bibfield  {title} {\bibinfo
  {title} {Diagnosis of {I}nteraction-driven {T}opological {P}hase via {E}xact
  {D}iagonalization},\ }\href {https://doi.org/10.1103/PhysRevLett.117.066403}
  {\bibfield  {journal} {\bibinfo  {journal} {Phys. Rev. Lett.}\ }\textbf
  {\bibinfo {volume} {117}},\ \bibinfo {pages} {066403} (\bibinfo {year}
  {2016})}\BibitemShut {NoStop}%
\bibitem [{\citenamefont {Juli{\`a}-Farr{\'e}}\ \emph
  {et~al.}(2020)\citenamefont {Juli{\`a}-Farr{\'e}}, \citenamefont
  {M{\"u}ller}, \citenamefont {Lewenstein},\ and\ \citenamefont
  {Dauphin}}]{julia-farre_self-trapped_2020}%
  \BibitemOpen
  \bibfield  {author} {\bibinfo {author} {\bibfnamefont {S.}~\bibnamefont
  {Juli{\`a}-Farr{\'e}}}, \bibinfo {author} {\bibfnamefont {M.}~\bibnamefont
  {M{\"u}ller}}, \bibinfo {author} {\bibfnamefont {M.}~\bibnamefont
  {Lewenstein}},\ and\ \bibinfo {author} {\bibfnamefont {A.}~\bibnamefont
  {Dauphin}},\ }\bibfield  {title} {\bibinfo {title} {Self-{T}rapped {P}olarons
  and {T}opological {D}efects in a {T}opological {M}ott {I}nsulator},\ }\href
  {https://doi.org/10.1103/PhysRevLett.125.240601} {\bibfield  {journal}
  {\bibinfo  {journal} {Phys. Rev. Lett.}\ }\textbf {\bibinfo {volume} {125}},\
  \bibinfo {pages} {240601} (\bibinfo {year} {2020})}\BibitemShut {NoStop}%
\bibitem [{\citenamefont {Dauphin}\ \emph {et~al.}(2016)\citenamefont
  {Dauphin}, \citenamefont {M{\"u}ller},\ and\ \citenamefont
  {Martin-Delgado}}]{PhysRevA.93.043611}%
  \BibitemOpen
  \bibfield  {author} {\bibinfo {author} {\bibfnamefont {A.}~\bibnamefont
  {Dauphin}}, \bibinfo {author} {\bibfnamefont {M.}~\bibnamefont
  {M{\"u}ller}},\ and\ \bibinfo {author} {\bibfnamefont {M.~A.}\ \bibnamefont
  {Martin-Delgado}},\ }\bibfield  {title} {\bibinfo {title} {Quantum simulation
  of a topological {M}ott insulator with {R}ydberg atoms in a {L}ieb lattice},\
  }\href {https://doi.org/10.1103/PhysRevA.93.043611} {\bibfield  {journal}
  {\bibinfo  {journal} {Phys. Rev. A}\ }\textbf {\bibinfo {volume} {93}},\
  \bibinfo {pages} {043611} (\bibinfo {year} {2016})}\BibitemShut {NoStop}%
\bibitem [{\citenamefont {Garc\'{\i}a-Mart\'{\i}nez}\ \emph
  {et~al.}(2013)\citenamefont {Garc\'{\i}a-Mart\'{\i}nez}, \citenamefont
  {Grushin}, \citenamefont {Neupert}, \citenamefont {Valenzuela},\ and\
  \citenamefont {Castro}}]{Garcia-Martinez2013}%
  \BibitemOpen
  \bibfield  {author} {\bibinfo {author} {\bibfnamefont {N.~A.}\ \bibnamefont
  {Garc\'{\i}a-Mart\'{\i}nez}}, \bibinfo {author} {\bibfnamefont {A.~G.}\
  \bibnamefont {Grushin}}, \bibinfo {author} {\bibfnamefont {T.}~\bibnamefont
  {Neupert}}, \bibinfo {author} {\bibfnamefont {B.}~\bibnamefont
  {Valenzuela}},\ and\ \bibinfo {author} {\bibfnamefont {E.~V.}\ \bibnamefont
  {Castro}},\ }\bibfield  {title} {\bibinfo {title} {Interaction-driven phases
  in the half-filled spinless honeycomb lattice from exact diagonalization},\
  }\href {https://doi.org/10.1103/PhysRevB.88.245123} {\bibfield  {journal}
  {\bibinfo  {journal} {Phys. Rev. B}\ }\textbf {\bibinfo {volume} {88}},\
  \bibinfo {pages} {245123} (\bibinfo {year} {2013})}\BibitemShut {NoStop}%
\bibitem [{\citenamefont {Jia}\ \emph {et~al.}(2013)\citenamefont {Jia},
  \citenamefont {Guo}, \citenamefont {Chen}, \citenamefont {Shen},\ and\
  \citenamefont {Feng}}]{Jia2013}%
  \BibitemOpen
  \bibfield  {author} {\bibinfo {author} {\bibfnamefont {Y.}~\bibnamefont
  {Jia}}, \bibinfo {author} {\bibfnamefont {H.}~\bibnamefont {Guo}}, \bibinfo
  {author} {\bibfnamefont {Z.}~\bibnamefont {Chen}}, \bibinfo {author}
  {\bibfnamefont {S.-Q.}\ \bibnamefont {Shen}},\ and\ \bibinfo {author}
  {\bibfnamefont {S.}~\bibnamefont {Feng}},\ }\bibfield  {title} {\bibinfo
  {title} {Effect of interactions on two-dimensional {D}irac fermions},\ }\href
  {https://doi.org/10.1103/PhysRevB.88.075101} {\bibfield  {journal} {\bibinfo
  {journal} {Phys. Rev. B}\ }\textbf {\bibinfo {volume} {88}},\ \bibinfo
  {pages} {075101} (\bibinfo {year} {2013})}\BibitemShut {NoStop}%
\bibitem [{\citenamefont {Daghofer}\ and\ \citenamefont
  {Hohenadler}(2014)}]{Daghofer2014}%
  \BibitemOpen
  \bibfield  {author} {\bibinfo {author} {\bibfnamefont {M.}~\bibnamefont
  {Daghofer}}\ and\ \bibinfo {author} {\bibfnamefont {M.}~\bibnamefont
  {Hohenadler}},\ }\bibfield  {title} {\bibinfo {title} {Phases of correlated
  spinless fermions on the honeycomb lattice},\ }\href
  {https://doi.org/10.1103/PhysRevB.89.035103} {\bibfield  {journal} {\bibinfo
  {journal} {Phys. Rev. B}\ }\textbf {\bibinfo {volume} {89}},\ \bibinfo
  {pages} {035103} (\bibinfo {year} {2014})}\BibitemShut {NoStop}%
\bibitem [{\citenamefont {Guo}\ and\ \citenamefont {Jia}(2014)}]{Guo_2014}%
  \BibitemOpen
  \bibfield  {author} {\bibinfo {author} {\bibfnamefont {H.}~\bibnamefont
  {Guo}}\ and\ \bibinfo {author} {\bibfnamefont {Y.}~\bibnamefont {Jia}},\
  }\bibfield  {title} {\bibinfo {title} {Interaction-driven phases in a {D}irac
  semimetal: exact diagonalization results},\ }\href
  {https://doi.org/10.1088/0953-8984/26/47/475601} {\bibfield  {journal}
  {\bibinfo  {journal} {J. Phys.: Condens. Matter}\ }\textbf {\bibinfo {volume}
  {26}},\ \bibinfo {pages} {475601} (\bibinfo {year} {2014})}\BibitemShut
  {NoStop}%
\bibitem [{\citenamefont {Motruk}\ \emph {et~al.}(2015)\citenamefont {Motruk},
  \citenamefont {Grushin}, \citenamefont {de~Juan},\ and\ \citenamefont
  {Pollmann}}]{Motruk2015}%
  \BibitemOpen
  \bibfield  {author} {\bibinfo {author} {\bibfnamefont {J.}~\bibnamefont
  {Motruk}}, \bibinfo {author} {\bibfnamefont {A.~G.}\ \bibnamefont {Grushin}},
  \bibinfo {author} {\bibfnamefont {F.}~\bibnamefont {de~Juan}},\ and\ \bibinfo
  {author} {\bibfnamefont {F.}~\bibnamefont {Pollmann}},\ }\bibfield  {title}
  {\bibinfo {title} {Interaction-driven phases in the half-filled honeycomb
  lattice: {A}n infinite density matrix renormalization group study},\ }\href
  {https://doi.org/10.1103/PhysRevB.92.085147} {\bibfield  {journal} {\bibinfo
  {journal} {Phys. Rev. B}\ }\textbf {\bibinfo {volume} {92}},\ \bibinfo
  {pages} {085147} (\bibinfo {year} {2015})}\BibitemShut {NoStop}%
\bibitem [{\citenamefont {Capponi}\ and\ \citenamefont
  {L{\"a}uchli}(2015)}]{capponi_phase_2015}%
  \BibitemOpen
  \bibfield  {author} {\bibinfo {author} {\bibfnamefont {S.}~\bibnamefont
  {Capponi}}\ and\ \bibinfo {author} {\bibfnamefont {A.~M.}\ \bibnamefont
  {L{\"a}uchli}},\ }\bibfield  {title} {\bibinfo {title} {Phase diagram of
  interacting spinless fermions on the honeycomb lattice: A comprehensive exact
  diagonalization study},\ }\href {https://doi.org/10.1103/PhysRevB.92.085146}
  {\bibfield  {journal} {\bibinfo  {journal} {Phys. Rev. B}\ }\textbf {\bibinfo
  {volume} {92}},\ \bibinfo {pages} {085146} (\bibinfo {year}
  {2015})}\BibitemShut {NoStop}%
\bibitem [{\citenamefont {Scherer}\ \emph {et~al.}(2015)\citenamefont
  {Scherer}, \citenamefont {Scherer},\ and\ \citenamefont
  {Honerkamp}}]{Scherer2015}%
  \BibitemOpen
  \bibfield  {author} {\bibinfo {author} {\bibfnamefont {D.~D.}\ \bibnamefont
  {Scherer}}, \bibinfo {author} {\bibfnamefont {M.~M.}\ \bibnamefont
  {Scherer}},\ and\ \bibinfo {author} {\bibfnamefont {C.}~\bibnamefont
  {Honerkamp}},\ }\bibfield  {title} {\bibinfo {title} {Correlated spinless
  fermions on the honeycomb lattice revisited},\ }\href
  {https://doi.org/10.1103/PhysRevB.92.155137} {\bibfield  {journal} {\bibinfo
  {journal} {Phys. Rev. B}\ }\textbf {\bibinfo {volume} {92}},\ \bibinfo
  {pages} {155137} (\bibinfo {year} {2015})}\BibitemShut {NoStop}%
\bibitem [{\citenamefont {Sun}\ and\ \citenamefont
  {Fradkin}(2008)}]{Sun2008TRS}%
  \BibitemOpen
  \bibfield  {author} {\bibinfo {author} {\bibfnamefont {K.}~\bibnamefont
  {Sun}}\ and\ \bibinfo {author} {\bibfnamefont {E.}~\bibnamefont {Fradkin}},\
  }\bibfield  {title} {\bibinfo {title} {{Time-reversal symmetry breaking and
  spontaneous anomalous Hall effect in Fermi fluids}},\ }\href
  {https://doi.org/10.1103/PhysRevB.78.245122} {\bibfield  {journal} {\bibinfo
  {journal} {Phys. Rev. B}\ }\textbf {\bibinfo {volume} {78}},\ \bibinfo
  {pages} {245122} (\bibinfo {year} {2008})}\BibitemShut {NoStop}%
\bibitem [{\citenamefont {Vafek}\ and\ \citenamefont {Yang}(2010)}]{Vafek2010}%
  \BibitemOpen
  \bibfield  {author} {\bibinfo {author} {\bibfnamefont {O.}~\bibnamefont
  {Vafek}}\ and\ \bibinfo {author} {\bibfnamefont {K.}~\bibnamefont {Yang}},\
  }\bibfield  {title} {\bibinfo {title} {Many-body instability of {C}oulomb
  interacting bilayer graphene: {R}enormalization group approach},\ }\href
  {https://doi.org/10.1103/PhysRevB.81.041401} {\bibfield  {journal} {\bibinfo
  {journal} {Phys. Rev. B}\ }\textbf {\bibinfo {volume} {81}},\ \bibinfo
  {pages} {041401} (\bibinfo {year} {2010})}\BibitemShut {NoStop}%
\bibitem [{\citenamefont {D\'ora}\ \emph {et~al.}(2014)\citenamefont {D\'ora},
  \citenamefont {Herbut},\ and\ \citenamefont {Moessner}}]{Dora2014}%
  \BibitemOpen
  \bibfield  {author} {\bibinfo {author} {\bibfnamefont {B.}~\bibnamefont
  {D\'ora}}, \bibinfo {author} {\bibfnamefont {I.~F.}\ \bibnamefont {Herbut}},\
  and\ \bibinfo {author} {\bibfnamefont {R.}~\bibnamefont {Moessner}},\
  }\bibfield  {title} {\bibinfo {title} {Occurrence of nematic, topological,
  and {B}erry phases when a flat and a parabolic band touch},\ }\href
  {https://doi.org/10.1103/PhysRevB.90.045310} {\bibfield  {journal} {\bibinfo
  {journal} {Phys. Rev. B}\ }\textbf {\bibinfo {volume} {90}},\ \bibinfo
  {pages} {045310} (\bibinfo {year} {2014})}\BibitemShut {NoStop}%
\bibitem [{Note1()}]{Note1}%
  \BibitemOpen
  \bibinfo {note} {We hereafter set $\hbar =1$}\BibitemShut {NoStop}%
\bibitem [{\citenamefont {Thouless}\ \emph {et~al.}(1982)\citenamefont
  {Thouless}, \citenamefont {Kohmoto}, \citenamefont {Nightingale},\ and\
  \citenamefont {den Nijs}}]{thouless_quantized_1982}%
  \BibitemOpen
  \bibfield  {author} {\bibinfo {author} {\bibfnamefont {D.~J.}\ \bibnamefont
  {Thouless}}, \bibinfo {author} {\bibfnamefont {M.}~\bibnamefont {Kohmoto}},
  \bibinfo {author} {\bibfnamefont {M.~P.}\ \bibnamefont {Nightingale}},\ and\
  \bibinfo {author} {\bibfnamefont {M.}~\bibnamefont {den Nijs}},\ }\bibfield
  {title} {\bibinfo {title} {Quantized {H}all {C}onductance in a
  {T}wo-{D}imensional {P}eriodic {P}otential},\ }\href
  {https://doi.org/10.1103/PhysRevLett.49.405} {\bibfield  {journal} {\bibinfo
  {journal} {Phys. Rev. Lett.}\ }\textbf {\bibinfo {volume} {49}},\ \bibinfo
  {pages} {405} (\bibinfo {year} {1982})}\BibitemShut {NoStop}%
\bibitem [{\citenamefont {Haldane}(1988)}]{Haldane88}%
  \BibitemOpen
  \bibfield  {author} {\bibinfo {author} {\bibfnamefont {F.~D.~M.}\
  \bibnamefont {Haldane}},\ }\bibfield  {title} {\bibinfo {title} {Model for a
  {Q}uantum {H}all {E}ffect without {L}andau {L}evels: {C}ondensed-{M}atter
  {R}ealization of the "{P}arity {A}nomaly"},\ }\href
  {https://doi.org/10.1103/PhysRevLett.61.2015} {\bibfield  {journal} {\bibinfo
   {journal} {Phys. Rev. Lett.}\ }\textbf {\bibinfo {volume} {61}},\ \bibinfo
  {pages} {2015} (\bibinfo {year} {1988})}\BibitemShut {NoStop}%
\bibitem [{\citenamefont {Greiner}\ \emph {et~al.}(2002)\citenamefont
  {Greiner}, \citenamefont {Mandel}, \citenamefont {Esslinger}, \citenamefont
  {H{\"a}nsch},\ and\ \citenamefont {Bloch}}]{greiner_quantum_2002}%
  \BibitemOpen
  \bibfield  {author} {\bibinfo {author} {\bibfnamefont {M.}~\bibnamefont
  {Greiner}}, \bibinfo {author} {\bibfnamefont {O.}~\bibnamefont {Mandel}},
  \bibinfo {author} {\bibfnamefont {T.}~\bibnamefont {Esslinger}}, \bibinfo
  {author} {\bibfnamefont {T.~W.}\ \bibnamefont {H{\"a}nsch}},\ and\ \bibinfo
  {author} {\bibfnamefont {I.}~\bibnamefont {Bloch}},\ }\bibfield  {title}
  {\bibinfo {title} {Quantum phase transition from a superfluid to a {M}ott
  insulator in a gas of ultracold atoms},\ }\href
  {https://doi.org/10.1038/415039a} {\bibfield  {journal} {\bibinfo  {journal}
  {Nature}\ }\textbf {\bibinfo {volume} {415}},\ \bibinfo {pages} {39}
  (\bibinfo {year} {2002})}\BibitemShut {NoStop}%
\bibitem [{\citenamefont {Gross}\ and\ \citenamefont
  {Bloch}(2017)}]{gross_quantum_2017}%
  \BibitemOpen
  \bibfield  {author} {\bibinfo {author} {\bibfnamefont {C.}~\bibnamefont
  {Gross}}\ and\ \bibinfo {author} {\bibfnamefont {I.}~\bibnamefont {Bloch}},\
  }\bibfield  {title} {\bibinfo {title} {Quantum simulations with ultracold
  atoms in optical lattices},\ }\href {https://doi.org/10.1126/science.aal3837}
  {\bibfield  {journal} {\bibinfo  {journal} {Science}\ }\textbf {\bibinfo
  {volume} {357}},\ \bibinfo {pages} {995} (\bibinfo {year}
  {2017})}\BibitemShut {NoStop}%
\bibitem [{\citenamefont {Bloch}\ \emph {et~al.}(2012)\citenamefont {Bloch},
  \citenamefont {Dalibard},\ and\ \citenamefont
  {Nascimb{\`e}ne}}]{bloch_quantum_2012}%
  \BibitemOpen
  \bibfield  {author} {\bibinfo {author} {\bibfnamefont {I.}~\bibnamefont
  {Bloch}}, \bibinfo {author} {\bibfnamefont {J.}~\bibnamefont {Dalibard}},\
  and\ \bibinfo {author} {\bibfnamefont {S.}~\bibnamefont {Nascimb{\`e}ne}},\
  }\bibfield  {title} {\bibinfo {title} {Quantum simulations with ultracold
  quantum gases},\ }\href {https://doi.org/10.1038/nphys2259} {\bibfield
  {journal} {\bibinfo  {journal} {Nat. Phys.}\ }\textbf {\bibinfo {volume}
  {8}},\ \bibinfo {pages} {267} (\bibinfo {year} {2012})}\BibitemShut {NoStop}%
\bibitem [{\citenamefont {Gross}\ and\ \citenamefont {Bakr}(2021)}]{Gross2021}%
  \BibitemOpen
  \bibfield  {author} {\bibinfo {author} {\bibfnamefont {C.}~\bibnamefont
  {Gross}}\ and\ \bibinfo {author} {\bibfnamefont {W.~S.}\ \bibnamefont
  {Bakr}},\ }\bibfield  {title} {\bibinfo {title} {Quantum gas microscopy for
  single atom and spin detection},\ }\href
  {https://doi.org/10.1038/s41567-021-01370-5} {\bibfield  {journal} {\bibinfo
  {journal} {Nat. Phys.}\ }\textbf {\bibinfo {volume} {17}},\ \bibinfo {pages}
  {1316} (\bibinfo {year} {2021})}\BibitemShut {NoStop}%
\bibitem [{\citenamefont {Asteria}\ \emph {et~al.}(2021)\citenamefont
  {Asteria}, \citenamefont {Zahn}, \citenamefont {Kosch}, \citenamefont
  {Sengstock},\ and\ \citenamefont {Weitenberg}}]{asteria_quantum_2021}%
  \BibitemOpen
  \bibfield  {author} {\bibinfo {author} {\bibfnamefont {L.}~\bibnamefont
  {Asteria}}, \bibinfo {author} {\bibfnamefont {H.~P.}\ \bibnamefont {Zahn}},
  \bibinfo {author} {\bibfnamefont {M.~N.}\ \bibnamefont {Kosch}}, \bibinfo
  {author} {\bibfnamefont {K.}~\bibnamefont {Sengstock}},\ and\ \bibinfo
  {author} {\bibfnamefont {C.}~\bibnamefont {Weitenberg}},\ }\bibfield  {title}
  {\bibinfo {title} {Quantum gas magnifier for sub-lattice-resolved imaging of
  {3D} quantum systems},\ }\href {https://doi.org/10.1038/s41586-021-04011-2}
  {\bibfield  {journal} {\bibinfo  {journal} {Nature}\ }\textbf {\bibinfo
  {volume} {599}},\ \bibinfo {pages} {571} (\bibinfo {year}
  {2021})}\BibitemShut {NoStop}%
\bibitem [{\citenamefont {L{\"u}hmann}\ \emph {et~al.}(2014)\citenamefont
  {L{\"u}hmann}, \citenamefont {J{\"u}rgensen}, \citenamefont {Weinberg},
  \citenamefont {Simonet}, \citenamefont {Soltan-Panahi},\ and\ \citenamefont
  {Sengstock}}]{luhmann_quantum_2014}%
  \BibitemOpen
  \bibfield  {author} {\bibinfo {author} {\bibfnamefont {D.-S.}\ \bibnamefont
  {L{\"u}hmann}}, \bibinfo {author} {\bibfnamefont {O.}~\bibnamefont
  {J{\"u}rgensen}}, \bibinfo {author} {\bibfnamefont {M.}~\bibnamefont
  {Weinberg}}, \bibinfo {author} {\bibfnamefont {J.}~\bibnamefont {Simonet}},
  \bibinfo {author} {\bibfnamefont {P.}~\bibnamefont {Soltan-Panahi}},\ and\
  \bibinfo {author} {\bibfnamefont {K.}~\bibnamefont {Sengstock}},\ }\bibfield
  {title} {\bibinfo {title} {Quantum phases in tunable state-dependent
  hexagonal optical lattices},\ }\href
  {https://doi.org/10.1103/PhysRevA.90.013614} {\bibfield  {journal} {\bibinfo
  {journal} {Phys. Rev. A}\ }\textbf {\bibinfo {volume} {90}},\ \bibinfo
  {pages} {013614} (\bibinfo {year} {2014})}\BibitemShut {NoStop}%
\bibitem [{\citenamefont {Tarruell}\ \emph {et~al.}(2012)\citenamefont
  {Tarruell}, \citenamefont {Greif}, \citenamefont {Uehlinger}, \citenamefont
  {Jotzu},\ and\ \citenamefont {Esslinger}}]{tarruell_creating_2012}%
  \BibitemOpen
  \bibfield  {author} {\bibinfo {author} {\bibfnamefont {L.}~\bibnamefont
  {Tarruell}}, \bibinfo {author} {\bibfnamefont {D.}~\bibnamefont {Greif}},
  \bibinfo {author} {\bibfnamefont {T.}~\bibnamefont {Uehlinger}}, \bibinfo
  {author} {\bibfnamefont {G.}~\bibnamefont {Jotzu}},\ and\ \bibinfo {author}
  {\bibfnamefont {T.}~\bibnamefont {Esslinger}},\ }\bibfield  {title} {\bibinfo
  {title} {Creating, moving and merging {D}irac points with a {F}ermi gas in a
  tunable honeycomb lattice},\ }\href {https://doi.org/10.1038/nature10871}
  {\bibfield  {journal} {\bibinfo  {journal} {Nature}\ }\textbf {\bibinfo
  {volume} {483}},\ \bibinfo {pages} {302} (\bibinfo {year}
  {2012})}\BibitemShut {NoStop}%
\bibitem [{\citenamefont {Jaksch}\ and\ \citenamefont
  {Zoller}(2003)}]{Jaksch2003}%
  \BibitemOpen
  \bibfield  {author} {\bibinfo {author} {\bibfnamefont {D.}~\bibnamefont
  {Jaksch}}\ and\ \bibinfo {author} {\bibfnamefont {P.}~\bibnamefont
  {Zoller}},\ }\bibfield  {title} {\bibinfo {title} {{Creation of effective
  magnetic fields in optical lattices: the Hofstadter butterfly for cold
  neutral atoms}},\ }\href {https://doi.org/10.1088/1367-2630/5/1/356}
  {\bibfield  {journal} {\bibinfo  {journal} {New J. Phys.}\ }\textbf {\bibinfo
  {volume} {5}},\ \bibinfo {pages} {56} (\bibinfo {year} {2003})}\BibitemShut
  {NoStop}%
\bibitem [{\citenamefont {Eckardt}\ \emph {et~al.}(2005)\citenamefont
  {Eckardt}, \citenamefont {Weiss},\ and\ \citenamefont
  {Holthaus}}]{eckardt_superfluid-insulator_2005}%
  \BibitemOpen
  \bibfield  {author} {\bibinfo {author} {\bibfnamefont {A.}~\bibnamefont
  {Eckardt}}, \bibinfo {author} {\bibfnamefont {C.}~\bibnamefont {Weiss}},\
  and\ \bibinfo {author} {\bibfnamefont {M.}~\bibnamefont {Holthaus}},\
  }\bibfield  {title} {\bibinfo {title} {Superfluid-{I}nsulator {T}ransition in
  a {P}eriodically {D}riven {O}ptical {L}attice},\ }\href
  {https://doi.org/10.1103/PhysRevLett.95.260404} {\bibfield  {journal}
  {\bibinfo  {journal} {Phys. Rev. Lett.}\ }\textbf {\bibinfo {volume} {95}},\
  \bibinfo {pages} {260404} (\bibinfo {year} {2005})}\BibitemShut {NoStop}%
\bibitem [{\citenamefont {Lignier}\ \emph {et~al.}(2007)\citenamefont
  {Lignier}, \citenamefont {Sias}, \citenamefont {Ciampini}, \citenamefont
  {Singh}, \citenamefont {Zenesini}, \citenamefont {Morsch},\ and\
  \citenamefont {Arimondo}}]{lignier_dynamical_2007}%
  \BibitemOpen
  \bibfield  {author} {\bibinfo {author} {\bibfnamefont {H.}~\bibnamefont
  {Lignier}}, \bibinfo {author} {\bibfnamefont {C.}~\bibnamefont {Sias}},
  \bibinfo {author} {\bibfnamefont {D.}~\bibnamefont {Ciampini}}, \bibinfo
  {author} {\bibfnamefont {Y.}~\bibnamefont {Singh}}, \bibinfo {author}
  {\bibfnamefont {A.}~\bibnamefont {Zenesini}}, \bibinfo {author}
  {\bibfnamefont {O.}~\bibnamefont {Morsch}},\ and\ \bibinfo {author}
  {\bibfnamefont {E.}~\bibnamefont {Arimondo}},\ }\bibfield  {title} {\bibinfo
  {title} {Dynamical {C}ontrol of {M}atter-{W}ave {T}unneling in {P}eriodic
  {P}otentials},\ }\href {https://doi.org/10.1103/PhysRevLett.99.220403}
  {\bibfield  {journal} {\bibinfo  {journal} {Phys. Rev. Lett.}\ }\textbf
  {\bibinfo {volume} {99}},\ \bibinfo {pages} {220403} (\bibinfo {year}
  {2007})}\BibitemShut {NoStop}%
\bibitem [{\citenamefont {Ma}\ \emph {et~al.}(2011)\citenamefont {Ma},
  \citenamefont {Tai}, \citenamefont {Preiss}, \citenamefont {Bakr},
  \citenamefont {Simon},\ and\ \citenamefont
  {Greiner}}]{ma_photon-assisted_2011}%
  \BibitemOpen
  \bibfield  {author} {\bibinfo {author} {\bibfnamefont {R.}~\bibnamefont
  {Ma}}, \bibinfo {author} {\bibfnamefont {M.~E.}\ \bibnamefont {Tai}},
  \bibinfo {author} {\bibfnamefont {P.~M.}\ \bibnamefont {Preiss}}, \bibinfo
  {author} {\bibfnamefont {W.~S.}\ \bibnamefont {Bakr}}, \bibinfo {author}
  {\bibfnamefont {J.}~\bibnamefont {Simon}},\ and\ \bibinfo {author}
  {\bibfnamefont {M.}~\bibnamefont {Greiner}},\ }\bibfield  {title} {\bibinfo
  {title} {Photon-{A}ssisted {T}unneling in a {B}iased {S}trongly {C}orrelated
  {B}ose {G}as},\ }\href {https://doi.org/10.1103/PhysRevLett.107.095301}
  {\bibfield  {journal} {\bibinfo  {journal} {Phys. Rev. Lett.}\ }\textbf
  {\bibinfo {volume} {107}},\ \bibinfo {pages} {095301} (\bibinfo {year}
  {2011})}\BibitemShut {NoStop}%
\bibitem [{\citenamefont {Miyake}\ \emph {et~al.}(2013)\citenamefont {Miyake},
  \citenamefont {Siviloglou}, \citenamefont {Kennedy}, \citenamefont {Burton},\
  and\ \citenamefont {Ketterle}}]{miyake_realizing_2013}%
  \BibitemOpen
  \bibfield  {author} {\bibinfo {author} {\bibfnamefont {H.}~\bibnamefont
  {Miyake}}, \bibinfo {author} {\bibfnamefont {G.~A.}\ \bibnamefont
  {Siviloglou}}, \bibinfo {author} {\bibfnamefont {C.~J.}\ \bibnamefont
  {Kennedy}}, \bibinfo {author} {\bibfnamefont {W.~C.}\ \bibnamefont
  {Burton}},\ and\ \bibinfo {author} {\bibfnamefont {W.}~\bibnamefont
  {Ketterle}},\ }\bibfield  {title} {\bibinfo {title} {Realizing the {H}arper
  {H}amiltonian with {L}aser-{A}ssisted {T}unneling in {O}ptical {L}attices},\
  }\href {https://doi.org/10.1103/PhysRevLett.111.185302} {\bibfield  {journal}
  {\bibinfo  {journal} {Phys. Rev. Lett.}\ }\textbf {\bibinfo {volume} {111}},\
  \bibinfo {pages} {185302} (\bibinfo {year} {2013})}\BibitemShut {NoStop}%
\bibitem [{\citenamefont {Gallagher}(1994)}]{gallagher_rydberg_1994}%
  \BibitemOpen
  \bibfield  {author} {\bibinfo {author} {\bibfnamefont {T.~F.}\ \bibnamefont
  {Gallagher}},\ }\href {https://doi.org/10.1017/cbo9780511524530} {\emph
  {\bibinfo {title} {Rydberg Atoms}}}\ (\bibinfo  {publisher} {Cambridge
  University Press},\ \bibinfo {year} {1994})\BibitemShut {NoStop}%
\bibitem [{\citenamefont {Sibalic}(2018)}]{sibalic_rydberg_2018}%
  \BibitemOpen
  \bibfield  {author} {\bibinfo {author} {\bibfnamefont {N.}~\bibnamefont
  {Sibalic}},\ }\href {https://doi.org/10.1088/978-0-7503-1635-4} {\emph
  {\bibinfo {title} {Rydberg Physics}}}\ (\bibinfo  {publisher} {{IOP}
  Publishing},\ \bibinfo {year} {2018})\BibitemShut {NoStop}%
\bibitem [{\citenamefont {Adams}\ \emph {et~al.}(2019)\citenamefont {Adams},
  \citenamefont {Pritchard},\ and\ \citenamefont
  {Shaffer}}]{adams_rydberg_2019}%
  \BibitemOpen
  \bibfield  {author} {\bibinfo {author} {\bibfnamefont {C.~S.}\ \bibnamefont
  {Adams}}, \bibinfo {author} {\bibfnamefont {J.~D.}\ \bibnamefont
  {Pritchard}},\ and\ \bibinfo {author} {\bibfnamefont {J.~P.}\ \bibnamefont
  {Shaffer}},\ }\bibfield  {title} {\bibinfo {title} {Rydberg atom quantum
  technologies},\ }\href {https://doi.org/10.1088/1361-6455/ab52ef} {\bibfield
  {journal} {\bibinfo  {journal} {J. Phys. B}\ }\textbf {\bibinfo {volume}
  {53}},\ \bibinfo {pages} {012002} (\bibinfo {year} {2019})}\BibitemShut
  {NoStop}%
\bibitem [{\citenamefont {Singer}\ \emph {et~al.}(2004)\citenamefont {Singer},
  \citenamefont {Reetz-Lamour}, \citenamefont {Amthor}, \citenamefont
  {Marcassa},\ and\ \citenamefont {Weidemüller}}]{Singer2004}%
  \BibitemOpen
  \bibfield  {author} {\bibinfo {author} {\bibfnamefont {K.}~\bibnamefont
  {Singer}}, \bibinfo {author} {\bibfnamefont {M.}~\bibnamefont
  {Reetz-Lamour}}, \bibinfo {author} {\bibfnamefont {T.}~\bibnamefont
  {Amthor}}, \bibinfo {author} {\bibfnamefont {L.~G.}\ \bibnamefont
  {Marcassa}},\ and\ \bibinfo {author} {\bibfnamefont {M.}~\bibnamefont
  {Weidemüller}},\ }\bibfield  {title} {\bibinfo {title} {{Suppression of
  excitation and spectral broadening induced by interactions in a cold gas of
  Rydberg atoms}},\ }\href {https://doi.org/10.1103/PhysRevLett.93.163001}
  {\bibfield  {journal} {\bibinfo  {journal} {Phys. Rev. Lett.}\ }\textbf
  {\bibinfo {volume} {93}},\ \bibinfo {pages} {163001} (\bibinfo {year}
  {2004})}\BibitemShut {NoStop}%
\bibitem [{\citenamefont {Tong}\ \emph {et~al.}(2004)\citenamefont {Tong},
  \citenamefont {Farooqi}, \citenamefont {Stanojevic}, \citenamefont
  {Krishnan}, \citenamefont {Zhang}, \citenamefont {Côté}, \citenamefont
  {Eyler},\ and\ \citenamefont {Gould}}]{Tong2004}%
  \BibitemOpen
  \bibfield  {author} {\bibinfo {author} {\bibfnamefont {D.}~\bibnamefont
  {Tong}}, \bibinfo {author} {\bibfnamefont {S.~M.}\ \bibnamefont {Farooqi}},
  \bibinfo {author} {\bibfnamefont {J.}~\bibnamefont {Stanojevic}}, \bibinfo
  {author} {\bibfnamefont {S.}~\bibnamefont {Krishnan}}, \bibinfo {author}
  {\bibfnamefont {Y.~P.}\ \bibnamefont {Zhang}}, \bibinfo {author}
  {\bibfnamefont {R.}~\bibnamefont {Côté}}, \bibinfo {author} {\bibfnamefont
  {E.~E.}\ \bibnamefont {Eyler}},\ and\ \bibinfo {author} {\bibfnamefont
  {P.~L.}\ \bibnamefont {Gould}},\ }\bibfield  {title} {\bibinfo {title}
  {{Local blockade of Rydberg excitation in an ultracold gas}},\ }\href
  {https://doi.org/10.1103/PhysRevLett.93.063001} {\bibfield  {journal}
  {\bibinfo  {journal} {Phys. Rev. Lett.}\ }\textbf {\bibinfo {volume} {93}},\
  \bibinfo {pages} {063001} (\bibinfo {year} {2004})}\BibitemShut {NoStop}%
\bibitem [{\citenamefont {Vogt}\ \emph {et~al.}(2006)\citenamefont {Vogt},
  \citenamefont {Viteau}, \citenamefont {Zhao}, \citenamefont {Chotia},
  \citenamefont {Comparat},\ and\ \citenamefont {Pillet}}]{Vogt2006}%
  \BibitemOpen
  \bibfield  {author} {\bibinfo {author} {\bibfnamefont {T.}~\bibnamefont
  {Vogt}}, \bibinfo {author} {\bibfnamefont {M.}~\bibnamefont {Viteau}},
  \bibinfo {author} {\bibfnamefont {J.}~\bibnamefont {Zhao}}, \bibinfo {author}
  {\bibfnamefont {A.}~\bibnamefont {Chotia}}, \bibinfo {author} {\bibfnamefont
  {D.}~\bibnamefont {Comparat}},\ and\ \bibinfo {author} {\bibfnamefont
  {P.}~\bibnamefont {Pillet}},\ }\bibfield  {title} {\bibinfo {title} {{Dipole
  blockade at Förster resonances in high resolution laser excitation of
  Rydberg states of cesium atoms}},\ }\href
  {https://doi.org/10.1103/PhysRevLett.97.083003} {\bibfield  {journal}
  {\bibinfo  {journal} {Phys. Rev. Lett.}\ }\textbf {\bibinfo {volume} {97}},\
  \bibinfo {pages} {083003} (\bibinfo {year} {2006})}\BibitemShut {NoStop}%
\bibitem [{\citenamefont {Heidemann}\ \emph
  {et~al.}(2007{\natexlab{b}})\citenamefont {Heidemann}, \citenamefont
  {Raitzsch}, \citenamefont {Bendkowsky}, \citenamefont {Butscher},
  \citenamefont {Löw}, \citenamefont {Santos},\ and\ \citenamefont
  {Pfau}}]{Heidemann2007}%
  \BibitemOpen
  \bibfield  {author} {\bibinfo {author} {\bibfnamefont {R.}~\bibnamefont
  {Heidemann}}, \bibinfo {author} {\bibfnamefont {U.}~\bibnamefont {Raitzsch}},
  \bibinfo {author} {\bibfnamefont {V.}~\bibnamefont {Bendkowsky}}, \bibinfo
  {author} {\bibfnamefont {B.}~\bibnamefont {Butscher}}, \bibinfo {author}
  {\bibfnamefont {R.}~\bibnamefont {Löw}}, \bibinfo {author} {\bibfnamefont
  {L.}~\bibnamefont {Santos}},\ and\ \bibinfo {author} {\bibfnamefont
  {T.}~\bibnamefont {Pfau}},\ }\bibfield  {title} {\bibinfo {title} {{Evidence
  for coherent collective Rydberg excitation in the strong blockade regime}},\
  }\href {https://doi.org/10.1103/PhysRevLett.99.163601} {\bibfield  {journal}
  {\bibinfo  {journal} {Phys. Rev. Lett.}\ }\textbf {\bibinfo {volume} {99}},\
  \bibinfo {pages} {163601} (\bibinfo {year} {2007}{\natexlab{b}})}\BibitemShut
  {NoStop}%
\bibitem [{\citenamefont {Raitzsch}\ \emph {et~al.}(2008)\citenamefont
  {Raitzsch}, \citenamefont {Bendkowsky}, \citenamefont {Heidemann},
  \citenamefont {Butscher}, \citenamefont {Löw},\ and\ \citenamefont
  {Pfau}}]{Raitzsch2008}%
  \BibitemOpen
  \bibfield  {author} {\bibinfo {author} {\bibfnamefont {U.}~\bibnamefont
  {Raitzsch}}, \bibinfo {author} {\bibfnamefont {V.}~\bibnamefont
  {Bendkowsky}}, \bibinfo {author} {\bibfnamefont {R.}~\bibnamefont
  {Heidemann}}, \bibinfo {author} {\bibfnamefont {B.}~\bibnamefont {Butscher}},
  \bibinfo {author} {\bibfnamefont {R.}~\bibnamefont {Löw}},\ and\ \bibinfo
  {author} {\bibfnamefont {T.}~\bibnamefont {Pfau}},\ }\bibfield  {title}
  {\bibinfo {title} {{Echo experiments in a strongly interacting Rydberg
  gas}},\ }\href {https://doi.org/10.1103/PhysRevLett.100.013002} {\bibfield
  {journal} {\bibinfo  {journal} {Phys. Rev. Lett.}\ }\textbf {\bibinfo
  {volume} {100}},\ \bibinfo {pages} {013002} (\bibinfo {year}
  {2008})}\BibitemShut {NoStop}%
\bibitem [{\citenamefont {Urban}\ \emph {et~al.}(2009)\citenamefont {Urban},
  \citenamefont {Johnson}, \citenamefont {Henage}, \citenamefont {Isenhower},
  \citenamefont {Yavuz}, \citenamefont {Walker},\ and\ \citenamefont
  {Saffman}}]{urban_observation_2009}%
  \BibitemOpen
  \bibfield  {author} {\bibinfo {author} {\bibfnamefont {E.}~\bibnamefont
  {Urban}}, \bibinfo {author} {\bibfnamefont {T.~A.}\ \bibnamefont {Johnson}},
  \bibinfo {author} {\bibfnamefont {T.}~\bibnamefont {Henage}}, \bibinfo
  {author} {\bibfnamefont {L.}~\bibnamefont {Isenhower}}, \bibinfo {author}
  {\bibfnamefont {D.~D.}\ \bibnamefont {Yavuz}}, \bibinfo {author}
  {\bibfnamefont {T.~G.}\ \bibnamefont {Walker}},\ and\ \bibinfo {author}
  {\bibfnamefont {M.}~\bibnamefont {Saffman}},\ }\bibfield  {title} {\bibinfo
  {title} {Observation of {R}ydberg blockade between two atoms},\ }\href
  {https://doi.org/10.1038/nphys1178} {\bibfield  {journal} {\bibinfo
  {journal} {Nat. Phys.}\ }\textbf {\bibinfo {volume} {5}},\ \bibinfo {pages}
  {110} (\bibinfo {year} {2009})}\BibitemShut {NoStop}%
\bibitem [{\citenamefont {Ga{\"e}tan}\ \emph {et~al.}(2009)\citenamefont
  {Ga{\"e}tan}, \citenamefont {Miroshnychenko}, \citenamefont {Wilk},
  \citenamefont {Chotia}, \citenamefont {Viteau}, \citenamefont {Comparat},
  \citenamefont {Pillet}, \citenamefont {Browaeys},\ and\ \citenamefont
  {Grangier}}]{gaetan_observation_2009}%
  \BibitemOpen
  \bibfield  {author} {\bibinfo {author} {\bibfnamefont {A.}~\bibnamefont
  {Ga{\"e}tan}}, \bibinfo {author} {\bibfnamefont {Y.}~\bibnamefont
  {Miroshnychenko}}, \bibinfo {author} {\bibfnamefont {T.}~\bibnamefont
  {Wilk}}, \bibinfo {author} {\bibfnamefont {A.}~\bibnamefont {Chotia}},
  \bibinfo {author} {\bibfnamefont {M.}~\bibnamefont {Viteau}}, \bibinfo
  {author} {\bibfnamefont {D.}~\bibnamefont {Comparat}}, \bibinfo {author}
  {\bibfnamefont {P.}~\bibnamefont {Pillet}}, \bibinfo {author} {\bibfnamefont
  {A.}~\bibnamefont {Browaeys}},\ and\ \bibinfo {author} {\bibfnamefont
  {P.}~\bibnamefont {Grangier}},\ }\bibfield  {title} {\bibinfo {title}
  {Observation of collective excitation of two individual atoms in the
  {R}ydberg blockade regime},\ }\href {https://doi.org/10.1038/nphys1183}
  {\bibfield  {journal} {\bibinfo  {journal} {Nat. Phys.}\ }\textbf {\bibinfo
  {volume} {5}},\ \bibinfo {pages} {115} (\bibinfo {year} {2009})}\BibitemShut
  {NoStop}%
\bibitem [{\citenamefont {Lukin}\ \emph {et~al.}(2001)\citenamefont {Lukin},
  \citenamefont {Fleischhauer}, \citenamefont {Cote}, \citenamefont {Duan},
  \citenamefont {Jaksch}, \citenamefont {Cirac},\ and\ \citenamefont
  {Zoller}}]{Lukin2001}%
  \BibitemOpen
  \bibfield  {author} {\bibinfo {author} {\bibfnamefont {M.~D.}\ \bibnamefont
  {Lukin}}, \bibinfo {author} {\bibfnamefont {M.}~\bibnamefont {Fleischhauer}},
  \bibinfo {author} {\bibfnamefont {R.}~\bibnamefont {Cote}}, \bibinfo {author}
  {\bibfnamefont {L.~M.}\ \bibnamefont {Duan}}, \bibinfo {author}
  {\bibfnamefont {D.}~\bibnamefont {Jaksch}}, \bibinfo {author} {\bibfnamefont
  {J.~I.}\ \bibnamefont {Cirac}},\ and\ \bibinfo {author} {\bibfnamefont
  {P.}~\bibnamefont {Zoller}},\ }\bibfield  {title} {\bibinfo {title} {{Dipole
  Blockade and Quantum Information Processing in Mesoscopic Atomic
  Ensembles}},\ }\href {https://doi.org/10.1103/PhysRevLett.87.037901}
  {\bibfield  {journal} {\bibinfo  {journal} {Phys. Rev. Lett.}\ }\textbf
  {\bibinfo {volume} {87}},\ \bibinfo {pages} {037901} (\bibinfo {year}
  {2001})}\BibitemShut {NoStop}%
\bibitem [{\citenamefont {Xu}\ \emph {et~al.}(2021)\citenamefont {Xu},
  \citenamefont {Venkatramani}, \citenamefont {Cant\'u}, \citenamefont
  {\ifmmode~\check{S}\else \v{S}\fi{}umarac}, \citenamefont {Kl\"usener},
  \citenamefont {Lukin},\ and\ \citenamefont {Vuleti\ifmmode~\acute{c}\else
  \'{c}\fi{}}}]{Xu2021}%
  \BibitemOpen
  \bibfield  {author} {\bibinfo {author} {\bibfnamefont {W.}~\bibnamefont
  {Xu}}, \bibinfo {author} {\bibfnamefont {A.~V.}\ \bibnamefont
  {Venkatramani}}, \bibinfo {author} {\bibfnamefont {S.~H.}\ \bibnamefont
  {Cant\'u}}, \bibinfo {author} {\bibfnamefont {T.}~\bibnamefont
  {\ifmmode~\check{S}\else \v{S}\fi{}umarac}}, \bibinfo {author} {\bibfnamefont
  {V.}~\bibnamefont {Kl\"usener}}, \bibinfo {author} {\bibfnamefont {M.~D.}\
  \bibnamefont {Lukin}},\ and\ \bibinfo {author} {\bibfnamefont
  {V.}~\bibnamefont {Vuleti\ifmmode~\acute{c}\else \'{c}\fi{}}},\ }\bibfield
  {title} {\bibinfo {title} {Fast {P}reparation and {D}etection of a {R}ydberg
  {Q}ubit {U}sing {A}tomic {E}nsembles},\ }\href
  {https://doi.org/10.1103/PhysRevLett.127.050501} {\bibfield  {journal}
  {\bibinfo  {journal} {Phys. Rev. Lett.}\ }\textbf {\bibinfo {volume} {127}},\
  \bibinfo {pages} {050501} (\bibinfo {year} {2021})}\BibitemShut {NoStop}%
\bibitem [{\citenamefont {Shavitt}\ and\ \citenamefont
  {Redmon}(1980)}]{shavitt_quasidegenerate_1980}%
  \BibitemOpen
  \bibfield  {author} {\bibinfo {author} {\bibfnamefont {I.}~\bibnamefont
  {Shavitt}}\ and\ \bibinfo {author} {\bibfnamefont {L.~T.}\ \bibnamefont
  {Redmon}},\ }\bibfield  {title} {\bibinfo {title} {Quasidegenerate
  perturbation theories. a canonical van {V}leck formalism and its relationship
  to other approaches},\ }\href {https://doi.org/10.1063/1.440050} {\bibfield
  {journal} {\bibinfo  {journal} {J. Chem. Phys.}\ }\textbf {\bibinfo {volume}
  {73}},\ \bibinfo {pages} {5711} (\bibinfo {year} {1980})}\BibitemShut
  {NoStop}%
\bibitem [{\citenamefont {Stoudenmire}\ and\ \citenamefont
  {White}(2012)}]{Stoudenmire2012}%
  \BibitemOpen
  \bibfield  {author} {\bibinfo {author} {\bibfnamefont {E.}~\bibnamefont
  {Stoudenmire}}\ and\ \bibinfo {author} {\bibfnamefont {S.~R.}\ \bibnamefont
  {White}},\ }\bibfield  {title} {\bibinfo {title} {Studying two-dimensional
  systems with the density matrix renormalization group},\ }\href
  {https://doi.org/10.1146/annurev-conmatphys-020911-125018} {\bibfield
  {journal} {\bibinfo  {journal} {Annu. Rev. Condens. Matter Phys.}\ }\textbf
  {\bibinfo {volume} {3}},\ \bibinfo {pages} {111} (\bibinfo {year}
  {2012})}\BibitemShut {NoStop}%
\bibitem [{\citenamefont {Rivas}\ \emph {et~al.}(2013)\citenamefont {Rivas},
  \citenamefont {Viyuela},\ and\ \citenamefont {Martin-Delgado}}]{Rivas2013}%
  \BibitemOpen
  \bibfield  {author} {\bibinfo {author} {\bibfnamefont {A.}~\bibnamefont
  {Rivas}}, \bibinfo {author} {\bibfnamefont {O.}~\bibnamefont {Viyuela}},\
  and\ \bibinfo {author} {\bibfnamefont {M.~A.}\ \bibnamefont
  {Martin-Delgado}},\ }\bibfield  {title} {\bibinfo {title} {Density-matrix
  {C}hern insulators: Finite-temperature generalization of topological
  insulators},\ }\href {https://doi.org/10.1103/PhysRevB.88.155141} {\bibfield
  {journal} {\bibinfo  {journal} {Phys. Rev. B}\ }\textbf {\bibinfo {volume}
  {88}},\ \bibinfo {pages} {155141} (\bibinfo {year} {2013})}\BibitemShut
  {NoStop}%
\bibitem [{\citenamefont {Viyuela}\ \emph {et~al.}(2014)\citenamefont
  {Viyuela}, \citenamefont {Rivas},\ and\ \citenamefont
  {Martin-Delgado}}]{Viyuela2014}%
  \BibitemOpen
  \bibfield  {author} {\bibinfo {author} {\bibfnamefont {O.}~\bibnamefont
  {Viyuela}}, \bibinfo {author} {\bibfnamefont {A.}~\bibnamefont {Rivas}},\
  and\ \bibinfo {author} {\bibfnamefont {M.~A.}\ \bibnamefont
  {Martin-Delgado}},\ }\bibfield  {title} {\bibinfo {title} {{Two-Dimensional
  Density-Matrix Topological Fermionic Phases: Topological Uhlmann Numbers}},\
  }\href {https://doi.org/10.1103/PhysRevLett.113.076408} {\bibfield  {journal}
  {\bibinfo  {journal} {Phys. Rev. Lett.}\ }\textbf {\bibinfo {volume} {113}},\
  \bibinfo {pages} {076408} (\bibinfo {year} {2014})}\BibitemShut {NoStop}%
\bibitem [{\citenamefont {Huang}\ and\ \citenamefont
  {Arovas}(2014)}]{Huang2014}%
  \BibitemOpen
  \bibfield  {author} {\bibinfo {author} {\bibfnamefont {Z.}~\bibnamefont
  {Huang}}\ and\ \bibinfo {author} {\bibfnamefont {D.~P.}\ \bibnamefont
  {Arovas}},\ }\bibfield  {title} {\bibinfo {title} {{Topological Indices for
  Open and Thermal Systems Via Uhlmann's Phase}},\ }\href
  {https://doi.org/10.1103/PhysRevLett.113.076407} {\bibfield  {journal}
  {\bibinfo  {journal} {Phys. Rev. Lett.}\ }\textbf {\bibinfo {volume} {113}},\
  \bibinfo {pages} {076407} (\bibinfo {year} {2014})}\BibitemShut {NoStop}%
\bibitem [{\citenamefont {Budich}\ and\ \citenamefont
  {Diehl}(2015)}]{Budich2015}%
  \BibitemOpen
  \bibfield  {author} {\bibinfo {author} {\bibfnamefont {J.~C.}\ \bibnamefont
  {Budich}}\ and\ \bibinfo {author} {\bibfnamefont {S.}~\bibnamefont {Diehl}},\
  }\bibfield  {title} {\bibinfo {title} {Topology of density matrices},\ }\href
  {https://doi.org/10.1103/PhysRevB.91.165140} {\bibfield  {journal} {\bibinfo
  {journal} {Phys. Rev. B}\ }\textbf {\bibinfo {volume} {91}},\ \bibinfo
  {pages} {165140} (\bibinfo {year} {2015})}\BibitemShut {NoStop}%
\bibitem [{\citenamefont {Bardyn}\ \emph {et~al.}(2018)\citenamefont {Bardyn},
  \citenamefont {Wawer}, \citenamefont {Altland}, \citenamefont
  {Fleischhauer},\ and\ \citenamefont {Diehl}}]{Bardyn2018}%
  \BibitemOpen
  \bibfield  {author} {\bibinfo {author} {\bibfnamefont {C.-E.}\ \bibnamefont
  {Bardyn}}, \bibinfo {author} {\bibfnamefont {L.}~\bibnamefont {Wawer}},
  \bibinfo {author} {\bibfnamefont {A.}~\bibnamefont {Altland}}, \bibinfo
  {author} {\bibfnamefont {M.}~\bibnamefont {Fleischhauer}},\ and\ \bibinfo
  {author} {\bibfnamefont {S.}~\bibnamefont {Diehl}},\ }\bibfield  {title}
  {\bibinfo {title} {{Probing the Topology of Density Matrices}},\ }\href
  {https://doi.org/10.1103/PhysRevX.8.011035} {\bibfield  {journal} {\bibinfo
  {journal} {Phys. Rev. X}\ }\textbf {\bibinfo {volume} {8}},\ \bibinfo {pages}
  {011035} (\bibinfo {year} {2018})}\BibitemShut {NoStop}%
\bibitem [{\citenamefont
  {Schollw{\"o}ck}(2011)}]{schollwock_density-matrix_2011}%
  \BibitemOpen
  \bibfield  {author} {\bibinfo {author} {\bibfnamefont {U.}~\bibnamefont
  {Schollw{\"o}ck}},\ }\bibfield  {title} {\bibinfo {title} {The density-matrix
  renormalization group in the age of matrix product states},\ }\href
  {https://doi.org/10.1016/j.aop.2010.09.012} {\bibfield  {journal} {\bibinfo
  {journal} {Ann. Phys.}\ }\textbf {\bibinfo {volume} {326}},\ \bibinfo {pages}
  {96} (\bibinfo {year} {2011})}\BibitemShut {NoStop}%
\bibitem [{\citenamefont {Hauschild}\ and\ \citenamefont
  {Pollmann}(2018)}]{tenpy}%
  \BibitemOpen
  \bibfield  {author} {\bibinfo {author} {\bibfnamefont {J.}~\bibnamefont
  {Hauschild}}\ and\ \bibinfo {author} {\bibfnamefont {F.}~\bibnamefont
  {Pollmann}},\ }\bibfield  {title} {\bibinfo {title} {{Efficient numerical
  simulations with {T}ensor {N}etworks: {T}ensor {N}etwork {P}ython
  ({TeNPy})}},\ }\href {https://doi.org/10.21468/SciPostPhysLectNotes.5}
  {\bibfield  {journal} {\bibinfo  {journal} {SciPost Phys. Lect. Notes}\ ,\
  \bibinfo {pages} {5}} (\bibinfo {year} {2018})}\BibitemShut {NoStop}%
\bibitem [{Note2()}]{Note2}%
  \BibitemOpen
  \bibinfo {note} {The tunneling amplitude depends on the atomic mass but can
  be further fine tuned by changing the lattice depth.}\BibitemShut {Stop}%
\bibitem [{\citenamefont {Weber}\ \emph {et~al.}(2017)\citenamefont {Weber},
  \citenamefont {Tresp}, \citenamefont {Menke}, \citenamefont {Urvoy},
  \citenamefont {Firstenberg}, \citenamefont {B{\"u}chler},\ and\ \citenamefont
  {Hofferberth}}]{Weber2017}%
  \BibitemOpen
  \bibfield  {author} {\bibinfo {author} {\bibfnamefont {S.}~\bibnamefont
  {Weber}}, \bibinfo {author} {\bibfnamefont {C.}~\bibnamefont {Tresp}},
  \bibinfo {author} {\bibfnamefont {H.}~\bibnamefont {Menke}}, \bibinfo
  {author} {\bibfnamefont {A.}~\bibnamefont {Urvoy}}, \bibinfo {author}
  {\bibfnamefont {O.}~\bibnamefont {Firstenberg}}, \bibinfo {author}
  {\bibfnamefont {H.~P.}\ \bibnamefont {B{\"u}chler}},\ and\ \bibinfo {author}
  {\bibfnamefont {S.}~\bibnamefont {Hofferberth}},\ }\bibfield  {title}
  {\bibinfo {title} {{Tutorial: Calculation of Rydberg interaction
  potentials}},\ }\href {https://doi.org/10.1088/1361-6455/aa743a} {\bibfield
  {journal} {\bibinfo  {journal} {J. Phys. B: At. Mol. Opt. Phys.}\ }\textbf
  {\bibinfo {volume} {50}},\ \bibinfo {pages} {133001} (\bibinfo {year}
  {2017})}\BibitemShut {NoStop}%
\bibitem [{\citenamefont {Beterov}\ \emph {et~al.}(2009)\citenamefont
  {Beterov}, \citenamefont {Ryabtsev}, \citenamefont {Tretyakov},\ and\
  \citenamefont {Entin}}]{Beterov2009}%
  \BibitemOpen
  \bibfield  {author} {\bibinfo {author} {\bibfnamefont {I.~I.}\ \bibnamefont
  {Beterov}}, \bibinfo {author} {\bibfnamefont {I.~I.}\ \bibnamefont
  {Ryabtsev}}, \bibinfo {author} {\bibfnamefont {D.~B.}\ \bibnamefont
  {Tretyakov}},\ and\ \bibinfo {author} {\bibfnamefont {V.~M.}\ \bibnamefont
  {Entin}},\ }\bibfield  {title} {\bibinfo {title} {{Quasiclassical
  calculations of blackbody-radiation-induced depopulation rates and effective
  lifetimes of Rydberg $nS$, $nP$, and $nD$ alkali-metal atoms with
  $n\ensuremath{\le}80$}},\ }\href {https://doi.org/10.1103/PhysRevA.79.052504}
  {\bibfield  {journal} {\bibinfo  {journal} {Phys. Rev. A}\ }\textbf {\bibinfo
  {volume} {79}},\ \bibinfo {pages} {052504} (\bibinfo {year}
  {2009})}\BibitemShut {NoStop}%
\bibitem [{\citenamefont {Motruk}\ and\ \citenamefont
  {Pollmann}(2017)}]{Motruk2017}%
  \BibitemOpen
  \bibfield  {author} {\bibinfo {author} {\bibfnamefont {J.}~\bibnamefont
  {Motruk}}\ and\ \bibinfo {author} {\bibfnamefont {F.}~\bibnamefont
  {Pollmann}},\ }\bibfield  {title} {\bibinfo {title} {Phase transitions and
  adiabatic preparation of a fractional {C}hern insulator in a boson cold-atom
  model},\ }\href {https://doi.org/10.1103/PhysRevB.96.165107} {\bibfield
  {journal} {\bibinfo  {journal} {Phys. Rev. B}\ }\textbf {\bibinfo {volume}
  {96}},\ \bibinfo {pages} {165107} (\bibinfo {year} {2017})}\BibitemShut
  {NoStop}%
\bibitem [{\citenamefont {He}\ \emph {et~al.}(2017)\citenamefont {He},
  \citenamefont {Grusdt}, \citenamefont {Kaufman}, \citenamefont {Greiner},\
  and\ \citenamefont {Vishwanath}}]{He2017}%
  \BibitemOpen
  \bibfield  {author} {\bibinfo {author} {\bibfnamefont {Y.-C.}\ \bibnamefont
  {He}}, \bibinfo {author} {\bibfnamefont {F.}~\bibnamefont {Grusdt}}, \bibinfo
  {author} {\bibfnamefont {A.}~\bibnamefont {Kaufman}}, \bibinfo {author}
  {\bibfnamefont {M.}~\bibnamefont {Greiner}},\ and\ \bibinfo {author}
  {\bibfnamefont {A.}~\bibnamefont {Vishwanath}},\ }\bibfield  {title}
  {\bibinfo {title} {Realizing and adiabatically preparing bosonic integer and
  fractional quantum {H}all states in optical lattices},\ }\href
  {https://doi.org/10.1103/PhysRevB.96.201103} {\bibfield  {journal} {\bibinfo
  {journal} {Phys. Rev. B}\ }\textbf {\bibinfo {volume} {96}},\ \bibinfo
  {pages} {201103} (\bibinfo {year} {2017})}\BibitemShut {NoStop}%
\bibitem [{\citenamefont {Alba}\ \emph {et~al.}(2011)\citenamefont {Alba},
  \citenamefont {Fernandez-Gonzalvo}, \citenamefont {Mur-Petit}, \citenamefont
  {Pachos},\ and\ \citenamefont {Garcia-Ripoll}}]{PhysRevLett.107.235301}%
  \BibitemOpen
  \bibfield  {author} {\bibinfo {author} {\bibfnamefont {E.}~\bibnamefont
  {Alba}}, \bibinfo {author} {\bibfnamefont {X.}~\bibnamefont
  {Fernandez-Gonzalvo}}, \bibinfo {author} {\bibfnamefont {J.}~\bibnamefont
  {Mur-Petit}}, \bibinfo {author} {\bibfnamefont {J.~K.}\ \bibnamefont
  {Pachos}},\ and\ \bibinfo {author} {\bibfnamefont {J.~J.}\ \bibnamefont
  {Garcia-Ripoll}},\ }\bibfield  {title} {\bibinfo {title} {Seeing topological
  order in time-of-flight measurements},\ }\href
  {https://doi.org/10.1103/PhysRevLett.107.235301} {\bibfield  {journal}
  {\bibinfo  {journal} {Phys. Rev. Lett.}\ }\textbf {\bibinfo {volume} {107}},\
  \bibinfo {pages} {235301} (\bibinfo {year} {2011})}\BibitemShut {NoStop}%
\bibitem [{\citenamefont {Dauphin}\ and\ \citenamefont
  {Goldman}(2013)}]{dauphin2013}%
  \BibitemOpen
  \bibfield  {author} {\bibinfo {author} {\bibfnamefont {A.}~\bibnamefont
  {Dauphin}}\ and\ \bibinfo {author} {\bibfnamefont {N.}~\bibnamefont
  {Goldman}},\ }\bibfield  {title} {\bibinfo {title} {Extracting the {C}hern
  {N}umber from the {D}ynamics of a {F}ermi {G}as: {I}mplementing a {Q}uantum
  {H}all {B}ar for {C}old {A}toms},\ }\href
  {https://doi.org/10.1103/PhysRevLett.111.135302} {\bibfield  {journal}
  {\bibinfo  {journal} {Phys. Rev. Lett.}\ }\textbf {\bibinfo {volume} {111}},\
  \bibinfo {pages} {135302} (\bibinfo {year} {2013})}\BibitemShut {NoStop}%
\bibitem [{\citenamefont {Tran}\ \emph {et~al.}(2017)\citenamefont {Tran},
  \citenamefont {Dauphin}, \citenamefont {Grushin}, \citenamefont {Zoller},\
  and\ \citenamefont {Goldman}}]{Tran2017}%
  \BibitemOpen
  \bibfield  {author} {\bibinfo {author} {\bibfnamefont {D.~T.}\ \bibnamefont
  {Tran}}, \bibinfo {author} {\bibfnamefont {A.}~\bibnamefont {Dauphin}},
  \bibinfo {author} {\bibfnamefont {A.~G.}\ \bibnamefont {Grushin}}, \bibinfo
  {author} {\bibfnamefont {P.}~\bibnamefont {Zoller}},\ and\ \bibinfo {author}
  {\bibfnamefont {N.}~\bibnamefont {Goldman}},\ }\bibfield  {title} {\bibinfo
  {title} {Probing topology by {\textquotedblleft}heating{\textquotedblright}:
  Quantized circular dichroism in ultracold atoms},\ }\href
  {https://doi.org/10.1126/sciadv.1701207} {\bibfield  {journal} {\bibinfo
  {journal} {Sci. Adv.}\ }\textbf {\bibinfo {volume} {3}},\ \bibinfo {pages}
  {e1701207} (\bibinfo {year} {2017})}\BibitemShut {NoStop}%
\bibitem [{\citenamefont {Goldman}\ \emph {et~al.}(2013)\citenamefont
  {Goldman}, \citenamefont {Dalibard}, \citenamefont {Dauphin}, \citenamefont
  {Gerbier}, \citenamefont {Lewenstein}, \citenamefont {Zoller},\ and\
  \citenamefont {Spielman}}]{Goldman2013b}%
  \BibitemOpen
  \bibfield  {author} {\bibinfo {author} {\bibfnamefont {N.}~\bibnamefont
  {Goldman}}, \bibinfo {author} {\bibfnamefont {J.}~\bibnamefont {Dalibard}},
  \bibinfo {author} {\bibfnamefont {A.}~\bibnamefont {Dauphin}}, \bibinfo
  {author} {\bibfnamefont {F.}~\bibnamefont {Gerbier}}, \bibinfo {author}
  {\bibfnamefont {M.}~\bibnamefont {Lewenstein}}, \bibinfo {author}
  {\bibfnamefont {P.}~\bibnamefont {Zoller}},\ and\ \bibinfo {author}
  {\bibfnamefont {I.~B.}\ \bibnamefont {Spielman}},\ }\bibfield  {title}
  {\bibinfo {title} {Direct imaging of topological edge states in cold-atom
  systems},\ }\href {https://doi.org/10.1073/pnas.1300170110} {\bibfield
  {journal} {\bibinfo  {journal} {Proc. Natl. Acad. Sci. U.S.A.}\ }\textbf
  {\bibinfo {volume} {110}},\ \bibinfo {pages} {6736} (\bibinfo {year}
  {2013})}\BibitemShut {NoStop}%
\bibitem [{\citenamefont {Asteria}\ \emph
  {et~al.}(2019{\natexlab{b}})\citenamefont {Asteria}, \citenamefont {Tran},
  \citenamefont {Ozawa}, \citenamefont {Tarnowski}, \citenamefont {Rem},
  \citenamefont {Fl{\"a}schner}, \citenamefont {Sengstock}, \citenamefont
  {Goldman},\ and\ \citenamefont {Weitenberg}}]{Asteria_2018}%
  \BibitemOpen
  \bibfield  {author} {\bibinfo {author} {\bibfnamefont {L.}~\bibnamefont
  {Asteria}}, \bibinfo {author} {\bibfnamefont {D.~T.}\ \bibnamefont {Tran}},
  \bibinfo {author} {\bibfnamefont {T.}~\bibnamefont {Ozawa}}, \bibinfo
  {author} {\bibfnamefont {M.}~\bibnamefont {Tarnowski}}, \bibinfo {author}
  {\bibfnamefont {B.~S.}\ \bibnamefont {Rem}}, \bibinfo {author} {\bibfnamefont
  {N.}~\bibnamefont {Fl{\"a}schner}}, \bibinfo {author} {\bibfnamefont
  {K.}~\bibnamefont {Sengstock}}, \bibinfo {author} {\bibfnamefont
  {N.}~\bibnamefont {Goldman}},\ and\ \bibinfo {author} {\bibfnamefont
  {C.}~\bibnamefont {Weitenberg}},\ }\bibfield  {title} {\bibinfo {title}
  {Measuring quantized circular dichroism in ultracold topological matter},\
  }\href {https://doi.org/10.1038/s41567-019-0417-8} {\bibfield  {journal}
  {\bibinfo  {journal} {Nat. Phys.}\ }\textbf {\bibinfo {volume} {15}},\
  \bibinfo {pages} {449} (\bibinfo {year} {2019}{\natexlab{b}})}\BibitemShut
  {NoStop}%
\bibitem [{\citenamefont {Goldman}\ \emph {et~al.}(2016)\citenamefont
  {Goldman}, \citenamefont {Jotzu}, \citenamefont {Messer}, \citenamefont
  {G{\ifmmode\ddot{o}\else\"{o}\fi}rg}, \citenamefont {Desbuquois},\ and\
  \citenamefont {Esslinger}}]{Goldman_2016}%
  \BibitemOpen
  \bibfield  {author} {\bibinfo {author} {\bibfnamefont {N.}~\bibnamefont
  {Goldman}}, \bibinfo {author} {\bibfnamefont {G.}~\bibnamefont {Jotzu}},
  \bibinfo {author} {\bibfnamefont {M.}~\bibnamefont {Messer}}, \bibinfo
  {author} {\bibfnamefont {F.}~\bibnamefont
  {G{\ifmmode\ddot{o}\else\"{o}\fi}rg}}, \bibinfo {author} {\bibfnamefont
  {R.}~\bibnamefont {Desbuquois}},\ and\ \bibinfo {author} {\bibfnamefont
  {T.}~\bibnamefont {Esslinger}},\ }\bibfield  {title} {\bibinfo {title}
  {{Creating topological interfaces and detecting chiral edge modes in a
  two-dimensional optical lattice}},\ }\href
  {https://doi.org/10.1103/PhysRevA.94.043611} {\bibfield  {journal} {\bibinfo
  {journal} {Phys. Rev. A}\ }\textbf {\bibinfo {volume} {94}},\ \bibinfo
  {pages} {043611} (\bibinfo {year} {2016})}\BibitemShut {NoStop}%
\bibitem [{\citenamefont {Zurek}\ \emph {et~al.}(2005)\citenamefont {Zurek},
  \citenamefont {Dorner},\ and\ \citenamefont {Zoller}}]{Zurek_2005}%
  \BibitemOpen
  \bibfield  {author} {\bibinfo {author} {\bibfnamefont {W.~H.}\ \bibnamefont
  {Zurek}}, \bibinfo {author} {\bibfnamefont {U.}~\bibnamefont {Dorner}},\ and\
  \bibinfo {author} {\bibfnamefont {P.}~\bibnamefont {Zoller}},\ }\bibfield
  {title} {\bibinfo {title} {Dynamics of a {Q}uantum {P}hase {T}ransition},\
  }\href {https://doi.org/10.1103/PhysRevLett.95.105701} {\bibfield  {journal}
  {\bibinfo  {journal} {Phys. Rev. Lett.}\ }\textbf {\bibinfo {volume} {95}},\
  \bibinfo {pages} {105701} (\bibinfo {year} {2005})}\BibitemShut {NoStop}%
\bibitem [{\citenamefont {del Campo}\ and\ \citenamefont
  {Zurek}(2014)}]{DelCampo_2014}%
  \BibitemOpen
  \bibfield  {author} {\bibinfo {author} {\bibfnamefont {A.}~\bibnamefont {del
  Campo}}\ and\ \bibinfo {author} {\bibfnamefont {W.~H.}\ \bibnamefont
  {Zurek}},\ }\bibfield  {title} {\bibinfo {title} {Universality of phase
  transition dynamics: {T}opological defects from symmetry breaking},\ }\href
  {https://doi.org/10.1142/S0217751X1430018X} {\bibfield  {journal} {\bibinfo
  {journal} {Int. J. Mod. Phys. A}\ }\textbf {\bibinfo {volume} {29}},\
  \bibinfo {pages} {1430018} (\bibinfo {year} {2014})}\BibitemShut {NoStop}%
\bibitem [{\citenamefont {Keesling}\ \emph
  {et~al.}(2019{\natexlab{b}})\citenamefont {Keesling}, \citenamefont {Omran},
  \citenamefont {Levine}, \citenamefont {Bernien}, \citenamefont {Pichler},
  \citenamefont {Choi}, \citenamefont {Samajdar}, \citenamefont {Schwartz},
  \citenamefont {Silvi}, \citenamefont {Sachdev}, \citenamefont {Zoller},
  \citenamefont {Endres}, \citenamefont {Greiner}, \citenamefont
  {Vuleti{\'c}},\ and\ \citenamefont {Lukin}}]{KibbleZurekRydberg}%
  \BibitemOpen
  \bibfield  {author} {\bibinfo {author} {\bibfnamefont {A.}~\bibnamefont
  {Keesling}}, \bibinfo {author} {\bibfnamefont {A.}~\bibnamefont {Omran}},
  \bibinfo {author} {\bibfnamefont {H.}~\bibnamefont {Levine}}, \bibinfo
  {author} {\bibfnamefont {H.}~\bibnamefont {Bernien}}, \bibinfo {author}
  {\bibfnamefont {H.}~\bibnamefont {Pichler}}, \bibinfo {author} {\bibfnamefont
  {S.}~\bibnamefont {Choi}}, \bibinfo {author} {\bibfnamefont {R.}~\bibnamefont
  {Samajdar}}, \bibinfo {author} {\bibfnamefont {S.}~\bibnamefont {Schwartz}},
  \bibinfo {author} {\bibfnamefont {P.}~\bibnamefont {Silvi}}, \bibinfo
  {author} {\bibfnamefont {S.}~\bibnamefont {Sachdev}}, \bibinfo {author}
  {\bibfnamefont {P.}~\bibnamefont {Zoller}}, \bibinfo {author} {\bibfnamefont
  {M.}~\bibnamefont {Endres}}, \bibinfo {author} {\bibfnamefont
  {M.}~\bibnamefont {Greiner}}, \bibinfo {author} {\bibfnamefont
  {V.}~\bibnamefont {Vuleti{\'c}}},\ and\ \bibinfo {author} {\bibfnamefont
  {M.~D.}\ \bibnamefont {Lukin}},\ }\bibfield  {title} {\bibinfo {title}
  {Quantum {K}ibble--{Z}urek mechanism and critical dynamics on a programmable
  {R}ydberg simulator},\ }\href {https://doi.org/10.1038/s41586-019-1070-1}
  {\bibfield  {journal} {\bibinfo  {journal} {Nature}\ }\textbf {\bibinfo
  {volume} {568}},\ \bibinfo {pages} {207} (\bibinfo {year}
  {2019}{\natexlab{b}})}\BibitemShut {NoStop}%
\bibitem [{\citenamefont {Julià-Farré}\ \emph {et~al.}(2022)\citenamefont
  {Julià-Farré}, \citenamefont {Dauphin},\ and\ \citenamefont
  {Cardarelli}}]{repo}%
  \BibitemOpen
  \bibfield  {author} {\bibinfo {author} {\bibfnamefont {S.}~\bibnamefont
  {Julià-Farré}}, \bibinfo {author} {\bibfnamefont {A.}~\bibnamefont
  {Dauphin}},\ and\ \bibinfo {author} {\bibfnamefont {L.}~\bibnamefont
  {Cardarelli}},\ }\href
  {https://sergijulia94.github.io/topological_mott_insulator} {\bibinfo {title}
  {{Hartree-Fock toolbox for the Topological Mott Insulator}}} (\bibinfo {year}
  {2022})\BibitemShut {NoStop}%
\end{thebibliography}%

\appendix


\section{Hartree-Fock method}
\label{sec:apphf}

\subsection{Hartree-Fock expansion and self-consistent loop}
\label{sec:app02_hf}

Here we discuss how we find the mean-field parameters of the Hamiltonian in equation \eqref{eq:hamiltonianmf} of the main text, which results from the Hartree-Fock decoupling,
\begin{equation} 
\begin{split}
\hat{n}_i\hat{n}_j \simeq &-\xi_{ij}\cdop_j\cop_i-\xi_{ij}^*\cdop_i\cop_j+\abs{\xi_{ij}}^2\\&+\bar{n}_i\nop_j+\bar{n}_j\nop_i-\bar{n}_i\bar{n}_j,
\end{split}
\label{eq:hfe}
\end{equation}
with $\bar{n}_i=\langle \nop_i\rangle$, and $\xi_{ij}=\langle\cdop_i\cop_j\rangle$.
Such Hamiltonian can be brought into the convenient form
\begin{equation}\label{eq:H_hf_app}
    \hat{H}_\textrm{HF}= \sum_{i,j=1}^d \left[ h_{ij}(\bar{n},\xi)\cdop_i\cop_j + \textrm{H.c.} \right]+C(\bar{n},\xi),
\end{equation}
where the sum is over a finite number $d$ of lattice sites, $C$ is a scalar term resulting from the Hartree-Fock decoupling, and $(\bar{n},\xi)$ refers to the set of Hartree-Fock parameters, on which the Hamiltonian matrix elements $h_{ij}$ depend self-consistently.
The general case in which no constraints are imposed on the mean-field parameters is commonly known as the unrestricted Hartree-Fock method. Below we outline this method, which we have used in this work to study inhomogeneous solutions in real space, shown in Figs.~\ref{fig:06_doping},\ref{fig:09_temp_defects}. In the next subsection we also discuss  the restricted Hartree-Fock method in which one constrains the Hartree-Fock values to be periodic within a certain unit cell, resulting in a numerical simplification of the algorithm.

Starting from an initial guess for the value of these Hartree-Fock parameters, the Hamiltonian matrix $h_{ij}$ can be diagonalized by means of a Bogoliubov transformation $U$ defined by $(\hat{f}_1^\dagger, \hat{f}_1, \dots, \hat{f}_d^\dagger, \hat{f}_d)^{T} = U(\cdop_1,\ \cop_1, \ \dots, \ \cdop_d,\ \cop_d)^{T}$, such that the Hamiltonian takes a diagonal form:
\begin{equation}
    \hat{H}_\textrm{HF}=\sum_{i=1}^d E_i\hat{f}_i^\dagger\hat{f}_i+C'.
\end{equation}
Notice that, in the mean-field approximation, it is feasible to consider large system sizes $d$ because the numerical complexity scales polynomially in $d$ (diagonalization of a $d\times d$ matrix $h_{ij}$) in contrast with the exponential scaling of the general many-body case.
Finally, in the equilibrium state at temperature $T$, the occupation number of the Bogoliubov modes $\hat{f}_i$ is given by the Fermi distribution
\begin{equation}\label{eq:exp_bogo}
    \langle\hat{f}^\dagger_i\hat{f}_j\rangle=\delta_{ij}f(E_i,\mu,T)=\frac{\delta_{ij}}{1+\exp[(E_i-\mu)/k_{\textrm{B}}T]},
\end{equation}
where $\mu$ is the chemical potential that is used to fix the total particle number trough the condition $\sum_i f(E_i,\mu,T)=N$. At half filling and $T=0$ one gets $\langle\hat{f}^\dagger_i\hat{f}_i\rangle=1$ for the $N/2$ lower energy states, and $0$ for the other ones. 

Therefore, any expectation value in the original fermionic basis can be computed using Eq.~\eqref{eq:exp_bogo} together with the Bogoliubov transformation $U$. In particular, one can compute the new values of the Hartree-Fock parameters $\bar{n}_i$ and $\xi_{ij}$, which give a new Hamiltonian matrix $h'_{ij}$ that can be diagonalized again following the procedure described above. 
This process is iterated until convergence of the Hartree-Fock parameters is achieved. In order to avoid metastable solutions, one needs to compare the free energies $F_\textrm{HF}$ of different converged solutions. The free energy is defined as
\begin{equation}
\begin{split}
F_{\textrm{HF}}=\sum_i &\left\{ \frac{\mu}{1+\exp(\frac{E_i-\mu}{k_BT})}\right.\\
&\left. -k_BT\ln \left[\exp\left(-\frac{E_i-\mu}{k_BT}\right)+1\right]\right\}+C',
\end{split}
\end{equation}
which, at $T=0$, is simply given by the expectation value of $\Hop_{\textrm{HF}}$.
Importantly, in order to converge to solutions that break Hamiltonian symmetries, it is crucial that the initial  Hartree-Fock parameters already break them. For instance, the QAH phase requires initial complex values of $\xi_{ij}$ to break time-reversal symmetry, that is $\xi_{ij}$ should be initialized to numbers with a finite imaginary part, and charge ordered phases require spatially inhomogeneous distributions of $\bar{n}_i$.

\subsection{Restricted Hartree-Fock with eight-site unit cell}
The checkerboard lattice is a bipartite two-dimensional Bravais lattice that can be uniquely defined with a two-site unit-cell coordinate, accounting for sub-lattices A and B, and two unit vectors. The Fourier-transformed free Hamiltonian $\Hop_0$ of Eq.~\eqref{eq:hamiltonianfree} has thus a two-dimensional matrix form in $\mathbf{k}$-space, accounting for hopping and interactions between the two sites A and B of the unit cell.

The inclusion of density-density interactions in $\Hop_{\textrm{int}}$ of Eq.~\eqref{eq:hamiltonianint} breaks the block diagonal structure of $\Hop_0$ in $\mathbf{k}$-space, as interactions represent scattering processes which couple modes with different $\mathbf{k}$s.
However, when working with the Hartree-Fock Hamiltonian $\Hop_\textrm{HF}$ described in the previous section, one can artificially impose a certain spatial periodicity of the Hartree-Fock parameters, and recover a block diagonal structure of $\Hop_\textrm{HF}$ in $\mathbf{k}$-space. 

\begin{figure}[b]
\centering
\includegraphics[width=.8\columnwidth]{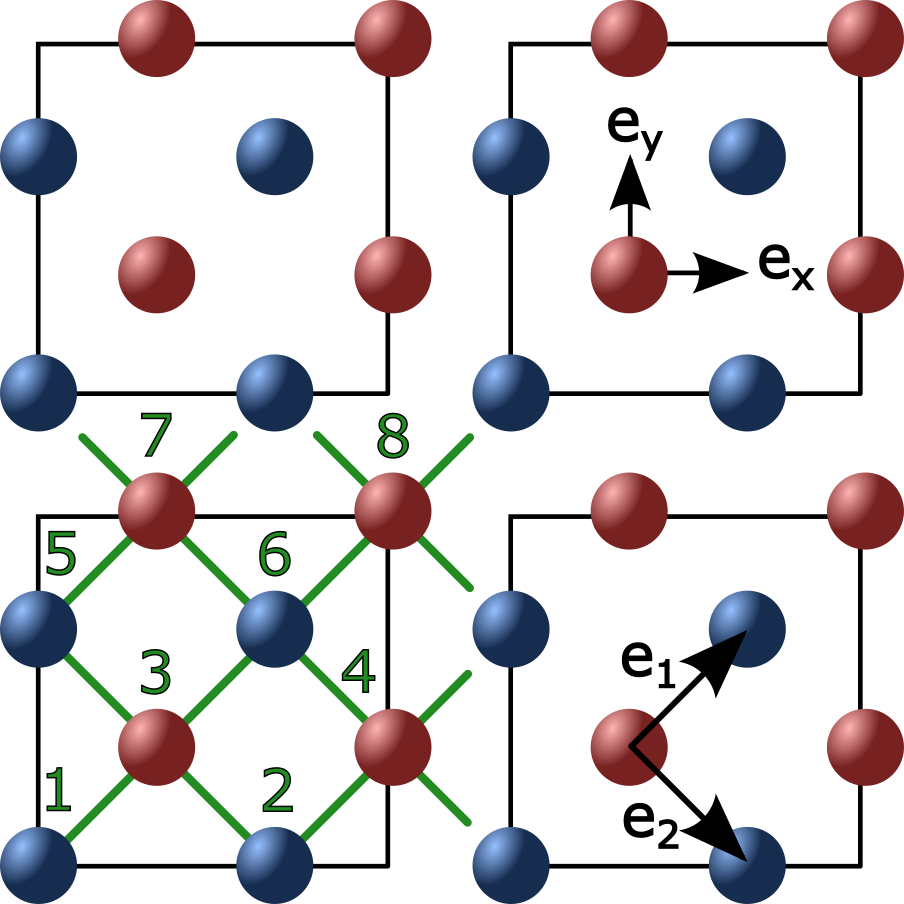}
\caption[A caption]{Eight-site cells. The square lattice unit vectors are $[\mathbf{e_1}, \mathbf{e_2}]$.
The eight-sites unit-cell Bravais vectors are $[4\mathbf{e_x}, 4\mathbf{e_y}]$.
The green segments identify the first-neighboring pairs, composed of inter- and intra-cells entries.
The sites labelling, arbitrary, helps to keep track of all the terms in the derivation of $\Hop_t$ in \textit{k}-space, presented in Eq.~\eqref{eq:HtFT}.}
\label{fig:12}
\end{figure}

This is known as the restricted Hartree-Fock method, and is typically used at particle fillings commensurate with the lattice size, where one expects that interactions preserve a certain translational symmetry.
In this method, it is important to do a proper choice of the unit cell size. A too small cell size may lead to constrictions: in $\mathbf{k}$-space, two-body energy terms acting on a distance larger than the extent of the cell become identical to existing shorter-range terms and renormalize them.
Also, charge orders with a periodicity on larger length scales can not be captured.
In particular, in order to resolve the charge density distribution associated to dominant $\tilde{V}_3$ interactions, shown in Fig.~\ref{fig:03_eightsites} of the main text, a four-site square cell is not sufficient.
Therefore, in this work, we consider the eight-site cell depicted in Fig.~\ref{fig:12}. In what follows, we use this unit cell to express $\Hop_{\textrm{HF}}$ in $\mathbf{k}$-space and also to derive the Hartree-Fock parameters.

The Fourier transform of the real-space annihilation operator is defined as
\begin{equation}
\hat{c}_{i} \to \hat{c}_{mn} = \frac{1}{\sqrt{N_{\textrm{uc}}}} \sum_{\kv \in \mathrm{FBZ}}\exp (\im \kv \cdot \mathbf{r}_{mn}) \hat{c}(\kv),
\end{equation}
where $N_{\textrm{uc}}$ is the total number of unit cells, in this case corresponding to the number of sites, and $\mathbf{r}_{ mn}$ is a linear combination of the Bravais vectors $[\mathbf{e}_{1}, \mathbf{e}_{2}]$ spanning over the lattice sites:
\begin{equation}
\mathbf{r}_{mn} = m\mathbf{e}_{1} + n\mathbf{e}_{2}, \quad \mathrm{with} \quad m,n \in \mathbb{Z}.
\end{equation}
In model~\eqref{eq:hamiltonianM4}, we include inter-site interactions up to fourth-neighbors, each one yielding a hopping term of the same neighboring order in Hartree-Fock approximation.
The eight-sites lattice cell in Figure~\ref{fig:12} accommodates all hopping terms distinctly.
The geometry of the cell correctly allows for a complete tiling of the lattice via $[\mathbf{a}_{1}, \mathbf{a}_{2}]$, the two unit-cell Bravais vectors:
\begin{equation}
\mathbf{a}_{1} = 4\mathbf{e}_{x}, \quad \mathbf{a}_{2} = 4\mathbf{e}_{y}.
\end{equation}
In the unit-cell framework, one can map the spatial coordinates $\mathbf{r}_{mn}$ onto a combined coordinate of the cell index $mn$ and an additional index $\alpha$ that identify the cell sites, as indicated by the enumeration in Fig.~\ref{fig:12}:
\begin{equation}
\mathbf{r}_{mn} \to \mathbf{R}_{\alpha mn} = m\mathbf{a}_{1} + n\mathbf{a}_{2} + \mathbf{r}_{\alpha}, \quad \mathrm{with} \quad m,n \in \mathbb{Z}.
\end{equation}
Ultimately, we can rewrite the Fourier transform as:
\begin{equation}
\hat{c}_{i} \to \hat{c}_{\alpha mn} = \frac{1}{\sqrt{N_{\textrm{uc}}}} \sum_{\kv \in \mathrm{FBZ}}\exp (\im \kv \cdot \mathbf{R}_{\alpha mn} ) \cop_{\alpha}(\kv).
\end{equation}

\subsubsection{Bare hopping t, first-neighbors}
\label{sec:app02_baret}

Let us now expressly illustrate, as an example, the mathematical steps yielding the nearest-neighbors hopping in \textit{k}-space; the procedure is identical for other elements in the Hamiltonian.
Expanding the kinetic energy term $\Hop_t = - t \sum_{{<i,j>}} \left( \cdi \cj + \hc \right)$, we have:
\begin{widetext}
\begin{align}
\label{eq:HtFT}
\Hop_t & = -t \sum_{{<\alpha mn,\beta m'n'>}} \left( \frac{1}{\sqrt{N_{\textrm{uc}}}} \sum_{\mathbf{p} \in \mathrm{FBZ}}\exp (- \im \mathbf{p} \cdot \mathbf{R}_{\alpha mn}) \cda(\mathbf{p})
\frac{1}{\sqrt{N_{\textrm{uc}}}} \sum_{\mathbf{q} \in \mathrm{FBZ}}\exp (\im \mathbf{q} \cdot \mathbf{R}_{\beta m'n'}) \cb(\mathbf{q})
+ \hc \right) \nonumber \\
& = -t \frac{1}{N_{\textrm{uc}}} \sum_{\mathbf{p},\mathbf{q} \in \mathrm{FBZ}} \sum_{{<\alpha, \beta >}} \sum_{mn} \left( \cda(\mathbf{p}) \cb(\mathbf{q}) e^{ \im (\mathbf{q}-\mathbf{p}) \cdot \mathbf{R}_{\textrm{1} mn} } e^{- \im \mathbf{q} \cdot \mathbf{r}_{\alpha\beta} } + \hc \right),
\end{align}
\end{widetext}
where $<,>$ symbolizes the restriction over all first-neighboring pairs and $\mathbf{r}_{\beta \alpha} = \mathbf{R}_{\beta mn} - \mathbf{R}_{\alpha mn}$ is the associated vector, also illustrated by the green segments in Fig.~\ref{fig:12}; here, $\mathbf{R}_{\textrm{1} mn}$ points at cell $mn$, whose origin coordinate we arbitrarily fixed to the bottom-left site $\alpha=1$.
The sum over \textit{mn} gives a Dirac delta function in $\mathbf{q}-\mathbf{p}$, which reduces the \textit{k}-dependency to $\mathbf{q}$ only:
\begin{equation}
\Hop_t = -t \sum_{\mathbf{q} \in \mathrm{FBZ}} \sum_{{<\alpha, \beta >}}
\left( \cda(\mathbf{q}) \cb(\mathbf{q}) e^{- \im \mathbf{q} \cdot \mathbf{r}_{\alpha\beta} } + \hc \right).
\end{equation}
Ultimately, dropping the explicit $\mathbf{q}$-dependency of the operators to ease readability and following the labelling in Fig.~\ref{fig:12}, we obtain
\begin{align}
\label{eq:Htk}
\Hop_t = - t
& \sum_{\mathbf{q} \in \mathrm{FBZ}} 
\Big[ e^{-\im \mathbf{q} \cdot \mathbf{e}_{1}} \Big( \cdop_{7} \cop_{5} 
+ \cdop_{2} \cop_{7} 
+ \cdop_{3} \cop_{1} 
+ \cdop_{6} \cop_{3}
+ ...
\Big) \nonumber \\
&+ e^{-\im \mathbf{q} \cdot \mathbf{e}_{2}} \Big( \cdop_{3} \cop_{5} 
+ \cdop_{2} \cop_{3} 
+ \cdop_{7} \cop_{1} 
+ \cdop_{6} \cop_{7}
+ ...
\Big) + \hc \Big]
\end{align}

\subsubsection{Expectation values, first-neighbors}

In real-space coordinates, the Hartree-Fock parameters $\xi_{ij}$ could be in principle different from one another.
Assuming translational invariance over the unit cells, the number of first-neighboring pairs reduces to the sixteen elements shown in Figure~\ref{fig:12}.
The expectation values are calculated as follows:
\begin{align}
& \xi_{\alpha\beta} =  \sum_{\mathrm{cells}} \expval{ \cdop_{\alpha} \cop_{\beta} } \nonumber \\
& = \frac{1}{N} \sum_{\mathbf{p},\mathbf{q} \in \mathrm{FBZ}} \sum_{mn} \expval{\cda(\mathbf{p}) \cb(\mathbf{q})} e^{ \im (\mathbf{q}-\mathbf{p}) \cdot \mathbf{R}_{\textrm{1} mn} } e^{- \im \mathbf{q} \cdot \mathbf{r}_{\alpha\beta} } \nonumber \\
& = \sum_{\mathbf{q} \in \mathrm{FBZ}} \expval{ \cdop_{\alpha}(\mathbf{q}) \cop_{\beta}(\mathbf{q})} e^{- \im \mathbf{q} \cdot  \mathbf{r}_{\alpha\beta}}.
\end{align}

\subsubsection{First-neighbors interaction}

Lastly, using Eq.~\eqref{eq:hfe}, we can derive the \textit{k}-space form of the first-neighbors density-density interaction, in the mean-field approximation.
The operator in real-space coordinate reads:
\begin{equation}
\Hop_{\Vgen_1} = \Vgen_1 \sum_{{<i,j>}} \bar{n}_i\nopj + \bar{n}_j\nopi - \bar{n}_i\bar{n}_j - \xi_{ij}\cdj\ci - \xi_{ij}^*\cdi\cj + |\xi_{ij}|^2.
\end{equation}
The mean-field approximation introduces constant and diagonal on-site terms and renormalizes the bare hopping.
After Fourier transformation, we obtain:
\begin{align*}
\Hop_t = - t \sum_{\mathbf{q} \in \mathrm{FBZ}} 
& e^{-\im \mathbf{q} \cdot \mathbf{e}_{1}} \Big[ \cdop_{7} \cop_{5} (1 + \Vgen_1 \xi_{57}) + ... \Big] + \nonumber \\
& e^{-\im \mathbf{q} \cdot \mathbf{e}_{2}} \Big[ \cdop_{3} \cop_{5} (1 + \Vgen_1 \xi_{53}) + ... \Big] + \hc
\end{align*}


\section{Quadratic band touching}
\label{sec:app01_qbt}

\begin{figure}
\centering
\includegraphics[width=\columnwidth]{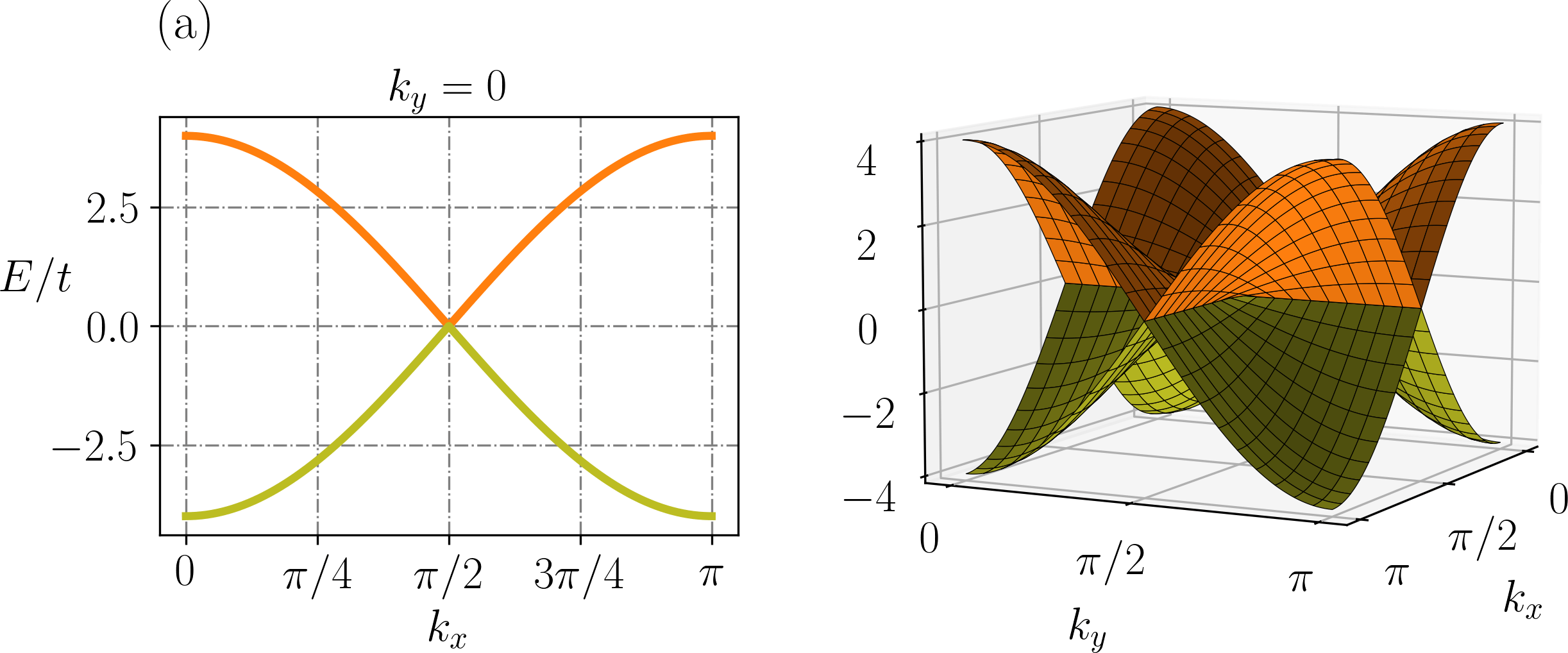}
\includegraphics[width=\columnwidth]{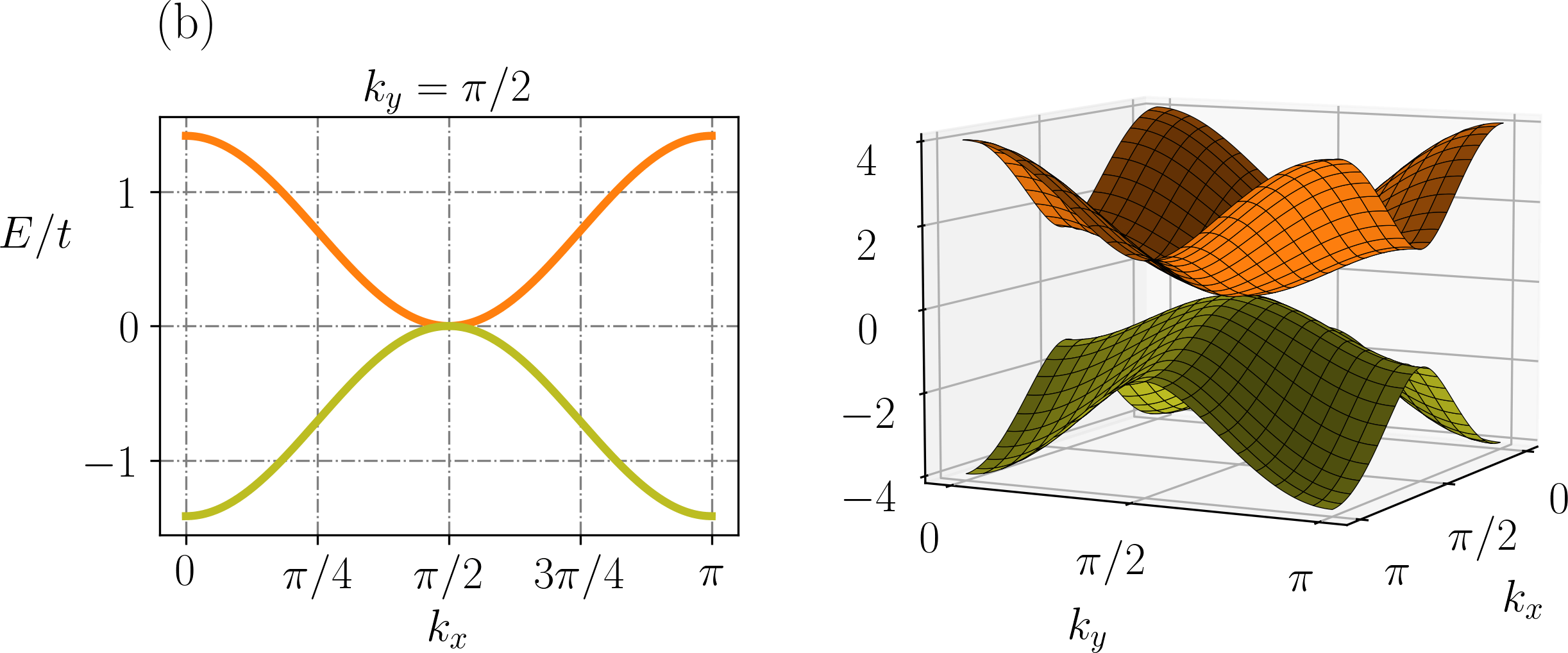}
\caption[A caption]{Single-particle band structure. (a) Valence and conduction bands of the free Fermi-Hubbard model with bare hopping $t$. Left panel: cut of the right panel at $k_y=0$, evidencing a linear dispersion at the touching point. (b) The quadratic band touching appears when a $\pi$-flux is introduced. Left panel: cut at $k_y=\pi/2$.}
\label{fig:11_band_touching}
\end{figure}

As discussed in Section~\ref{sec:tmiqbt}, numerical evidences~\cite{raghu_topological_2008,Sun2009,zhu_interaction-driven_2016,Sun2009,dauphin_rydberg-atom_2012,PhysRevA.93.043611,zeng_tuning_2018,Sur2018,PhysRevLett.117.066403,julia-farre_self-trapped_2020,Garcia-Martinez2013,Jia2013,Daghofer2014,Guo_2014,Motruk2015,capponi_phase_2015,Scherer2015,Sun2008TRS,Sun2009,Vafek2010,Dora2014} suggest that a topological Mott insulating phase emerges in the presence of a quadratic band touching.
The free Fermi-Hubbard model on a checkerboard lattice with only nearest-neighbors hopping does not possess this property in the band structure.
In fact, conduction and valence bands touch but present a linear dispersion; this can be changed by introducing a bipartite second-neighbors hopping on the two sub-lattices $A$ and $B$, $J_x^A/t=J_y^B/t=+0.5$ and $J_y^A/t=J_x^B/t=-0.5$, as shown in Fig.~\ref{fig:01_intro} in the main text.
This specific design of the hopping introduces a so-called $\pi$-flux on square second-neighbors plaquettes of both sub-lattices, which gives the quadratic band touching.
We can see this analytically by studying the model Hamiltonian $\Hop_0$ [Eq.~\eqref{eq:hamiltonianfree}] on a two-site unit cell.
In this case, one can show that the dispersion relation reads
\begin{align}
    &E_{k_x,k_y} = -[ S_x\cos(2k_x) + S_y\cos(2k_y)] \\[2mm]
    &\pm \sqrt{[D_x \cos(2k_x) + D_y\cos(2k_y)]^2 + 16t^2\cos(k_x)^2\cos(k_y)^2} \nonumber
\end{align}
where $S_x = J^A_x + J^B_x$, $D_x = J^A_x - J^B_x$ and likewise for $S_y,\ D_y$, and where $k_{x,y}=\mathbf{k}\cdot\mathbf{e_{1,2}}$.
For $J^A_{\mu}=J^B_{\mu}=0$, $S_{\mu}=D_{\mu}=0$ [see Fig.~\ref{fig:11_band_touching}(a)] the system reverts to the bare Fermi-Hubbard gas with a linear dispersion, around the points where conduction and valence bands touch:
\begin{equation}
    E_{k_x,k_y} = \pm 4t \cos(k_x) \cos(k_y).
\end{equation}
The particular case $J^A_x=J^B_y=0.5t=-J^A_y=-J^B_x$ considered in the article, gives $S_{\mu}=0$, $D_{\mu}=\pm t$:
\begin{align}
    &E_{k_x,k_y} = \\
    &= \pm t \sqrt{[\cos(2k_x) - \cos(2k_y)]^2 + 16 \cos(k_x)^2\cos(k_y)^2}. \nonumber
\end{align}
In this case, the bands touch at a zero energy point in $k_x,k_y=\pm\pi/2$.
The quadratic behaviour, shown in Fig.~\ref{fig:11_band_touching}(b), is easily derived, for instance fixing $k_y=\pi/2$:
\begin{equation}
    E_{k_x,k_y=\pi/2} \simeq \pm t \sqrt{2} k_x^2.
\end{equation}

\section{Van Vleck's perturbation theory for effective Rydberg potential}
\label{sec:app03_vanvleck}

Here we outline the derivation of the effective Rydberg potential, Eq.~\eqref{eq:vv4th}.
Given a Hamiltonian $\Hop =\Hzero+\beta \Hop_I$, where $\Hzero$ is diagonal, $\beta \Hop_I$ is the interaction and $\beta \ll 1$, the Van Vleck's formalism~\cite{shavitt_quasidegenerate_1980,dauphin_rydberg-atom_2012} provides an analytical expression of the Hamiltonian matrix in a block-diagonal form up to a desired order in the parameter $\beta$.
The method requires a bipartition of the Hilbert space onto a model space $\mathcal{P}$ and its orthogonal complement $\mathcal{Q}$.
The model space includes the subset of $\Hzero$ eigenstates of which one wants to study the interaction-induced hybridization and obtain the energy level corrections.
We define the associated projection operators
\begin{equation}
\hP = \sum_{p\in\mathcal{P}} \dyad{p}, \quad \hQ=\sum_{q\in\mathcal{Q}} \dyad{q}
\end{equation}
and use them to split operators, in particular the Hamiltonian, onto a block-diagonal part,
\begin{equation}
\Hop^D = \hP \Hop \hP + \hQ \Hop \hQ = \Hzero + \beta \Hop_I^D,
\end{equation}
and a block-anti-diagonal term,
\begin{equation}
\Hop^X = \hP \Hop \hQ + \hQ \Hop \hP = \beta \Hop_I^X.
\end{equation}
The task is to find the decoupling operator $\hU$ that block-diagonalizes the Hamiltonian: $\Hvv = \hU^{-1} \Hop \hU$.
The specificity of the Van Vleck's method, among other quasi-degenerate perturbation theories, consists in assuming that $\hU=e^{\hG}$, where $\hG \equiv \hG^X$ is a block-anti-diagonal anti-Hermitian matrix.
A series expansion of $\hG$ in increasing powers of $\beta$ can be obtained recursively:
\begin{align}
\label{eq:gmats}
\Gz = 0, \nonumber \\
\Gu = \hR & \Hop_I^X, \nonumber \\
\Gd = \hR & \left[\Hop_I^D,\Gu \right], \nonumber \\
\Gt = \hR & \left( \left[\Hop_I^D,\Gd \right] + \frac{1}{3} \left[\left[ \Hop_I^X,\Gu \right],\Gu \right] \right), \nonumber \\
\Gq = \hR & \Big( \left[\Hop_I^D,\Gt\right] + \frac{1}{3} \left[\left[ \Hop_I^X,\Gu \right], \Gd \right] \nonumber \\
& \quad + \frac{1}{3} \left[\left[\Hop_I^X, \Gd \right] ,\Gu\right] \Big).
\end{align}
Here $\hR$ is the resolvent operator,
\begin{equation}
\hR_{p \in \mathcal{P}} = \sum_{q}^{\mathcal{Q}}  \frac{\dyad{q}}{\varepsilon_{p}-\varepsilon_{q}},
\quad
\hR_{q \in \mathcal{Q}} = \sum_{p}^{\mathcal{P}}  \frac{\dyad{p}}{\varepsilon_{q}-\varepsilon_{p}},
\end{equation}
and $\varepsilon_{i}$ are the eigenenergies of $\Hzero$.
The equivalences in Eq.~\eqref{eq:gmats} are obtained using the rule
\begin{equation}
\hU^X \ket{\alpha} = -\hR_{\alpha} [\Hzero,\hU^X] \ket{\alpha},
\end{equation}
which also applies to $\hG$ thanks to its anti-diagonal form.
Finally, one can rearrange the resulting $\Hvv$ terms in a convenient form, $\Hvv = \Hzero + \sum_i \beta^i W^{(i)}$, where
\begin{align}
\label{eq:vvw}
&W^{(1)} = \Hop_I^D, \\
&W^{(2)} = \frac{1}{2} \left[\Hop_I^X,\Gu \right], \nonumber \\
&W^{(3)} = \frac{1}{2} \left[\Hop_I^X,\Gd \right], \nonumber \\
&W^{(4)} = \frac{1}{2} \left[\Hop_I^X,\Gt \right] - \frac{1}{24} \left[ \left[ \left[\Hop_I^X,\Gu \right],\Gu \right],\Gu \right]. \nonumber
\end{align}
Ultimately, the hybridized states of subspace $\mathcal{P}$ are found by diagonalizing the $\mathcal{P}$-block of $\Hvv$.

In this work, we consider a many-body system of cold atoms dressed to a Rydberg state.
The single-particle physics is well captured by the Hamiltonian $\Hop_{\mathrm{c}} = (\Omega \ket{r}\bra{g} + \hc) + \Delta \ket{r}\bra{r}$.
The Hamiltonian for a two laser-driven atoms pair in the basis $[ \ket{gg}, \ket{rg}, \ket{gr}, \ket{rr} ]$ is:
\begin{equation}
\Hop = 
\begin{pmatrix}
0 & \Omega & \Omega & 0 \\
\Omega & \Delta & 0 & \Omega \\
\Omega & 0 & \Delta & \Omega \\
0 & \Omega & \Omega & 2\Delta + \uvwr \\
\end{pmatrix}.
\end{equation}
The regime of far off-resonance is defined in the limit of frequency detuning much larger than the Rabi frequency, $\Omega / \Delta \ll 1$.
In this regime, the off-diagonal part of the Hamiltonian becomes perturbatively small.
If we rewrite the Hamiltonian matrix in energy units of $\Delta$ and in terms of the perturbation parameter $\beta = \Omega / \Delta$, we obtain:
\begin{equation*}
 \frac{\Hzero}{\Delta} = 
\left(
\begin{array}{cccc}
0 & 0 & 0 & 0 \\
0 & 1 & 0 & 0 \\
0 & 0 & 1 & 0 \\
0 & 0 & 0 & 2+\frac{\uvwr}{\Delta} \\
\end{array}\right),
\quad
 \frac{\Hop_I}{\Delta} = \beta
\left(
\begin{array}{cccc}
0 & 1 & 1 & 0 \\
1 & 0 & 0 & 1 \\
1 & 0 & 0 & 1 \\
0 & 1 & 1 & 0 \\
\end{array}
\right).
\end{equation*}
At this point, we can apply van Vleck's method to derive the corrections to the eigen-energies of the unperturbed system, defining the spaces $\mathcal{P}$ and $\mathcal{Q}$.
Since we are interested in obtaining the corrections to eigen-energy $E_{\ket{gg}}$ of the two-atom ground-state $\ket{gg}$, we define $\mathcal{P} : [\ket{gg}] \quad \mathcal{Q} : [\ket{rg},\ket{gr},\ket{rr}]$, and split accordingly the Hamiltonian onto its block-diagonal and block-anti-diagonal parts:
\begin{equation*}
 \frac{\Hop^D}{\Delta} = 
\left(
\begin{array}{cccc}
0 & 0 & 0 & 0 \\
0 & 1 & 0 & 1 \\
0 & 0 & 1 & 1 \\
0 & 1 & 1 & 2+\frac{\uvwr}{\Delta} \\
\end{array}\right),
\quad
 \frac{\Hop^X}{\Delta} = \beta
\left(
\begin{array}{cccc}
0 & 1 & 1 & 0 \\
1 & 0 & 0 & 0 \\
1 & 0 & 0 & 0 \\
0 & 0 & 0 & 0 \\
\end{array}
\right).
\end{equation*}
Applying the formalism introduced above, we can derive the corrections to $E_{\ket{gg}}$ to fourth order in $\beta = \Omega/\Delta$:
\begin{equation}
    \frac{V(r)}{\Delta} = -2\frac{\Omega^2}{\Delta^2} + 2\frac{\Omega^4}{\Delta^4} + 2\frac{\Omega^4}{\Delta^4} \frac{\uvwr}{\uvwr+2\Delta}.
\end{equation}
The spatial dependence is given by the last term.

\end{document}